\documentclass[mathleft
]{an}
\usepackage{graphicx}
\usepackage{times}
\begin{document}

\Pagespan{1}{}
\Yearpublication{0000}%
\Yearsubmission{0000}%
\Month{00}%
\Volume{0000}%
\Issue{00}%

\title{Finding the most variable stars in the Orion Belt
with the All Sky Automated Survey}

\author{Jos\'e A. Caballero\thanks{Corresponding author:
  \email{\tt caballero@astrax.fis.ucm.es}\newline}
\and M. Cornide
\and E. de Castro
}
\titlerunning{The most variable ASAS stars in the Orion Belt}
\authorrunning{Caballero et~al.}
\institute{
Departamento de Astrof\'{\i}sica y Ciencias de la Atm\'osfera, Facultad de
F\'{\i}sica, Universidad Complutense de Madrid, E-28040 Madrid, Spain}

\received{05 Sep 2009}
\accepted{04 Jan 2010}
\publonline{later}


\abstract{%
We look for high-amplitude variable young stars in the open clusters and
associations of the Orion Belt.  
We use public data from the ASAS-3 Photometric $V$-band Catalogue of the All
Sky Automated Survey, infrared photometry from the 2MASS and {\em IRAS}
catalogues, proper motions, and the Aladin sky atlas to obtain a list of the
most variable stars in a survey area of side 5\,deg centred on the bright
star Alnilam ($\epsilon$~Ori) in the centre of the Orion Belt.
We identify 32 highly-variable stars, of which 16 had not been reported
to vary before.
They are mostly variable young stars and candidates (16) and background giants
(8), but there are also field cataclysmic variables, contact binaries, and
eclipsing binary candidates.
Of the young stars, which typically are active Herbig Ae/Be and T~Tauri stars
with H$\alpha$ emission and infrared flux excess, we discover four new variables
and confirm the variability status of another two.
Some of them belong to the well known $\sigma$~Orionis cluster.
Besides, six of the eight giants are new variables, and three are new periodic
variables.}

\maketitle

\section{Introduction}

The \object{Ori~OB1\,b} subgroup association, mostly coincident 
with the asterism
of the Orion Belt, contains a rich population of very young stars (e.g. Haro \&
Moreno 1953; Walker 1969; Warren \& Hesser 1978; Brown et~al. 1994; Sterzik
et~al. 1995; de~Zeeuw et~al. 1999; Brice\~no et~al. 2001).  
The youngest regions, including \object{Alnitak} ($\zeta$~Ori), which
illuminates the \object{Flame Nebula} (NGC~2024), and \object{$\sigma$~Ori},
which illuminates the Horsehead Nebula (\object{Barnard~33} and 
\object{IC~434}), are to the east of the Belt and have ages of 1 to 5\,Ma.
Here lies the \object{$\sigma$~Orionis} cluster, one of the most suitable
sites for discovering and characterising substellar objects (B\'ejar et~al.
1999; Zapatero Osorio et~al. 2002; Sherry et~al. 2004; Caballero et~al. 2007).
The population of stars surrounding \object{Alnilam} ($\epsilon$~Ori) and
\object{Mintaka} ($\delta$~Ori), to the west of the Belt, are older (5--10\,Ma)
and spread over a wider area than the much denser {$\sigma$~Orionis cluster}
(Caballero \& Solano 2008).
Surrounding the Belt, there is another dispersed population of 10--20\,Ma-old
stars, mostly associated to the \object{Ori~OB1\,a} subgroup association
(Jeffries et~al. 2006; Caballero 2007) and its agglomerates, such as the
\object{25~Orionis} group or the \object{$\eta$~Orionis} overdensity (Brice\~no
et~al. 2007; Caballero \& Dinis~2008). 

H$\alpha$ and X-ray emission and infrared excess are the primary features for
the identification of very young stars as those in the Orion Belt. 
Optical and near-infrared variability, cosmic lithium abundances, surrounding
jets, or forbidden emission lines are other signposts used to confirm very
young ages in stars.
Photometric variability in young stars is mostly due to dark (cool) and bright
(hot)  spots
and flare activity, which are related to the strengthened magnetic activy due to
fast rotation, and interactions with circumstellar discs, such as intense,
variable accretion episodes or occultations (e.g. classical works: Joy 1945;
Walker 1956; Herbig 1977 -- theoretical works: Koenigl 1991; Collier Comeron \&
Campbell 1993; Shu et~al. 1994 -- modern observational works: Bouvier et~al.
1993; Herbst et~al. 1994; Th\'e et~al. 1994; Hartmann \& Kenyon 1996; Hamilton
et~al. 2001; Eiroa et~al. 2002).

The \object{Orion Nebula Cluster} in the \object{Ori~OB1\,d} subgroup association
has been a traditional target for studying the photometric variability of young
stars (Jones \& Walker 1988; Carpenter et~al. 2001; Herbst et~al. 2002; Stassun
et~al. 2006; Parihar et~al. 2009), but the surveys for variability in the Orion
Belt are relatively scarce.                         
The most thorough variability surveys in the Ori~OB1~b association have been
aimed at the low-mass pre-main sequence stars and brown dwarfs (Bailer-Jones \&
Mundt 2001; Zapatero Osorio et~al. 2003; Caballero et~al. 2004, 2006; Scholz \&
Eisl\"offel 2004, 2005; Brice\~no et~al. 2005; Scholz et~al. 2009), while the
brightest variable stars were first recognised as early as the beginning of the
20th century and have not received quite attention.
Only a few (multiple) young stars of important astrophysical interest had been
extensively monitored, such as \object{VV~Ori} (Miller Barr 1904; Struve \&
Luyten 1949; Terrell et~al. 2007), \object{$\sigma$~Ori~E} (Walborn \& Hesser
1976; Landstreet \& Borra 1978), or Mintaka~AE--D (Stebbins 1915; Koch
\& Hrivnak 1981).
Many of the intermediate-mass T~Tauri stars in the Orion Belt with variability
denominations (e.g. \object{BG~Ori}, \object{V505~Ori}) were compiled by 
Fedorovich (1960) and have never been monitored again. 
After a century of photometric searches in the region, there are still
serendipitous discoveries of bright, UX~Orionis-, A-type, variable stars
(Caballero et~al. 2008). 

Our aim is to find the most variable stars in the Orion Belt, giving especial
emphasis to the young Herbig~Ae/Be (HAeBe) and T~Tauri stars.
To accomplish that, we have used the ASAS-3 Photometric $V$-band Catalogue of
the All Sky Automated Survey (ASAS)\footnote{\tt
http://www.astrouw.edu.pl/asas/.}, which is ``a low cost project dedicated to
constant photometric monitoring of the whole available sky'' (Pojma\'nski 
1997, 2002).

\section{Analysis}
\label{analysis}

   \begin{table*}
      \caption[]{Identified variables in the Orion Belt.} 
         \label{table.variables}
         \begin{tabular}{l cc cccc l l l l}
            \hline
            \hline
            \noalign{\smallskip}
No. 	& $\alpha^{\rm 2MASS}$	& $\delta^{\rm 2MASS}$	& $\overline{V_0}$	& $\sigma(V_0)$	& $\overline{\delta V_0}$	& $N^\star_{\rm obs}$	& Spectral & Name  			& Class		& Variable	\\
    	& (J2000)		& (J2000)		& [mag]			& [mag] 	& [mag] 			& 		& type		&				&		&		\\
            \noalign{\smallskip}
            \hline
            \noalign{\smallskip}																								   
01 	& 05 28 59.57 		& --03 33 52.3 		& 13.621		& 0.412 	& 0.055 			& 190 		& ...		& \object{V1159 Ori}		& Dwarf nova 	& Known		\\ 
03 	& 05 31 01.46 		& --03 23 24.3 		& 12.492		& 0.341 	& 0.056 			& 294 		& ...		& \object{IRAS 05285--0325}	& Giant		& New 		\\ 
06 	& 05 33 22.45 		& --03 09 55.7 		& 11.185		& 0.085 	& 0.032 			& 207 		& ...		& No. 06			& Unknown	& New 		\\ 
07 	& 05 37 48.79 		& --03 07 48.7 		& 10.780		& 0.101 	& 0.021 			& 286 		& ...		& \object{IRAS 05353--0309}	& Giant		& New 		\\ 
11 	& 05 32 09.94 		& --02 49 46.8 		& 11.581		& 0.524 	& 0.032 			& 191 		& F5--8Vpe	& \object{RY Ori}		& HAeBe		& Known		\\ 
12 	& 05 38 33.68 		& --02 44 14.2 		& 12.885		& 0.368 	& 0.054 			& 187 		& K4e		& \object{TX Ori}		& T Tauri	& Known		\\ 
13 	& 05 38 25.87 		& --02 43 51.2 		& 13.059		& 0.356 	& 0.054 			& 142 		& K3e		& \object{TY Ori}		& T Tauri	& Known		\\ 
14 	& 05 43 54.26 		& --02 43 35.4 		& 11.628		& 0.135 	& 0.032 			& 222 		& ...		& \object{2M054354--0243.6}	& Contact binary& Known 	\\ 
15 	& 05 39 39.99 		& --02 43 09.7 		& 13.246		& 0.246 	& 0.054 			& 197 		& ...		& \object{RW Ori}		& T Tauri	& Known		\\ 
18 	& 05 38 48.04 		& --02 27 14.2 		& 13.340		& 0.295 	& 0.053 			& 224 		& K7e+M:	& \object{Mayrit 528005 AB}	& T Tauri	& New		\\ 
21 	& 05 36 20.91 		& --02 10 57.6 		& 12.607		& 0.192 	& 0.056 			& 252 		& F3e		& \object{PQ Ori}		& HAeBe		& Known		\\ 
29 	& 05 37 38.67 		& --01 46 16.6 		& 12.752		& 1.441 	& 0.054 			& 179 		& M8--9III:	& \object{X Ori}		& Giant		& Known 	\\ 
30 	& 05 37 56.59 		& --01 40 50.3 		& 10.734		& 0.250 	& 0.021 			& 273 		& ...		& \object{IRAS 05354--0142}	& Giant		& New 		\\ 
33 	& 05 30 02.89 		& --01 30 05.3 		& 12.786		& 0.263 	& 0.056 			& 176 		& ...		& No. 33			& Unknown	& New		\\ 
36 	& 05 34 25.82 		& --01 21 06.5 		& 12.667		& 0.325 	& 0.055 			& 170 		& ...		& \object{V469 Ori}		& T Tauri	& Known		\\ 
41 	& 05 26 53.53 		& --01 09 02.3 		& 13.478		& 0.480 	& 0.057 			& 144 		& ...		& \object{Kiso A--0903 135}	& T Tauri?	& New		\\ 
43 	& 05 37 59.04 		& --01 05 52.8 		& 13.554		& 0.371 	& 0.054 			& 200 		& ...		& \object{V993 Ori}		& T Tauri?	& Known 	\\ 
46 	& 05 31 26.41 		& --00 58 33.1 		& 10.454		& 0.076 	& 0.021 			& 192 		& F0		& \object{HD 290509}		& Unknown	& New		\\ 
48 	& 05 39 45.65 		& --00 55 50.9 		& 10.934		& 0.075 	& 0.021 			& 203 		& ...		& \object{GSC 04767--00071}	& T Tauri?	& Known 	\\ 
51 	& 05 36 24.29 		& --00 42 12.0 		& 13.399		& 0.388 	& 0.056 			& 227 		& K3e		& \object{PU Ori}		& T Tauri	& Known		\\ 
53 	& 05 34 48.91 		& --00 37 16.7 		& 13.384		& 0.298 	& 0.056 			& 178 		& ...		& \object{V472 Ori}		& T Tauri?	& Known		\\ 
54 	& 05 27 24.67 		& --00 35 11.2 		& 11.031		& 0.134 	& 0.033 			& 184 		& ...		& \object{IRAS 05248--0037}	& Giant		& New 		\\ 
55 	& 05 35 43.27 		& --00 34 36.7 		& 12.411		& 0.278 	& 0.057 			& 246 		& K4e		& \object{StHA 48}		& T Tauri	& New		\\ 
60 	& 05 26 34.28 		& --00 19 27.2 		& 12.353		& 0.137 	& 0.056 			& 181 		& ...		& No. 60			& Giant		& New 		\\ 
65 	& 05 36 28.55 		&  +00 04 45.7 		& 11.987		& 0.131 	& 0.056 			& 257 		& ...		& No. 65			& Unknown	& New 		\\ 
66 	& 05 43 29.25 		&  +00 04 58.9 		& 11.459		& 0.513 	& 0.032 			& 202 		& F0?		& \object{GT Ori}		& Giant		& Known 	\\ 
70 	& 05 44 18.80 		&  +00 08 40.4 		&  8.906		& 0.061 	& 0.019 			& 204 		& A7IIIe	& \object{V351 Ori}		& HAeBe		& Known		\\ 
72 	& 05 46 11.86 		&  +00 32 25.9 		& 13.390		& 0.439 	& 0.053 			& 167 		& ...		& \object{VSS VI--32}		& Unknown	& New		\\ 
73 	& 05 41 59.81 		&  +00 35 27.1 		& 12.698		& 0.435 	& 0.053 			& 185 		& ...		& No. 73			& Unknown	& New 		\\ 
74 	& 05 36 42.64 		&  +00 38 34.3 		& 11.010		& 0.101 	& 0.033 			& 254 		& A2		& \object{HD 290625}		& HAeBe		& New		\\ 
75 	& 05 33 47.85 		&  +00 55 36.3 		& 10.433		& 0.166 	& 0.021 			& 181 		& ...		& \object{IRAS 05312+0053}	& Giant		& New 		\\ 
78 	& 05 28 17.85 		&  +01 10 06.1 		& 11.219		& 0.165 	& 0.033 			& 173 		& F8--K0e	& \object{StHA 40}		& T Tauri	& Known		\\ 
            \hline
         \end{tabular}
   \end{table*}
%

\begin{figure}
\centering
\includegraphics[width=0.48\textwidth]{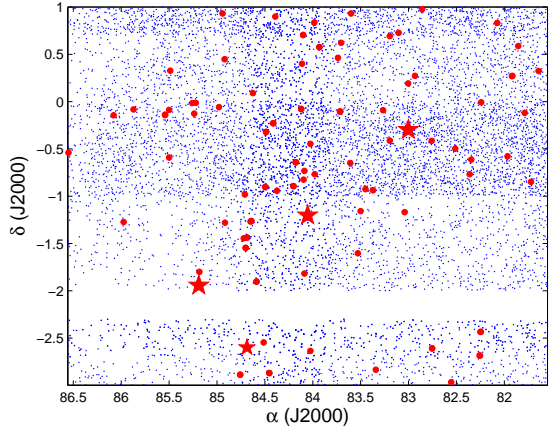}
\caption{Spatial location of the 10\,029 ASAS sources in the studied area with
light curves with more than 40 data points, marked with (blue) small dots.
North is up, east is to the left.
The 100 ASAS sources with possible variable light curves, the three Orion Belt
supergiant stars (Alnitak, Alnilam, and Mintaka, from east to west), and
$\sigma$~Ori are marked with (red) filled circles, big stars, and a small
star, respectively.
Note the blank strip where ASAS did not survey (between Alnitak and
$\sigma$~Orionis' declinations) and the overlapping between different ASAS
pointings (at about the Mintaka's declination).}
\label{AN1946fig1}
\end{figure}

First, we compiled 63\,276 ASAS light curves of objects in a square of side
5.0\,deg centred on Alnilam; see Fig.~\ref{AN1946fig1}.
We used the default parameters $N >$ 4 data points and $r <$ 15\,arcsec ($N$ is
the number of data points within a circle of radius $r$ centred on the
coordinates of an ASAS source).
If not for certain blank strip, the surveyed area would have been 25\,deg$^2$.
To maximise the scientific return and minimise possible further troubles with
light curves of poor quality, we discarded all ASAS light curves with $N \le$ 40
data points.
It left 10\,029 light curves (16\,\%) with $N >$ 40 for next phases of the
analysis. 
The ASAS data covered almost six years during the first decade of the 21st
century.

Of the five available apertures (0, 1, 2, 3, 4), the ``0'' aperture photometry
seemed to provide better data (i.e. less dispersion and number of grade ``C''
and ``D'' data points -- probably useless data or without photometric value
at all due to bad pixels, frame-edge and bright star/planet proximity, or
excessive faintness). 
According to Pojma\'nski (2002), the five apertures correspond to diameter sizes
of the aperture photometry from 2 to 6\,pixels; the ``0'' aperture
corresponds to the narrower photometry (Pojma\'nski, priv. comm.). 
From this point on, we used only the Johnson $V$ magnitude with the ASAS ``0''
aperture photometry, which we indicate with the symbol $V_0$.

\begin{figure}
\centering
\includegraphics[width=0.48\textwidth]{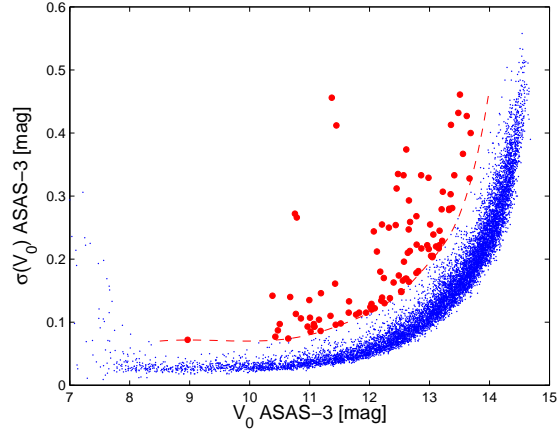}
\caption{Diagram showing $\sigma(V_0)$ vs. $\overline{V_0}$.
ASAS sources above the dashed line are variable star candidates in this work.
X~Ori is out of limits ($\overline{V_0} \sim$ 12.7\,mag, $\sigma(V_0) \sim$
1.4\,mag).} 
\label{AN1946fig2}
\end{figure}

Fig.~\ref{AN1946fig2} illustrates the selection of variable stars.
It is a $\sigma(V_0)$ vs. $\overline{V_0}$ diagram in which the variable star
candidates have larger values of standard deviation $\sigma(V_0)$ with respect
to non-variable stars of the same average magnitude $\overline{V_0}$.
ASAS stars brighter than $\overline{V_0} \approx$ 8.5\,mag are affected by
saturation in the detector and were not considered.
The lower envelope at the non-saturated, brightest end of the cloud of data
points in the diagram indicates the best photometric accuracy of the system,
which is of about 0.02\,mag. 
The obvious rise of the lower envelope at faint magnitudes is a combination of
the standard deviations (root-mean-squares) from Poisson noise in the targets
and from sky noise in the photometric aperture (e.g. Irwin et~al. 2007).
The line used as the selection criterion is the lower envelope shifted to larger
$\sigma(V_0)$s by magnitude-dependent amounts (roughly 0.06\,mag at
$\overline{V_0} \sim$ 8.5--11.5\,mag, 0.12\,mag at $\overline{V_0} \sim$
12.5\,mag, and 0.18\,mag at $\overline{V_0} \sim$ 13.5\,mag).
This choice was relatively arbitrary: more or less restrictive selection
criteria would have given significantly less or more variable star candidates,
respectively. 
Thus, the used selection criterion was a compromise to minimise the number of
false positives (variable star candidates that actually do not vary) and
maximise the number of actual variable stars, but always keeping a maneagable
number of objects to be followed up.

There were 100 ASAS light curves above the selection criterion in
Fig.~\ref{AN1946fig2}.
We were able to identify the 2MASS (Skrutskie et~al. 2006) and USNO-B1 (Monet
et~al. 2003) possible stellar counterparts of 99 of them. 
The unidenfied ASAS source was 054200--0154.7 ($V_0$ = 13.167\,mag,
$\sigma(V_0)$ = 0.245\,mag, N$_{\rm obs}$ = 49), which is actually a bright spot
in NGC~2024.
For the visual identification, we used the Aladin interactive sky atlas
(Bonnarel et~al. 2000).
The actual number of stars with light curves was 78, because there were 21
variable star candidates with two corresponding light curves. 
The separated datasets (with identical associated coordinates and magnitudes
within error bars) correspond to the same star in different observed fields
(Pojma\'nski 2002). 

We revisited the ASAS data for the 78 stars by retrieving all data points with
quality grades ``A'' and ``B''  in a circle of radius $r$ = 15\,arcsec centred
on the stars, and merging them into a unique light curve. 
``A'' and ``B''quality grades, which are tabulated by the ASAS catalogue
and derived from the average photometric quality of the frame for each
aperture, correspond to the best and mean data, respectively\footnote{The
quality grades ``A'' and ``B'' of a new ASAS frame depends on the average
dispersion between `old' and `new' magnitudes for stars in the range 8--11\,mag 
(Pojma\'nski, priv. comm.).}. 
Because of the large ASAS pixel size (of about 15\,$\mu$m) and typical full
width at half maximum, of no less than 1.4\,pixels (Pojma\'nski 2002), we
discarded all visual binaries of roughly the same brightness separated by less
than about 20\,arcsec. 
Besides, we also discarded relatively faint variable candidates at less than
1\,arcmin to very bright stars, including Alnilam, Mintaka, and $\sigma$~Ori,
which are seen by the naked eye and lead to a high sky background. 
After these removals, we kept 32 variable stars whose light curves were likely
not contaminated by other stars.
The 32 light curves are shown in Figs.~\ref{lightsaber.01-14}
to~\ref{lightsaber.65-78}. 

The identification number, 2MASS coordinates, average $\overline{V_0}$
magnitude, standard deviation $\sigma(V_0)$, mean error of magnitude
$\overline{\delta V_0}$, and number of finally used light curve data points,
$N^\star_{\rm obs}$, of the 32 identified variable stars are listed in
Table~\ref{table.variables}. 
The ratios between the standard deviations and mean errors vary between
$\sim$2.5 and 27; the expected ratio for a non-variable star should be lower
than 2 (e.g. Bailer-Jones \& Mundt 2001).
In Table~\ref{table.variables}, we also provide the spectral type from the
literature (if it has ever been given), name, variability class (see
Section~\ref{section.results}), and if the variability status was known or not.
For those stars that have never been mentioned in the literature, we use the
nomenclature ``No. {\em X}'', where {\em X} is the identification number.

For a typical number of light curve data points of $N^\star_{\rm obs} \sim$ 200
and a temporal coverage of almost six years, one may derive that there was an
ASAS visit every $\sim$11\,d for each of the 32 stars.
Actually, the typical time interval between consecutive ASAS visits was shorter
because there were long non-observing gaps without data.
The median time interval between consecutive ASAS visits was 2.0--3.0\,d,
depending on the star. 
Approximately 15--20, 20--33, 65--85, and 85--95\,\% of the time intervals
were shorter than 1, 2, 5, and 10\,d, respectively, and only in four cases,
coincident with the four Austral summers, the time intervals were longer than
100\,d. 

\begin{figure}
\centering
\includegraphics[width=0.48\textwidth]{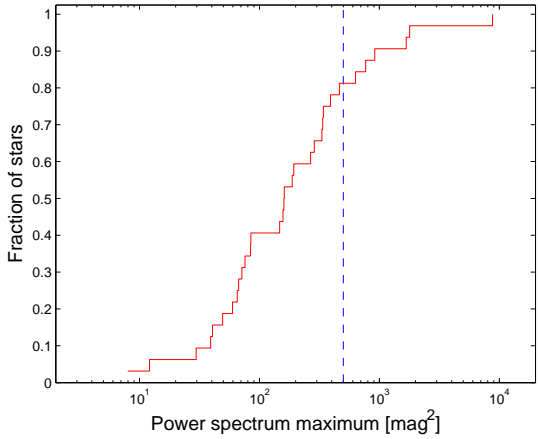}
\caption{Cumulative distribution function of power spectrum maxima of the light
curves of the 32 identified variables in the Orion Belt.
Stars to the right of the dashed vertical line at power spectrum of 500\,mag$^2$
are classified here as probable periodic variable stars.}
\label{AN1946fig3}
\end{figure}

We performed a rutinary time-series analysis on the 32 light curves by applying
the Scargle (1982) procedure.
Fig.~\ref{AN1946fig3} shows the cumulative distribution function of the
measured maxima of power spectrum.
Roughly 85\,\% of the stars have power spectrum maxima lower than 500\,mag$^2$,
where there seems to be an inflection point in the cumulative distribution
function.
The other 15\,\% (6) of the stars with larger power spectrum maxima are given in
Table~\ref{table.powerspectrum} and were classified by us as probable periodic
variable stars. 
We provide periods for two previously known periodic variable stars
(V1159~Ori and X~Ori) consistent with those in the literature, and found
probable periods for two previously unknwon variable giant stars (with {\em
IRAS} denomination), but failed to find any reliable periodicty for RY~Ori and
GT~Ori.
The six of them will be studied in detail in Section~\ref{section.results}.

\begin{figure}
\centering
\includegraphics[width=0.48\textwidth]{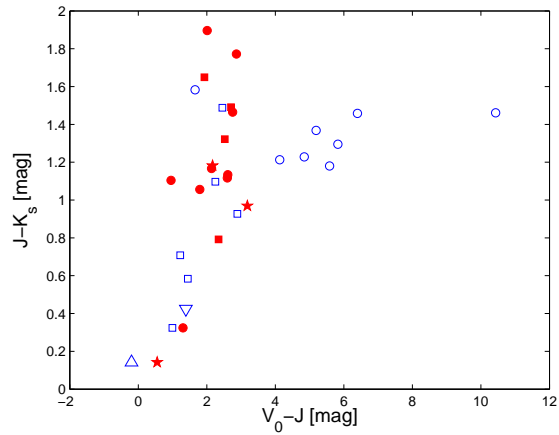}
\caption{Optical-infrared colour-colour diagram ($J-K_{\rm s}$ vs. $V_0-J$).
Filled (red) stars: new variable young stars;
filled (red) circles: previously known variable young stars;
filled (red) squares: variable young star candidates;
open (blue) circles: variable giants;
open (blue) up-triangle: dwarf nova (V1159~Ori);
open (blue) down-triangle: contact binary (2M054354--0243.6);
open (blue) squares: other new variable stars.}
\label{AN1946fig4}
\end{figure}

We performed a compilation of additional infrared data by collecting $JHK_{\rm
s}$ magnitudes from 2MASS and {\em IRAS} average non-colour corrected flux
densities with qualities ``3'' or ``2'' (high and moderate qualities,
respectively).
While all the 32 stars have a near-infrared counterpart in the 2MASS catalogue,
only 12 stars were identified in the {\em IRAS} catalogues (e.g. {\em IRAS}
catalogue of point sources, version 2.0; IPAC 1986).
Furthermore, only one star, the A7IIIe Herbig Ae/Be star V351~Ori, has
high/moderate quality measurements at the four {\em IRAS} passbands at 12, 25,
60, and 100\,$\mu$m.
A colour-colour diagram combining ASAS optical and 2MASS near-infrared
magnitudes is shown in Fig.~\ref{AN1946fig4}.

   \begin{table}
      \caption[]{Identified variables with power spectrum maxima larger than
      500\,mag$^2$.}  
         \label{table.powerspectrum}
         \begin{tabular}{l cc ll}
            \hline
            \hline
            \noalign{\smallskip}
No. 	& Power		& $P$		& Name		& Remarks		\\
    	& [mag$^2$]	& [d]		&		&			\\
            \noalign{\smallskip}
            \hline
            \noalign{\smallskip}
01 	&  767.19	&  44.345	& V1159 Ori	& $P$ = 44.2--46.8\,d	\\ %
03 	& 1785.4	&  78.226	& IRAS 05285--0325 & New periodic	\\ %
11 	&  632.39	& 600.38	& RY Ori	& No clear peak 	\\ %
29 	& 8779.7	& 416.05	& X Ori		& $P$ = 424.2$\pm$1.8\,d\\ %
30 	&  913.68	& 373.31	& IRAS 05354--0142 & New periodic	\\ %
66 	& 1671.2	& 920.32	& GT Ori	& Pulsating		\\ %
           \hline
         \end{tabular}
   \end{table}
%

   \begin{table*}
      \caption[]{Infrared photometry of identified variables.} 
         \label{variables.2}
         \begin{tabular}{l ccc cccc}
            \hline
            \hline
            \noalign{\smallskip}
No. 	& $J$ 			& $H$ 	        	& $K_{\rm s}$		& $F_{12}$ 		& $F_{25}$ 		& $F_{60}$ 	& $F_{100}$ 	\\ 
    	& [mag]	   		& [mag]	        	& [mag] 		& [Jy] 			& [Jy] 			& [Jy] 		& [Jy] 		\\ 
            \noalign{\smallskip}
            \hline
            \noalign{\smallskip}
01 	& 13.817 $\pm$ 0.027	& 13.781 $\pm$ 0.046	& 13.675 $\pm$ 0.050	& ...			& ...			& ...		& ...		\\ 
03 	&  6.097 $\pm$ 0.019	&  5.083 $\pm$ 0.027	&  4.639 $\pm$ 0.021	& 1.08 $\pm$ 0.08	& 0.39 $\pm$ 0.04	& $<$ 3		& $<$ 20	\\ 
06 	&  8.295 $\pm$ 0.034	&  7.552 $\pm$ 0.024	&  7.368 $\pm$ 0.026	& ...			& ...			& ...		& ...		\\ 
07 	&  5.590 $\pm$ 0.021	&  4.665 $\pm$ 0.076	&  4.222 $\pm$ 0.033	& 1.09 $\pm$ 0.07	& 0.47 $\pm$ 0.08	& $<$ 0.9	& $<$ 30	\\ 
11 	&  9.444 $\pm$ 0.023	&  8.885 $\pm$ 0.055	&  8.277 $\pm$ 0.031	& 0.75 $\pm$ 0.09	& 0.71 $\pm$ 0.09	& $<$ 2		& $<$ 20	\\ 
12 	& 10.131 $\pm$ 0.026	&  9.280 $\pm$ 0.024	&  8.666 $\pm$ 0.024	& 0.38 $\pm$ 0.04	& 0.50 $\pm$ 0.07	& $<$ 0.7	& $<$ 40	\\ 
13 	& 10.445 $\pm$ 0.027	&  9.726 $\pm$ 0.024	&  9.311 $\pm$ 0.028	& ...			& ...			& ...		& ...		\\ 
14 	& 10.240 $\pm$ 0.026	&  9.943 $\pm$ 0.026	&  9.816 $\pm$ 0.021	& ...			& ...			& ...		& ...		\\ 
15 	& 10.647 $\pm$ 0.027	&  9.920 $\pm$ 0.023	&  9.530 $\pm$ 0.019	& ...			& ...			& ...		& ...		\\ 
18 	& 10.156 $\pm$ 0.023	&  9.463 $\pm$ 0.026	&  9.187 $\pm$ 0.019	& ...			& ...			& ...		& ...		\\ 
21 	& 11.303 $\pm$ 0.022	& 11.065 $\pm$ 0.028	& 10.979 $\pm$ 0.023	& ...			& ...			& ...		& ...		\\ 
29 	&  2.322 $\pm$ 0.292	&  1.379 $\pm$ 0.306	&  0.861 $\pm$ 0.346	& 96 $\pm$ 4		& 49 $\pm$ 7		& 6.0 $\pm$ 0.9	& $<$ 20	\\ 
30 	&  5.152 $\pm$ 0.018	&  4.608 $\pm$ 0.076	&  3.972 $\pm$ 0.320	& 2.0 $\pm$ 0.2		& 0.63 $\pm$ 0.07	& $<$ 2		& $<$ 40	\\ 
33 	& 11.340 $\pm$ 0.024	& 10.878 $\pm$ 0.025	& 10.756 $\pm$ 0.021	& ...			& ...			& ...		& ...		\\ 
36 	& 10.655 $\pm$ 0.023	&  9.647 $\pm$ 0.030	&  8.759 $\pm$ 0.021	& ...			& ...			& ...		& ...		\\ 
41 	& 11.547 $\pm$ 0.027	& 10.652 $\pm$ 0.027	&  9.898 $\pm$ 0.019	& ...			& ...			& ...		& ...		\\ 
43 	& 10.846 $\pm$ 0.027	&  9.955 $\pm$ 0.024	&  9.355 $\pm$ 0.027	& ...			& ...			& ...		& ...		\\ 
46 	&  9.457 $\pm$ 0.023	&  9.247 $\pm$ 0.023	&  9.133 $\pm$ 0.021	& ...			& ...			& ...		& ...		\\ 
48 	&  8.589 $\pm$ 0.023	&  7.933 $\pm$ 0.055	&  7.797 $\pm$ 0.027	& ...			& ...			& ...		& ...		\\ 
51 	& 10.536 $\pm$ 0.023	&  9.551 $\pm$ 0.033	&  8.764 $\pm$ 0.019	& $<$ 0.4		& 0.38 $\pm$ 0.05	& $<$ 3		& $<$ 20	\\ 
53 	& 10.860 $\pm$ 0.022	& 10.043 $\pm$ 0.026	&  9.538 $\pm$ 0.019	& ...			& ...			& ...		& ...		\\ 
54 	&  5.206 $\pm$ 0.017	&  4.225 $\pm$ 0.202	&  3.911 $\pm$ 0.258	& 1.48 $\pm$ 0.09	& 0.49 $\pm$ 0.04	& $<$ 5		& 1.7 $\pm$ 0.3	\\ 
55 	& 10.243 $\pm$ 0.021	&  9.509 $\pm$ 0.033	&  9.061 $\pm$ 0.019	& ...			& ...			& ...		& ...		\\ 
60 	&  7.508 $\pm$ 0.023	&  6.596 $\pm$ 0.029	&  6.280 $\pm$ 0.021	& ...			& ...			& ...		& ...		\\ 
65 	& 10.762 $\pm$ 0.024	& 10.447 $\pm$ 0.025	& 10.054 $\pm$ 0.019	& ...			& ...			& ...		& ...		\\ 
66 	&  9.802 $\pm$ 0.027	&  9.043 $\pm$ 0.025	&  8.219 $\pm$ 0.026	& 0.87 $\pm$ 0.13	& 1.02 $\pm$ 0.17	& $<$ 5		& $<$ 40	\\ 
70 	&  7.950 $\pm$ 0.020	&  7.504 $\pm$ 0.040	&  6.846 $\pm$ 0.026	& 1.16 $\pm$ 0.07	& 4.1 $\pm$ 0.2		& 26 $\pm$ 3	& 21 $\pm$ 3	\\ 
72 	& 10.936 $\pm$ 0.024	& 10.046 $\pm$ 0.023	&  9.448 $\pm$ 0.019	& ...			& ...			& ...		& ...		\\ 
73 	& 10.448 $\pm$ 0.023	&  9.818 $\pm$ 0.023	&  9.351 $\pm$ 0.019	& ...			& ...			& ...		& ...		\\ 
74 	& 10.459 $\pm$ 0.024	& 10.342 $\pm$ 0.024	& 10.317 $\pm$ 0.023	& ...			& ...			& ...		& ...		\\ 
75 	&  6.304 $\pm$ 0.026	&  5.375 $\pm$ 0.027	&  5.091 $\pm$ 0.020	& 0.47 $\pm$ 0.03	& $<$ 0.2		& $<$ 0.4	& $<$ 20	\\ 
78 	&  9.422 $\pm$ 0.023	&  8.825 $\pm$ 0.040	&  8.366 $\pm$ 0.018	& 0.41 $\pm$ 0.06	& 0.67 $\pm$ 0.06	& 0.47 $\pm$ 0.05 & $<$ 1.1	\\ 
            \hline
         \end{tabular}
   \end{table*}
%

Besides, we also collected proper motions from Tycho-2 (H{\o}g et~al. 2000),
PPMX (R\"oser et~al. 2008), and USNO-B1 for the 32 stars. 
Given that only a few of them have appreciable proper motions, we do not list
them, but are mentioned in Section~\ref{section.results} when appropriated.

Finally, only 13 stars have variable star designations, and there were suspects
of variability for other three stars.
Besides, an important fraction of the targets have also been tabulated as
variable stars of the Southern Hemisphere by Pojma\'nski (2002), to which our
work complements.
Next, we classify the 32 variable stars based on spectroscopic, astrometric,
and photometric criteria.

\section{Identified variable stars}
\label{section.results}

\subsection{HAeBe and T~Tauri variable stars}

\subsubsection{Previously known variable young stars and candidates} 

This class is mostly made of probable accreting T~Tauri stars with intense
H$\alpha$ emission. 
Of the nine H$\alpha$ emitter stars, 
three belong to the $\sigma$~Orionis cluster (TX~Ori, TY~Ori, RW~Ori),
three have been associated to the Alnilam and Mintaka young star populations
(V469~Ori, PU~Ori, V472~Ori), 
and three have not been linked to any young region in particular but are bright
stars at the boundary between HAeBe and T~Tauri stars (RY~Ori, V351~Ori,
StHa~40 -- two of them may have edge-on discs). 
The true nature of the other three non-H$\alpha$ emitter stars (PQ~Ori,
V993~Ori, GSC~04767--00071) should be investigated with further spectroscopic
analyses. 
Besides, six of the stars in this class display X-ray emission, and other six,
infrared flux excess.
Only the three $\sigma$~Orionis members are known to simultaneously display
H$\alpha$ and X-ray emissions and infrared flux excess.

\begin{figure}
\centering
\includegraphics[width=0.48\textwidth]{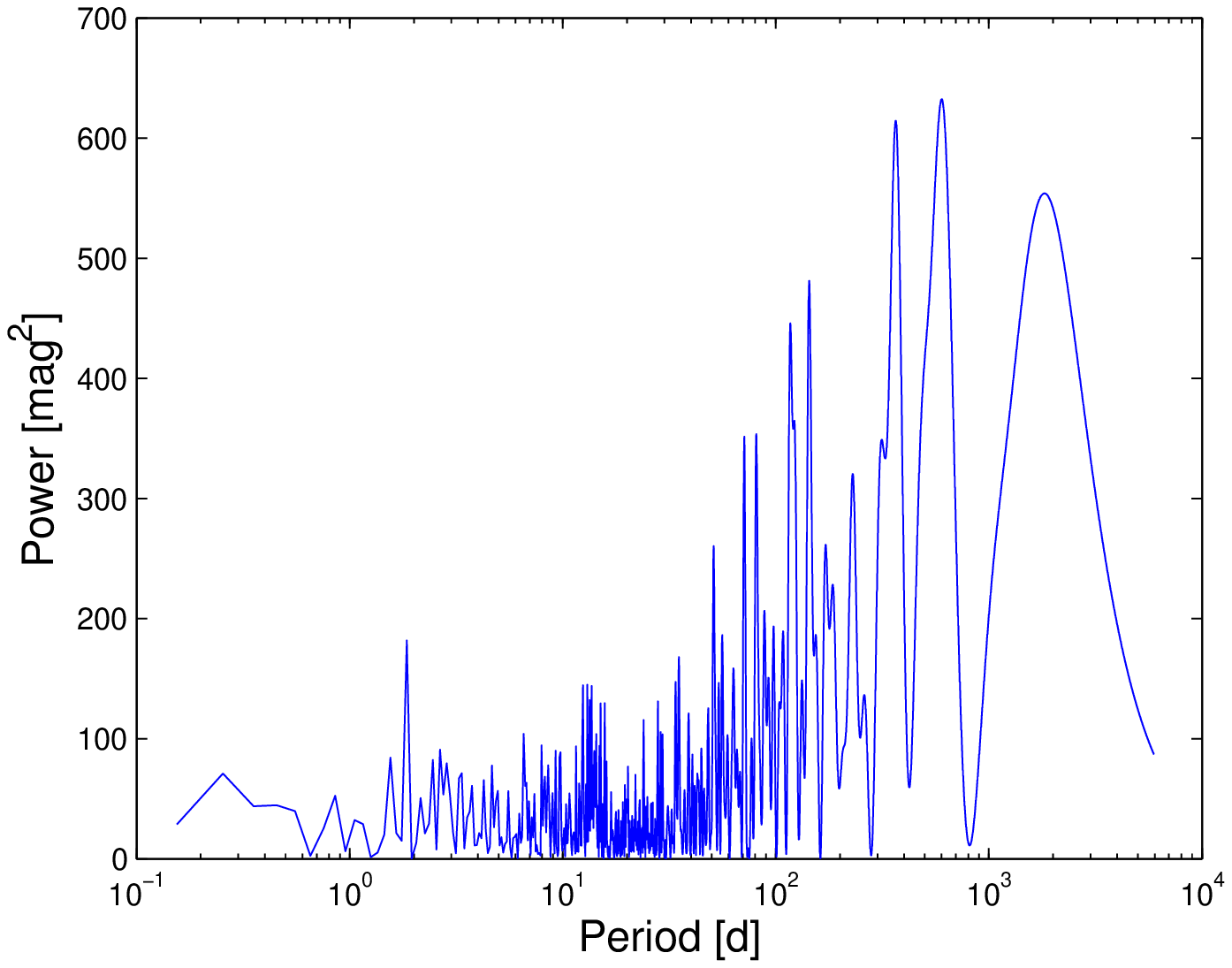}
\includegraphics[width=0.48\textwidth]{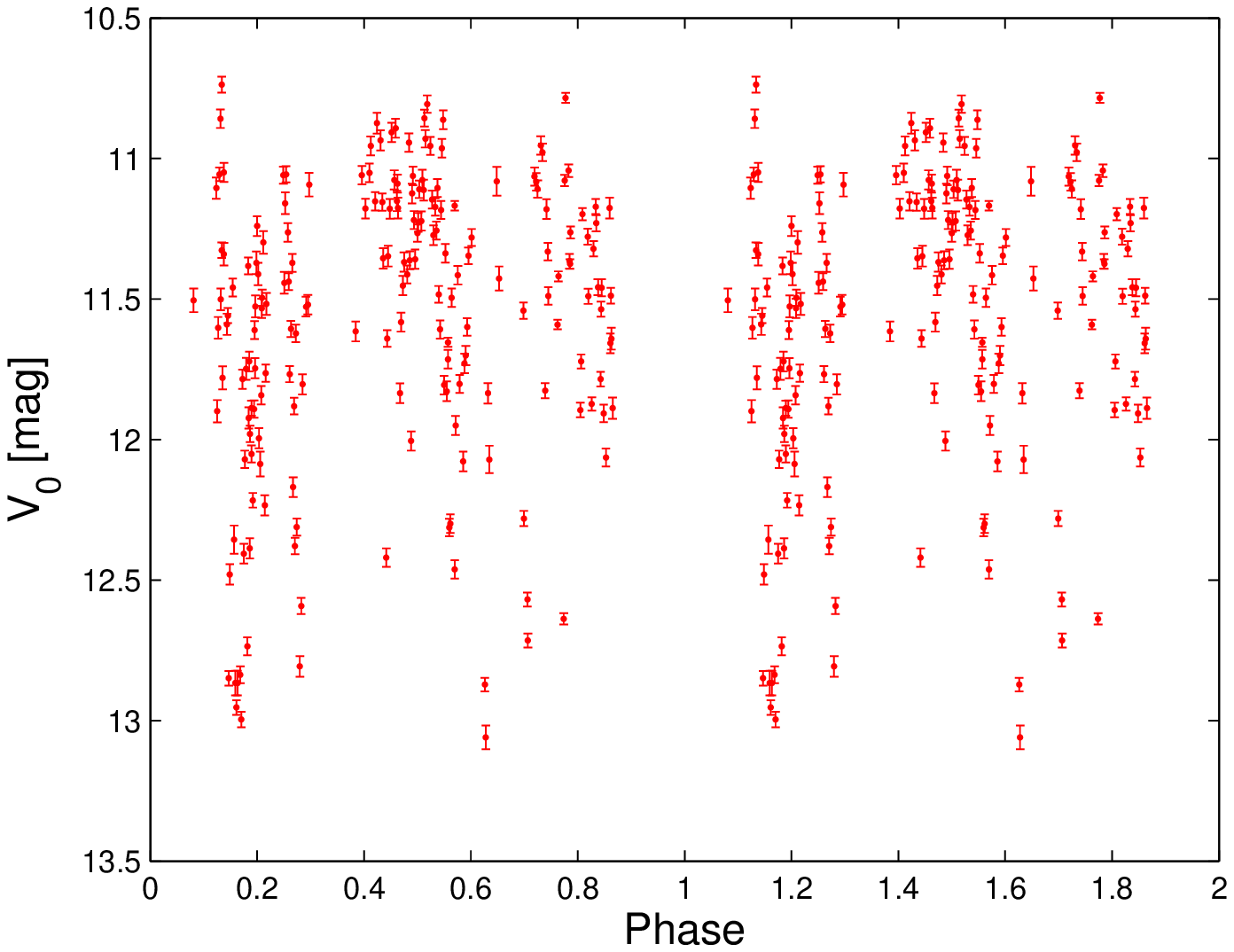}
\caption{Periodogram of the ASAS light curve ({\em top}) and phase-folded light
curve of RY~Ori to the period $P$ = 600.38\,d ({\em bottom}).} 
\label{forceblace.11}
\end{figure}

\paragraph{RY~Ori (No.~11).} 
It is a well investigated, emission-line star of the Orion population at the
boundary between HAeBe and T~Tauri stars (e.g. Haro \& Moreno 1953; Herbig \&
Bell 1988; Yudin \& Evans 1998; Eiroa et~al. 2001; Mora et~al. 2001; Hern\'andez
et~al. 2004; Acke \& van~den~Ancker 2006).   
Its photometric variability was detected very early (Pickering 1904).
According to Bibo \& Th\'e (1991), it is a UX~Ori-type star with a colour
reversal in the colour-magnitude diagram, indicative of an edge-on disc.
The disc of RY~Ori is also detected because of its long-time-known infrared
excess (Cohen 1973; Glass \& Penston 1974; Weintraub 1990; Weaver \& Gordon
1992).
Its photometric variability in the ASAS lightcurve is obvious, with small mean
photometric errors with respect to the peak-to-peak amplitude from 10.8 to
13.0\,mag ($\overline{\delta V_0}$ = 0.032\,mag).
RY~Ori is one of the six identified variable stars with power spectra maxima
larger than 500\,mag$^2$ (Table~\ref{table.powerspectrum} -- actually, it is the
star with the lowest power spectrum maximum).
The periodogram of the ASAS light curve, displayed in Fig.~\ref{forceblace.11},
shows no clear peak, with a number of power maxima increasing towards longer
periods. 
The phase-folded light curve at the period of the heighest peak, at $P \sim$
600\,d, does not show any evidence of real periodicity.
Actually, from the full-length ligh curve in Fig.~\ref{lightsaber.01-14}, the
time scale of the irregular variability of RY~Ori is of the order of a few
days, just like many confirmed young stars in this work.

\paragraph{TX~Ori (Mayrit~521199, No.~12) and TY~Ori (Mayrit~489196, No.~13).} 
They are two close ($\rho \sim$ 40\,arcsec) pre-main sequence
stars in the $\sigma$~Orionis cluster that have appeared in a number of young
and emission-line star catalogues (Herbig \& Kameswara Rao 1972; Cohen \& Kuhi
1979; Herbig \& Bell 1988; Ducourant et~al. 2005).
As in the case of many variable stars in this work, they were discovered as
variable stars by Pickering (1904).
Both TX~Ori and TY~Ori have been subject of comprehensive studies and data
compilations by Oliveira \& van~Loon (2004) and Caballero (2008).
Their light curves are quite similar, with minima and maxima at about $V_0 \sim$
12.5 and 14.0\,mag.
Remarkably, the TX~Ori light curve displayed an apparent dimming for
heliocentric Julian day\footnote{The values of heliocentric Julian day used in
the Figures are those tabulated by the ASAS catalogue. 
Actually, they are not heliocentric Julian days HJD, but the HJD minus the
constant 2\,450\,000.0 (do not confuse with the constant used to
transform the Julian day into the modified Julian day MJD, which is
2\,400\,000.5).} HJD $>$~3500.

\paragraph{RW~Ori (Mayrit~931117, No.~15).} 
It is another member in the $\sigma$~Orionis cluster, but it is much less known
than the stars above. 
Its photometric variability and H$\alpha$ emission were first detected by
Pickering (1906) and Haro \& Moreno (1953), respectively. 
Afterwards, it has been only investigated by Franciosini et~al. (2006 -- with
{\em XMM-Newton}, Her\-n\'an\-dez et~al. (2007 -- with {\em Spitzer}), and
Caballero (2008).
Its ASAS light curve varied between $V_0 \sim$ 12.8 and 14.0\,mag (the outlier
data point is probably unreliable).

\paragraph{PQ~Ori (No.~21).} 
In the discovery paper, McKnelly (1949) classified the star as a rapid,
irregular variable with an observed range of variability of 12.5--13.5\,mag
(photographic magnitude). 
Although its light curve displayed many maxima and minima, he was unable to 
determine ``any semblance of periodicity''.
Herbig \& Kameswara Rao (1975), Cohen \& Kuhi (1979), and Hern\'andez et~al. 
(2004) measured F0, F5:, and F3.0$\pm$1.5 spectral types and observed no 
emission lines. 
However, while the first authours classified it as a star in the Orion 
population, Hern\'andez et~al. (2004) did not find significant differences from 
F3 main-sequence stars nor near infrared excess, and categorized it as an object
with uncertain evolutionary status.
The variability patterns in the ASAS light curve of PQ~Ori, although ranging
between $V_0 \sim$ 12.3 and 13.3\,mag, do not resemble those of other young
stars in this work, and its $J-K_{\rm s}$ colour is quite blue with respect to
other early F stars in Ori~OB1~b. 
As a result, its field dwarf nature seems to us more plausible.

\paragraph{V469~Ori (No.~36).} 
It is an emission-line star with X-ray emission (Haro \& Moreno 1953; 
Wiramihardja et~al. 1989; Nakano et~al. 1999).
According to Caballero \& Solano (2008), the star is a member in the young
Ori~OB1~b stellar population surrounding Alnilam, and is proably associated to
the \object{[OS98]~29J}, \object{[OS98]~29H}, and \object{[OS98]~29K} remnant 
molecular clouds (Ogura \& Sugitani 1998).
Based on its $V_0 - K_{\rm s}$ colour and the empirical scale of effective 
temperatures of main-sequence stars of Alonso et~al. (1996), and extrapolating
it to the pre-main sequence, V469~Ori is likely a late Ge or early Ke star.
Its ASAS light curve looks like those of other T~Tauri stars.

\paragraph{V993~Ori (No.~43).}
It is a long-time-known variable star discovered by Luyten (1932).  
Caballero \& Solano (2008) classified it as a probable member of the Ori~OB1~b
young stellar population surrounding Alnilam.
It has a very red colour $J-K_{\rm s}$ = 1.26$\pm$0.04\,mag for its brightness
($H$ = 9.82$\pm$0.02\,mag), which may indicate the presence of a circumstellar
disc or, alternatively, that it is a background giant star.
It may be the optical/near infrared counterpart of {IRAS~Z05354--0107}, a source
in the {\em IRAS} Faint Source Reject Catalog (Moshir 1992).
There is no spectroscopy available to confirm or discard its possible T~Tauri
status.
The large peak-to-peak amplitude of its light curve of $V_0 \sim$ 13.0 to
fainter than 14.5\,mag, although sheds light on the photometric variability of
V993~Ori, is not conclusive.

\paragraph{GSC~04767--00071 (No.~48).}
Schirmer et~al. (2009) classified it as a chromospherically active star 
suspected of variability in the {\em ROTSE-1} database. 
The star GSC~04767--00071 is probably associated to the X-ray source
\object{1RXS~J053944.7--005612} (Voges et~al. 1999) and has a very low PPMX
proper motion, of ($\mu_\alpha \cos{\delta}$, $\mu_\delta$) = (--6$\pm$2,
--9$\pm$2)\,mas\,a$^{-1}$. 
The optical and near-infrared photometry supports membership in Ori~OB1~b.
It may be a very young solar analog.
The amplitude of photometric variability of its ASAS lightcurve, of $V_0 \sim$
10.8 to 11.1\,mag, is relatively small.

\paragraph{PU~Ori (No.~51).} 
It is a pre-main sequence star near the Alnilam supergiant with H$\alpha$ in
strong two-lobe emission (Haro \& Moreno 1953; Herbig \& Kameswara Rao 1972;
Cohen \& Kuhi 1979; Wiramihardja et~al. 1989), photometric variability
(Fedorovich 1960; Brice\~no et~al. 2005), mid-infrared flux excess at the {\em
IRAS} passbands (Weintraub 1990; Weaver \& Jones 1992), and forbidden emission
lines ([O~{\sc i}] $\lambda$6300.3\,\AA; Hirth et~al. 1997). 
It has a very red colour of $J-K_{\rm s}$ = 1.77$\pm$0.03\,mag (Caballero \&
Solano 2008). 
From our light curve, PU~Ori is an irregular variable with minimum and maximum
magnitudes $V_0 \sim$ 12.5 and 14.5\,mag.

\paragraph{V472~Ori (No.~53).} 
It is an H$\alpha$ emitter star (Haro \& Moreno 1953; Wiramihardja et~al. 1989)
of known photometric variability (Fedorovich 1960; Cieslinski et~al. 1997).
It may be the optical/near infrared counterpart of {IRAS~Z05322--0039}.
V472~Ori is a probable member in the Mintaka cluster (Caballero \& Solano 2008).
Its ASAS light curve has a peak-to-magnitude larger than 1\,mag.

\paragraph{V351~Ori (No.~70).} 
It is a HAeBe A7IIIe variable star of $\delta$~Scuti type discovered by
Hoffmeister (1934). 
It has been investigated in detail, especially after Kovalchuk (1984) observed a
flare-like event on its light curve at the $UBVR$ bands (Wouterloot \& Walmsley
1986; Sterken et~al. 1993; Rosenbush 1995; Marconi et~al. 2000; Balona et~al.
2002; Vieria et~al. 2003). 
It has been proposed that a temporarily strong accretion of matter onto the star
or extinction by circumstellar dust clouds takes place to explain the remarkable
V351~Ori change of behaviour from ``that of a [HAeBe] star with strong
photometric variations [...] to that of an almost non-variable star''
(van~den~Ancker et~al. 1996). 
The main period of pulsations of V351~Ori is only of the order of 0.06\,d with a
peak-to-peak amplitude of about 0.1\,mag (van~den~Ancker et~al. 1998; Marconi
et~al. 2001; Ripepi et~al. 2003). 
The pre-main sequence nature of the star is supported by the presence of
H$\alpha$ in emission, a pronounced inverse P~Cygni profile, S~{\sc i}
$\lambda$1296\,{\AA} and  O~{\sc i} $\lambda$1304\,{\AA} lines in emission, and
infrared excess (Valenti et~al. 2000 and references above).
With $V_0 \approx$ 8.91\,mag and 20--30\,Jy at the 60 and 100\,$\mu$m {\em IRAS}
passbands, it is by far the brightest star in our sample at all wavelengths, but
also the less variable one.
However, the amplitude of variability, between $V_0 \sim$ 8.80 and 9.15\,mag, is
slightly larger than previously measured.
It may be due to our long (ASAS) temporal coverage, probably longer than any
other previous monitoring.

\paragraph{StHa~40 (No.~78).}
It was first listed in the H$\alpha$-emission star catalogues of MacConnell 
(1982) and Stephenson (1986) and followed-up afterwards by Downes \& Keyes 
(1988), Torres et~al. (1995), Gregorio-Hetem \& Hetem (2002), and Maheswar 
et~al. (2003).
StHa~40 is a T~Tauri star with variable spectral type (from F8, through G2 and 
G5, to K0), variable H$\alpha$ emission (from pEW(H$\alpha$) = --14.5\,{\AA} to
a faint absorption), lithium absorption (from pEW(Li~{\sc i}) = +0.13 to 
+0.33\,{\AA}), and radial velocity (indicative of possible spectral binarity).
Besides, Gregorio-Hetem \& Hetem (2002) associated StHa~40 to a {\em ROSAT} 
X-ray source and derived basic stellar and circumstellar parameters.
Maheswar et~al. (2003) estimated a magnitude difference $\Delta V \sim$ 1.6\,mag
between two optical spectra taken almost three years apart and after comparison
with spectro-photometric standards.
The H$\alpha$ line was in emission [absorption] when the star was brighter, 
$V \sim$ 10.9\,mag [fainter, $V \sim$ 12.5\,mag].
However, while the brightest Maheswar et~al. (2003)'s estimation for $V$ agrees 
very well with our light curve (from were we measure $V_{0,{\rm bright}} 
\approx$ 11.0\,mag), none of the 173 StHa~40 data points covering almost six 
years were fainter than $V_{0,{\rm faint}} \approx$  11.8\,mag.

\subsubsection{Previously unknown variable young stars} 

Of the four stars in this class, only two (Mayrit~528005~AB and StHa~48) have
uncontrovertible signposts of youth. 
The other two stars (Kiso~A--0903~135 and HD~290625) have not been well
investigated yet and may not belong to the young Orion Belt stellar population.

\paragraph{Mayrit~528005~AB (No.~18).}
It is a hierarchical triple system in the $\sigma$~Orionis cluster, made of a
close binary ($\rho_{\rm A-B} =$ 0.40$\pm$0.08\,arcsec) resolved with adaptive
optics at $\rho_{\rm AB-C} \sim$ 7.6\,arcsec to the low-mass star
\object{Mayrit~530005} (S\,Ori~J053847.5--022711; Caballero 2005, 2009). 
The binary primary has a cosmic lithium abundance (Zapatero Osorio et~al. 2002),
double-peak H$\alpha$ (Caballero 2006), and strong X-ray emissions (Wolk 1996;
Fraciosini et~al. 2006; Caballero et~al. 2009), and an spectral energy
distribution typical of class~II objects (Hern\'andez et~al. 2007).
The ASAS light curve shows variability between $V_0 \sim$ 13 and 14\,mag, and an
apparent series of brightenings at HJD $\sim$~3000.

\paragraph{Kiso~A--0903~135 (No.~41).}
It is an emission-line star only referenced once by Kogure et~al. (1989), who
measured a weak H$\alpha$ emission intensity and placed it in the Orion region. 
Its optical (DENIS; Epchtein et~al. 1997) and near-infrared (2MASS) photometry
and null proper motion (within error bars; USNO-B1) are consistent with
membership in the Ori~OB1~b association. 
With $\sigma_{V_0}$ = 0.480\,mag, it is one of the most variable stars in the
study.
While its brightest magnitude is $V_0 \sim$ 12.8\,mag, the faintest one is $V_0
\sim$ 15.0\,mag or fainter.

\paragraph{StHa~48 (No.~55).}
It is a K4-type T~Tauri star with H$\alpha$ in emission and Li~{\sc i} in
absorption (Stephenson 1986; Maheswar et~al. 2003; Caballero \& Solano 2008).
The star, which is located in the surroundings of Alnilam, has a wide amplitude
of variability, with maximum and minimum brighteness at $V_0 \approx$ 11.9
and 13.3\,mag, respectively, and time scales of variation of a few days.

\paragraph{HD~290625 (No.~74).}
It is an A2 star classified as a probable member in the young stellar
population of Ori~OB1~b (Guetter 1979; Nesterov et~al. 1995).
The ASAS light curve displayed variations between $V_0 \approx$ 10.8 and 
11.3\,mag, approximately.
The Tycho-2 and 2MASS photometry and the low proper motion, of less than 
4\,mas\,a$^{-1}$ (Tycho-2, PPMX, USNO-B1), are consistent with the spectral type
and a long heliocentric distance.
Besides, lifes of A2 stars are relatively short, which confirm the HD~290625
youth.
However, the star is a couple of magnitudes fainter than expected for ``normal''
A2 Orion stars at $d \sim$ 0.4\,kpc.
Instead of resorting to another young population in the far background of Orion,
of which there is no evidence, we propose that HD~290625 is a variable HAeBe
star with a scattering edge-on disc in the Orion complex, just like
\object{UX~Ori} or the $\sigma$~Orionis star StHa~50 (\object{Mayrit~459340} --
Th\'e et~al. 1994; Caballero et~al. 2008) or, alternatively, a bright, blue,
subdwarf (there have been previous detections of such objects in the area:
Caballero \& Solano 2007; Vennes et~al. 2007).

\subsection{Variable background giants}

Unfortunately, because of the relative resemblance between the spectral energy
distributions of distant giant stars with extended cool envelopes and of young
stars with developed discs (due to the common strong infrared flux excess),
giant stars sometimes contaminate lists of member candidates in star-forming
regions. 
For example, the bright (in the infrared) giant star \object{IRAS~05358--0238}
was once classified as a Class~I object in the $\sigma$~Orionis cluster
(Oliveira \& van~Loon 2004; Hern\'andez et~al. 2007; Caballero 2008).
Since the surveyed area in this work is rich in star-forming regions (including
$\sigma$~Orionis), we had to be careful with disentangling both pre-main
sequence and giant star populations.

Of the eight identified variable background giants, only one (No.~60) has no
{\em IRAS} excesses or were not detected with the Spatial Infrared Imaging
Telescope onboard the {\em Midcourse Space Experiment} (8.3--21.3\,$\mu$m;
Kraemer et~al. 2003). 
There were also only three stars previously classified as giants: X~Ori, GT~Ori,
and IRAS~05354--0142 (for which variability had not ever been proposed).
Therefore, in this work we present six new variable giants.
Besides, we measure periods of photometric variability (with maxima of the power
spectrum above 500\,mag$^2$) between 78 and 920\,d for four of the eight giants,
while only one, X~Ori, had been reported before to display periodicity.

\subsubsection{Known background giants}
\label{knownbackgroundgiants}

\begin{figure}
\centering
\includegraphics[width=0.48\textwidth]{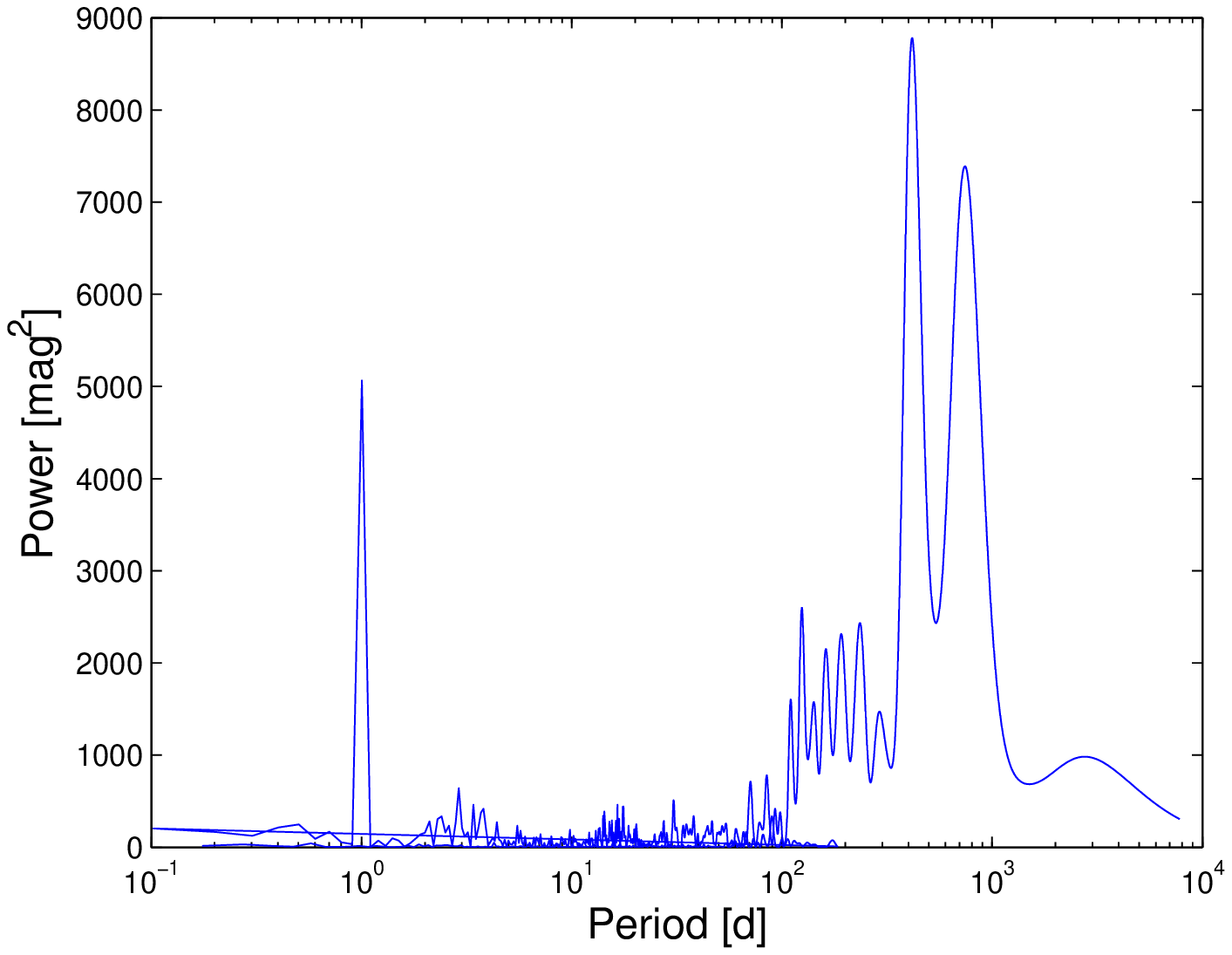}
\includegraphics[width=0.48\textwidth]{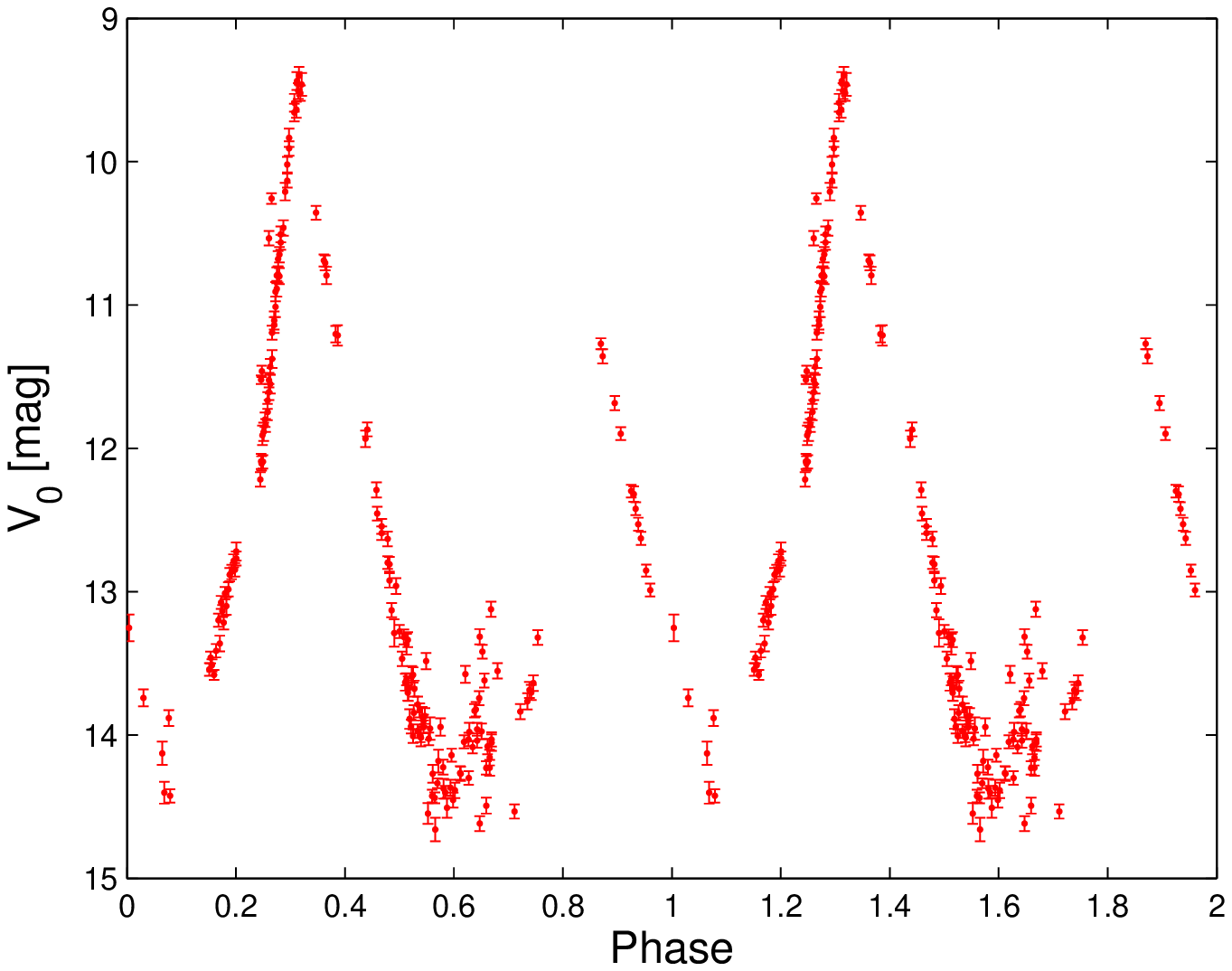}
\caption{Same as Fig.~\ref{forceblace.11}, but for X~Ori and a period $P$ =
416.05\,d.}
\label{forceblace.29}
\end{figure}

\paragraph{X Ori (No.~29).} 
This giant or superginat star was the reddest object in the Alnilam-Mintaka 
area studied by Caballero \& Solano (2008), but is located at about twice the
heliocentric distance to the Ori~OB1 association (Jura et~al. 1993).
It is an M8--9-type Mira~Ceti variable found by Wolf (1904) with $P$ =
424.75$\pm$1.77\,d (Templeton et~al. 2005) and silicate dust emission (Sloan \&
Price 1998; Speck et~al. 2000).
In particular, the 9.7 and 18\,$\mu$m silicate dust feature originates from the 
circumstellar envelopes of oxygen-rich asymptotic giant branch stars (Kwok 
et~al. 1997).
From our periodogram (Fig.~\ref{forceblace.29}), there are two clear peaks at $P
\approx$ 416\,d, consistent with literature values assuming a reasobale error of
2\,\%, and its harmonic at $P \approx$ 832\,d.
There is also a spurious (systematic) peak at $P$ = 1\,d.
The triangular shape of minima and maxima in the phase-folded light curve may be 
useful for modelling the astrophysical properties of the Mira star.

\begin{figure}
\centering
\includegraphics[width=0.48\textwidth]{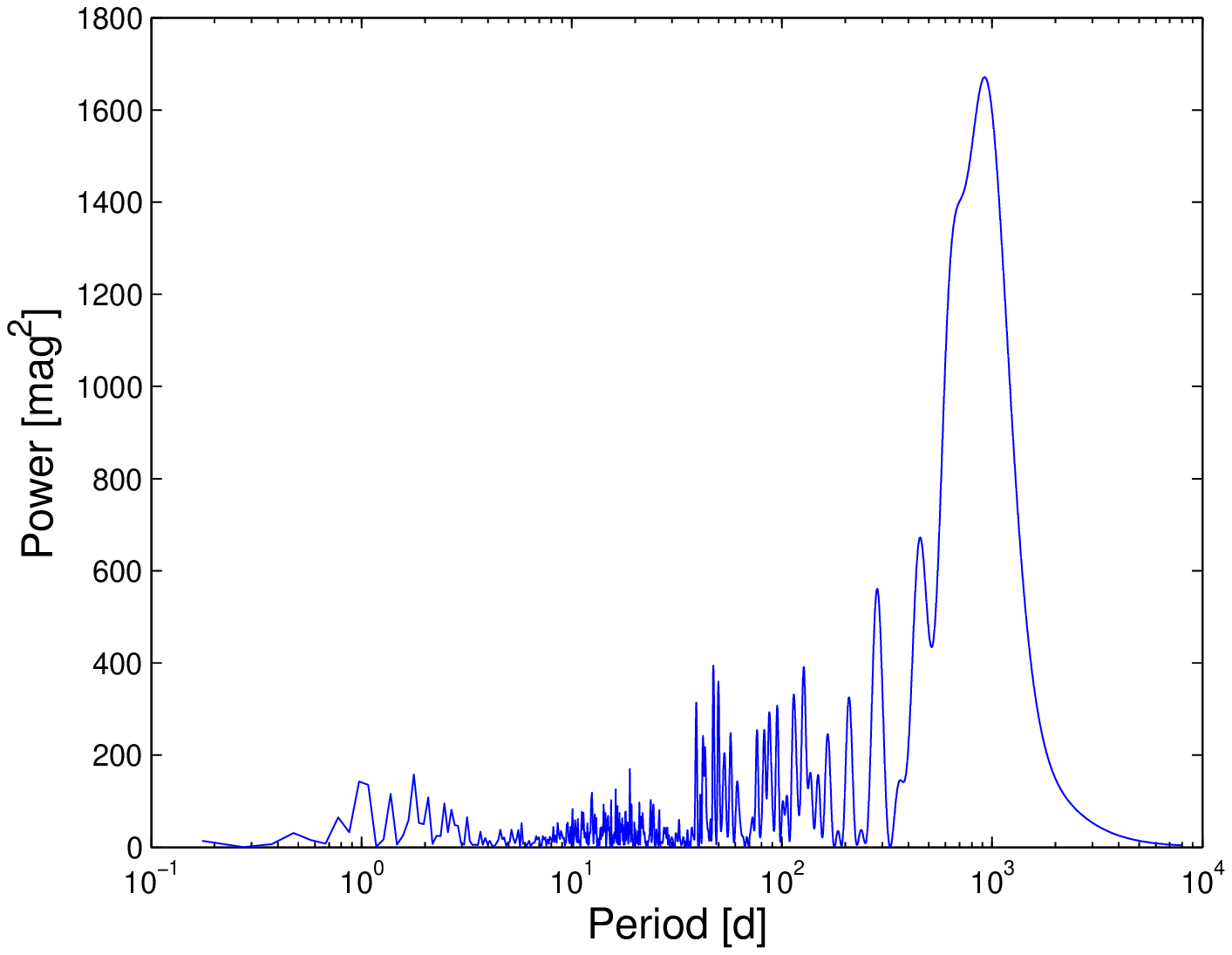}
\includegraphics[width=0.48\textwidth]{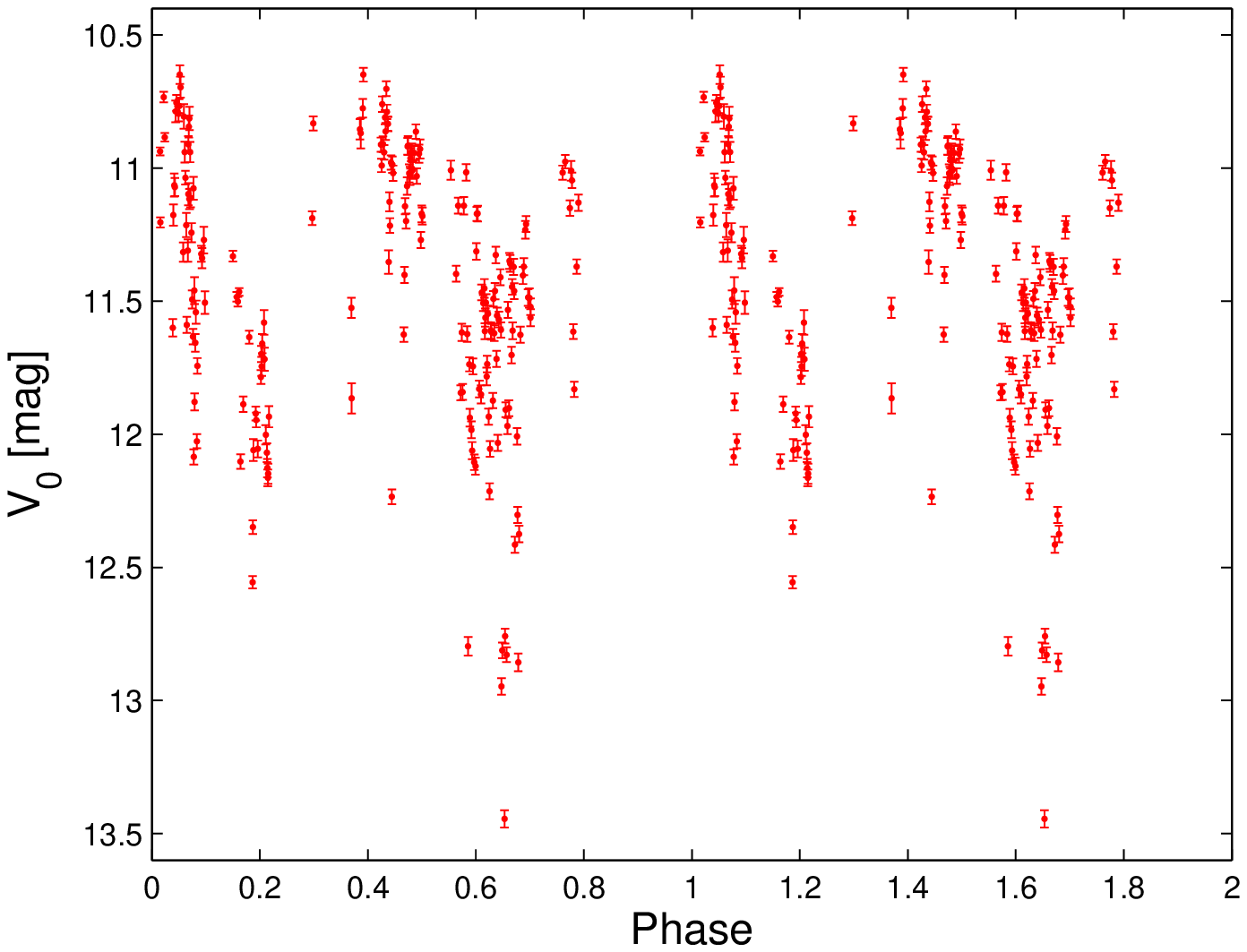}
\caption{Same as Fig.~\ref{forceblace.11}, but for GT~Ori and a period $P$ =
920.32\,d.}
\label{forceblace.66}
\end{figure}

\paragraph{GT~Ori (No.~66).}
It is a semi-regular pulsating star, originally proposed to be a likely Mira 
star with variations between 11.5 and 12.5\,mag by Hoffmeister (1934).
Afterwards, it has been monitored by several authors (e.g. Kurochkin 1949; 
Braune 2001).
The F0 spectral type determined by Nesterov et~al. (1995) is in contradiction
with the red optical and near-infrared colours, the detection of GT~Ori by
{\em IRAS} at the 12 and 25\,$\mu$m, and its low proper motion, of less than
3\,mas\,a$^{-1}$.
The photometric and astrometric information and the strong periodogram peak at
$P \sim$ 920\,d and the phase-folded light curve in Fig.~\ref{forceblace.66} are
more consistent with the semi-regular pulsating status and a probable K-type,
giant nature of GT~Ori. 
The large amplitude of variability of the star, of almost 3\,mag ($V_0 \approx$ 
10.6--13.5\,mag), and the overlapping of different (shorter) periods in the 
light curve are remarkable, as well as its abnormally blue $V_0-J$ colour for
its red $J-K_{\rm s}$ and {\em IRAS} flux excess emission.

\subsubsection{New variable giant candidates}
\label{newbackgroundgiants}

\begin{figure}
\centering
\includegraphics[width=0.48\textwidth]{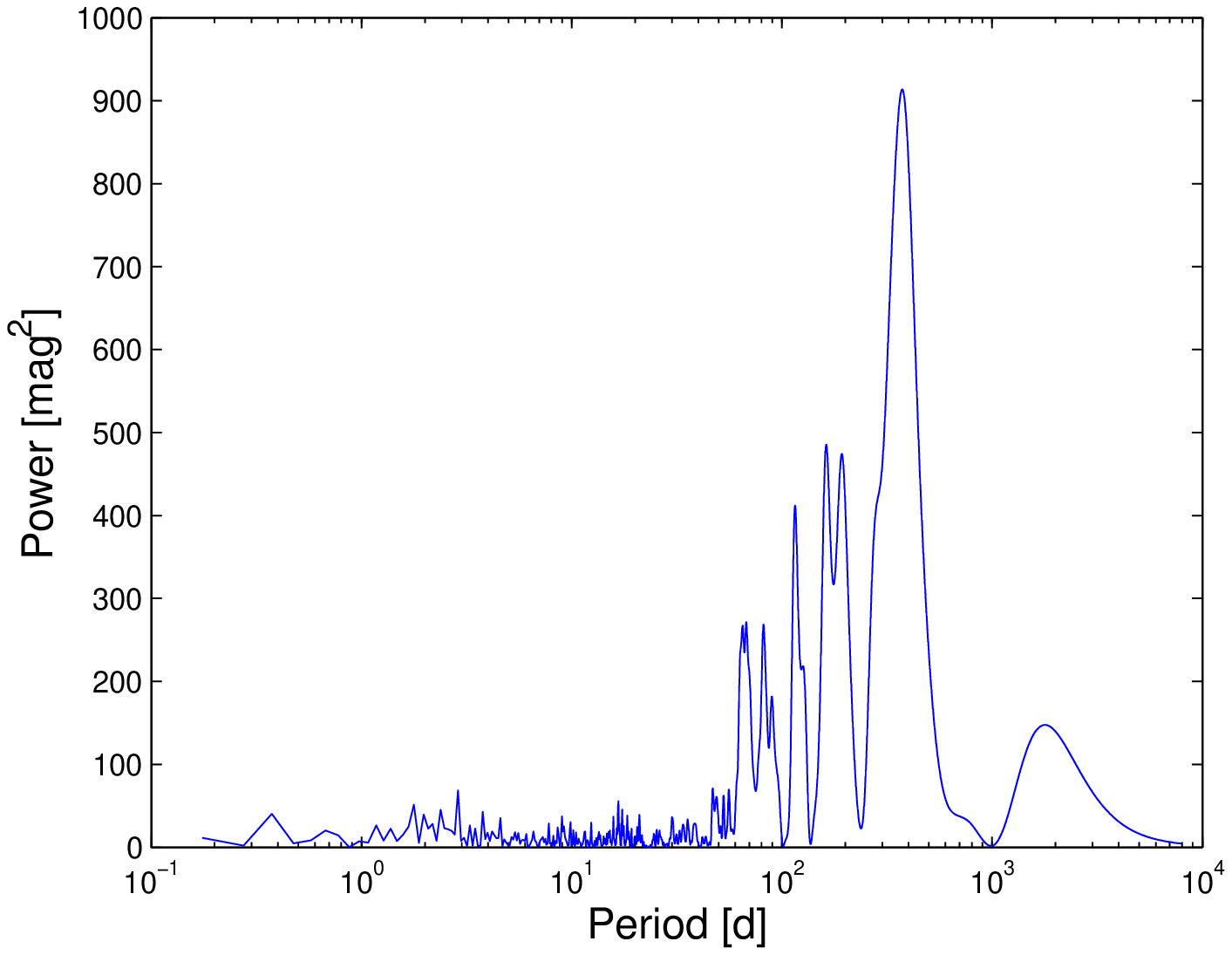}
\includegraphics[width=0.48\textwidth]{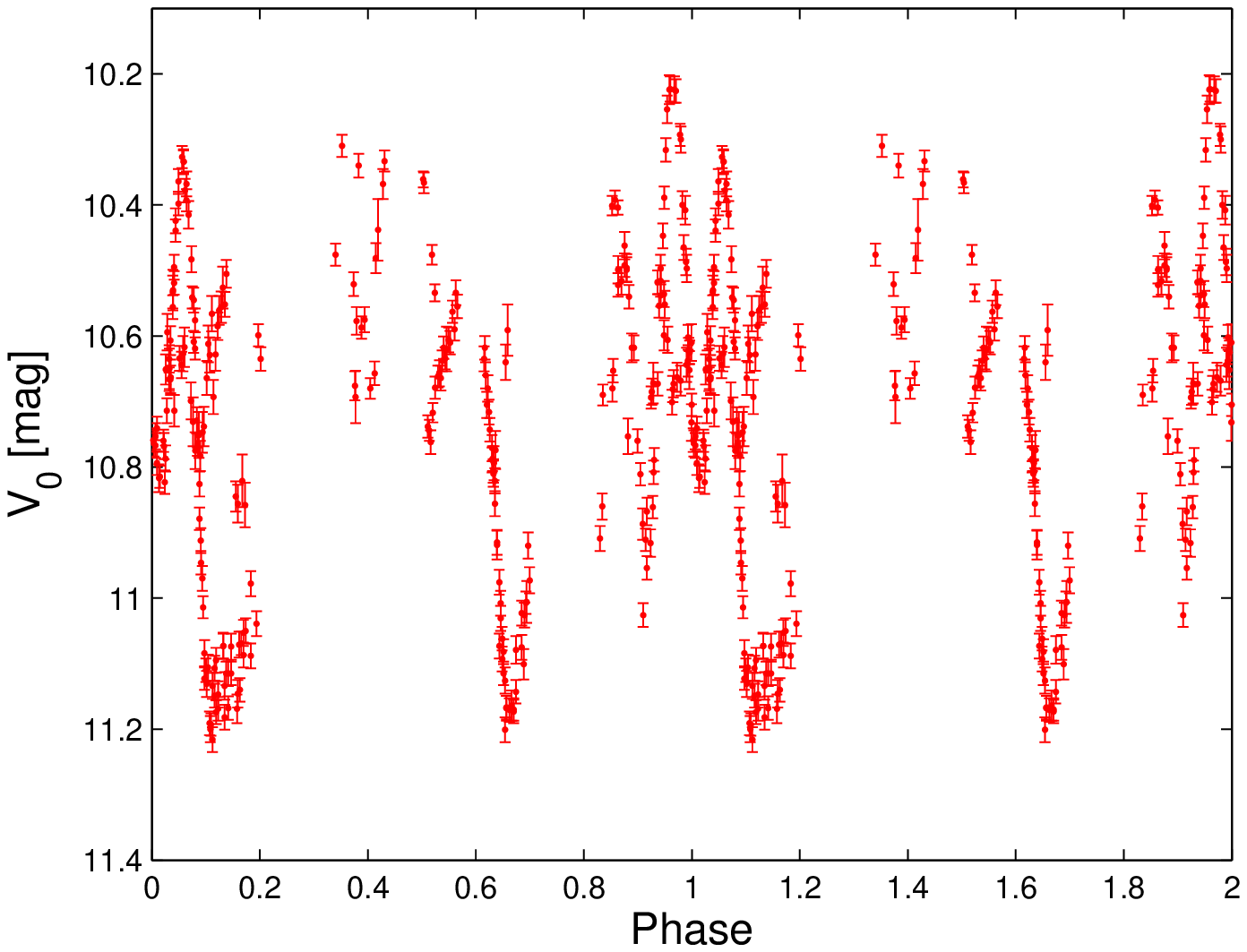}
\caption{Same as Fig.~\ref{forceblace.11}, but for IRAS~05354--0142 and a period
$P$ = 373.31\,d.}
\label{forceblace.30}
\end{figure}

\paragraph{IRAS 05354--0142 (No.~30).} 
With $V_T-K_{\rm s}$ = 7.1$\pm$0.3\,mag, it had the reddest optical-infrared
colour among the $\sim$1500 Tycho-2/2MASS stars investigated by Caballero \&
Solano (2008).  
According to them, ``the closeness of IRAS~0534--0142 to the \object{Ori~I--2}
globule may explain part of its reddening, but not all. 
It might be an S-type or a C-type giant with a very late spectral type
and very low effective temperature''. 
From our ASAS data, IRAS 05354--0142 is the brightest star in
Table~\ref{table.powerspectrum} ($V_0 \approx$ 10.2--11.2\,mag) and has the
fourth largest power spectrum. 
Besides, it is one the four most clearly variable stars in this study, together
with X~Ori, RY~Ori, and GT~Ori.
The light curve power spectrum peaks at $P$ = 373.31\,d, but there are also
secondary maxima at shorter periods (Fig.~\ref{forceblace.30}). 
The multi-pattern, phase-folded light curve indicates that a harmonic study is
necessary to disentangle the different oscillation modes of the giant. 
At about 3\,arcmin to the north lies the irregular variable star
\object{V1299~Ori}, found by Kaiser (1992).
It was the only SIMBAD object unidentified by Caballero \& Solano (2008) in
the Alnilam/Mintaka region. 
Tabulated coordinates are likely incorrect. 
IRAS 05354--0142 and V1299~Ori might be the same star, but the variability type
and amplitude given by Kaiser (1992) are inconsistent with the data presented
here.

\begin{figure}
\centering
\includegraphics[width=0.48\textwidth]{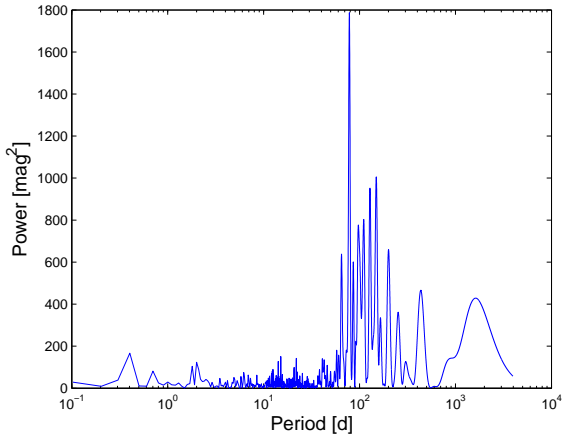}
\includegraphics[width=0.48\textwidth]{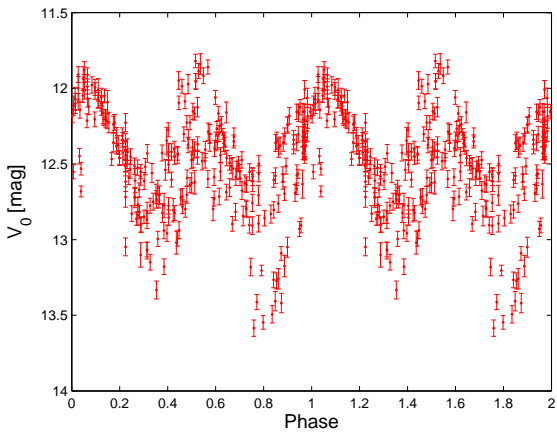}
\caption{Same as Fig.~\ref{forceblace.11}, but for IRAS~05285--0325 and a period
$P$ = 78.226\,d.}
\label{forceblace.03}
\end{figure}

\paragraph{IRAS~05285--0325 (No.~03), IRAS~05353--0309 (No.~07), 
IRAS~05248--0037 (No.~54), and IRAS 05312+0053 (No.~75).} 
They had been only reported by Kraemer et~al. (2003) as ``infrared'' sources and
have  ``high quality'' flux densities at the {\em IRAS} 12\,$\mu$m (and 
25\,$\mu$m) pass-bands and proper motion absolute values of a few 
milliarcseconds per year.
Of them, only one had a power spectra peak larger than 500\,mag$^2$ 
(Table~\ref{table.powerspectrum}).
The periodogram of IRAS~05285--0325 shows a forest of peaks associated to the 
different pulsation modes of the giant.
The main mode, which is quite clear in Fig.~\ref{forceblace.03}, corresponds to 
a period $P$ = 78.3$\pm$0.1\,d.

\paragraph{No.~60.} 
It is reported here for the first time and has not a measured {\em IRAS} flux
density.  
However, its very red optical and near-infrared colours, which puts the star in
the giant locus in Fig.~\ref{AN1946fig4} ($V_0-J \sim$ 4.8\,mag), and very low
proper motions make No.~60 to be a fair giant candidate.

\subsection{Miscellanea}

In this class, we include variable objects that are neither young stars nor
background giants.
Of the eight objects in this class, only two (the dwarf nova cataclysmic variable
V1159~Ori and the contact binary 2M054354--0243.6) were known to vary.
At least three of the other six stars, including the nearby early F-type star
HD~290509, might be eclipsing or interacting binaries.

\begin{figure}
\centering
\includegraphics[width=0.48\textwidth]{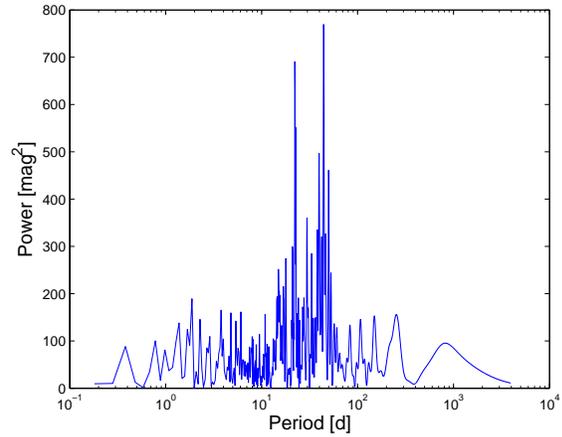}
\includegraphics[width=0.48\textwidth]{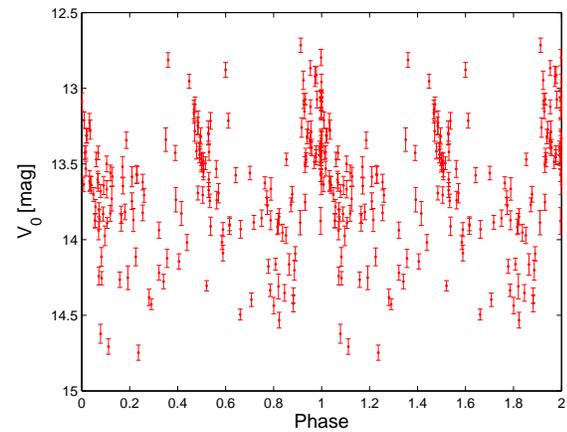}
\caption{Same as Fig.~\ref{forceblace.11}, but for V1159~Ori and a period $P$ =
44.345\,d.}
\label{forceblace.01}
\end{figure}

\paragraph{V1159~Ori (No.~01).} 
It is a well-known dwarf nova cataclysmic variable of ER~UMa (or RZ~LMi) type,
which is a subgroup of SU~UMa-type dwarf novae\footnote{A dwarf nova is a close
binary star system in which one of the components is a white dwarf, which
accretes matter from its companion.  
In contrast to classical novae, outbursts and super-outbursts in dwarf novae
result from instabilities in the accretion disc.
SU~UMa dwarf novae are characterised by two distinct types of outbursts in time
scales of days (more frequent) and weeks (less frequent and with larger
amplitude).} with short recurrence time of super-outburst ($\tau \le$
45\,d).  
Their super-cycles are stable both in length and outburst pattern 
(Osaki 1996; Patterson 1998; Downes et~al. 2001).
V1159~Ori was discovered by Wolf \& Wolf (1906), who gave it the name
``Var.,Orionis 36.1906'', and first investigated by Kippenhahn (1953) and
Jablonski \& Cieslinski (1992). 
Afterwards, its very short outburst cycle, superhumps, orbital periods, and high
energy (X-ray, ultraviolet) emission have been studied by a number of authors
(Nogami et~al. 1995; Robertson et~al. 1995; Patterson et~al. 1995; Thorstensen
et~al. 1997; Szkody et~al. 1999).
We can add little to the knowledge on V1159~Ori except for its supercycle
period.
In our periodogram (Fig.~\ref{forceblace.01}), we detect a clear peak (and its
harmonics) at $P$ = 44.3$\pm$0.1\,d, which is agreement with, or may
improve or complement, previous determinations at 44.2--46.8\,d  
(Robertson et~al. 1995; 
Richter 1995; 
Kato 2001; 
Pitts et~al. 2002). 

\paragraph{2M054354--0243.6 ([GGM2006]~12390250, No.~14).} 
It is a contact binary discovered by Gettel et~al. (2006), who measured an
heliocentric distance of $d$ = 288$\pm$5\,pc. 
Given the appreciable proper motion tabulated in the USNO-B1 and PPMX
catalogues, $\mu$ = 46\,mas\,a$^{-1}$, one can measure a transversal velocity of
about 63\,km\,s$^{-1}$, which is typical of Galactic disc stars. 
In our periodogram (not shown here), there are two faint peaks at 0.29 and
2.3\,d, but nothing at the 0.438118\,d measured by Gettel et~al. (2006).
Besides, the phase-folded light curve of 2M054354--0243.6 at this period does not
show any clear periodic behaviour.

\paragraph{No.~33 and No.~65.} 
They are reported here for the first time.
The stars display optical and near-infrared colours typical of early K dwarfs
and have no infrared excess.
The light curve magnitudes varied between $V_0 \approx$ 12.4 and 13.4--13.6\,mag 
(No.~33) and $V_0 \approx$ 11.7 and 12.4\,mag  (No.~65), with most of the data 
points concentrated towards the brightest limits.
This fact may be an indication that the stars are possibly new eclisping or 
interacting binaries, but the possibility of being stars with magnetic spots 
should not be ruled out.
That the periodograms (not shown here) do not display any strong peak puts a
limit on the maximum periods, which are possibly shorter than a few days.

\paragraph{HD~290509 (No.~46).}
It is an anodyne F0-type Tycho-2 star with only one entry in SIMBAD (Nesterov
et~al. 1995). 
The appreciable proper motion tabulated by PPMX, of (30.6$\pm$1.3,
--8.8$\pm$1.3)\,mas\,a$^{-1}$, indicates that it is a field star and not a young
star in Orion. 
Its optical and red colours match those of dwarfs of the same spectral type.
The ASAS light curve of HD~290509 resembles that of the star No.~33, but with a
shorter range of variation, between $V_0 \approx$ 10.35 and 10.60\,mag.
It might be another short-period eclipsing binary.

\paragraph{VSS~VI--32 (No.~72).}
It is an unpolarised star in the background of the \object{Messier~78} 
star-forming complex, only reported by Vrba et~al. (1976).
Its very red colours, of $V_0-J \sim$ 2.5\,mag and $J-K_{\rm s}$ = 
1.49$\pm$0.03\,mag, may not be intrinsic, but due to the high extinction in
the area.
Its variability class is unknown to us.

\paragraph{No.~06, No.~73.} 
They are also reported here for the first time and none of them have a measured
{\em IRAS} flux density.
They have very low proper motions and are the reddest (in $J-K_{\rm s}$) stars
in this section except for VSS~VI--32, but do not seem to be red enough to be
classified as giant star candidates. 
Their light curves are quite different, with No.~06's one varying only between
11.05 and 11.45\,mag, and No.~73's one between 12.0 and 14.0\,mag.
Neither the power spectrum analysis nor the light curve shapes provide any clear
indication of the nature of the two stars.

\section{Summary}

With the aim of discovering and confirming variable Herbig~Ae/Be and T~Tauri
stars among the young stellar population of the Orion Belt, we have analysed
over 60\,000 light curves from the ASAS-3 Photometric $V$-band Catalogue of
the All Sky Automated Survey in a squared area of side 5\,deg centred on the
supergiant star Alnilam ($\epsilon$~Ori, the middle star of the Orion~Belt). 

After consecutive filterings, visual inspections with the virtual observatory
tool Aladin, and cross-matches with infrared and astrometric catalogues, we kept
32 variable stars with average magnitudes $\overline{V_0}$ = 8.5--13.5\,mag for
a detailed analysis and data compilation. 
They are {\em the most variable stars in the Orion Belt} and complement the
results in Pojma\'nski (2002).
Based on spectroscopic, photometric, and astrometric information from catalogues
and the literature, we classified the 32 stars into:
\begin{itemize}

\item previously known variable young stars and candidates (12 stars).
The list includes well-investigated Herbig~Ae/Be and T~Tauri stars surrounding
the early-type stars $\sigma$~Ori, Alnilam, and Mintaka in the Ori~OB1~b
association. 
We confirm the variable status of two active stars suspected of variability
(GSC~04767--00071 and StHa~40) and measure a peak of the power spectrum larger
than 500\,mag$^2$ in the light curve of the emission-line, UX~Ori-type variable,
disc-host star RY~Ori.
However, the corresponding period of photometric variability seems unreliable.
Thus, the 12 stars appear to be irregular variables;

\item previously unknown variable young stars (4 stars).
They are two T~Tauri stars with Li~{\sc i} in absorption and H$\alpha$ in
emission (Mayrit~528005~AB and SHa~48) and two stars earlier classified as
probable members in the Ori~OB1~b association. 
Besides, one of them, Mayrit~528005~AB, displays infrared flux excess and
strong X-ray emission, and is a close binary within a hierarchical triple system
in the $\sigma$~Orionis cluster.
Again, we do not find periodic trends among these young stars;

\item known background giants (2 stars).
The two giants (or supergiants) are X~Ori, for which our period determination is
consistent with those provided in the literature, and GT~Ori, for which we
firstly provide a period of photometric variability at about 920\,d and
determine abnormal optical-near infrared colours;

\item new variable giant candidates (6 stars).
Of them, we remark the new, extremely red, periodic variables IRAS~05354--0142
($P \sim$ 370\,d -- with a lot of shorter periods) and IRAS~05285--0325 ($P
\sim$ 78\,d);

\item miscellanea objects (8 stars).
This class contains the cataclysmic variable V1159~Ori, for which we measure a
period $P$ = 44.3$\pm$0.1\,d, in agreement with previous determinations, the
contact binary 2M054354--0243.6, and six poorly investigated, new variable
stars. 

\end{itemize}

Some of these targets, including variable active Herbig~Ae/Be and T~Tauri stars
in the Ori~OB1 association and background giants, merit next spectroscopic
analyses and photometric monitoring.

\acknowledgements
We thank M.~Paegert for his helpful referee report and careful
reading of the manuscript, and G.~Pojma\'nski for his kind and prompt answer to
our inquiries.
JAC formerly was an ``investigador Juan de la Cierva'' at the Universidad
Complutense de Madrid and currently is an ``investigador Ram\'on y Cajal'' at
the Centro de Astrobiolog\'{\i}a (CSIC-INTA). 
This research has made use of the SIMBAD, operated at Centre de Donn\'ees
astronomiques de Strasbourg, France, and the NASA's Astrophysics Data System.
Financial support was provided by the Universidad Complutense de Madrid,
the Comunidad Aut\'onoma de Madrid, the Spanish Ministerio Educaci\'on y
Ciencia, and the European Social Fund under grants:
AyA2005-04286, AyA2005-24102-E,		
AyA2008-06423-C03-03, 			
AyA2008-00695,				
PRICIT S-0505/ESP-0237,			
and CSD2006-0070. 			

\appendix

\section{The most variable ASAS stars in the Orion Belt}

\begin{figure*}
\centering
\includegraphics[width=0.440\textwidth]{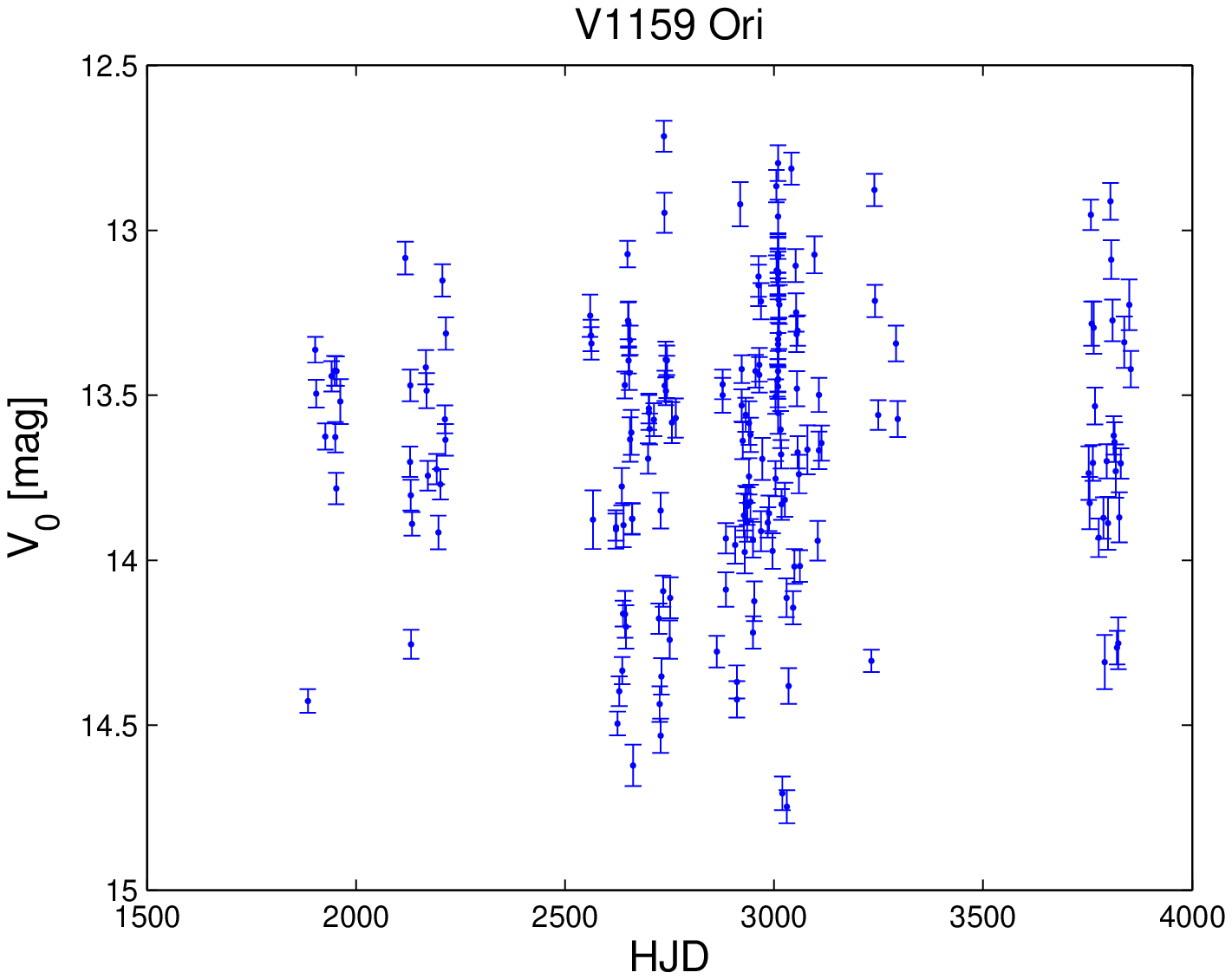}
\includegraphics[width=0.440\textwidth]{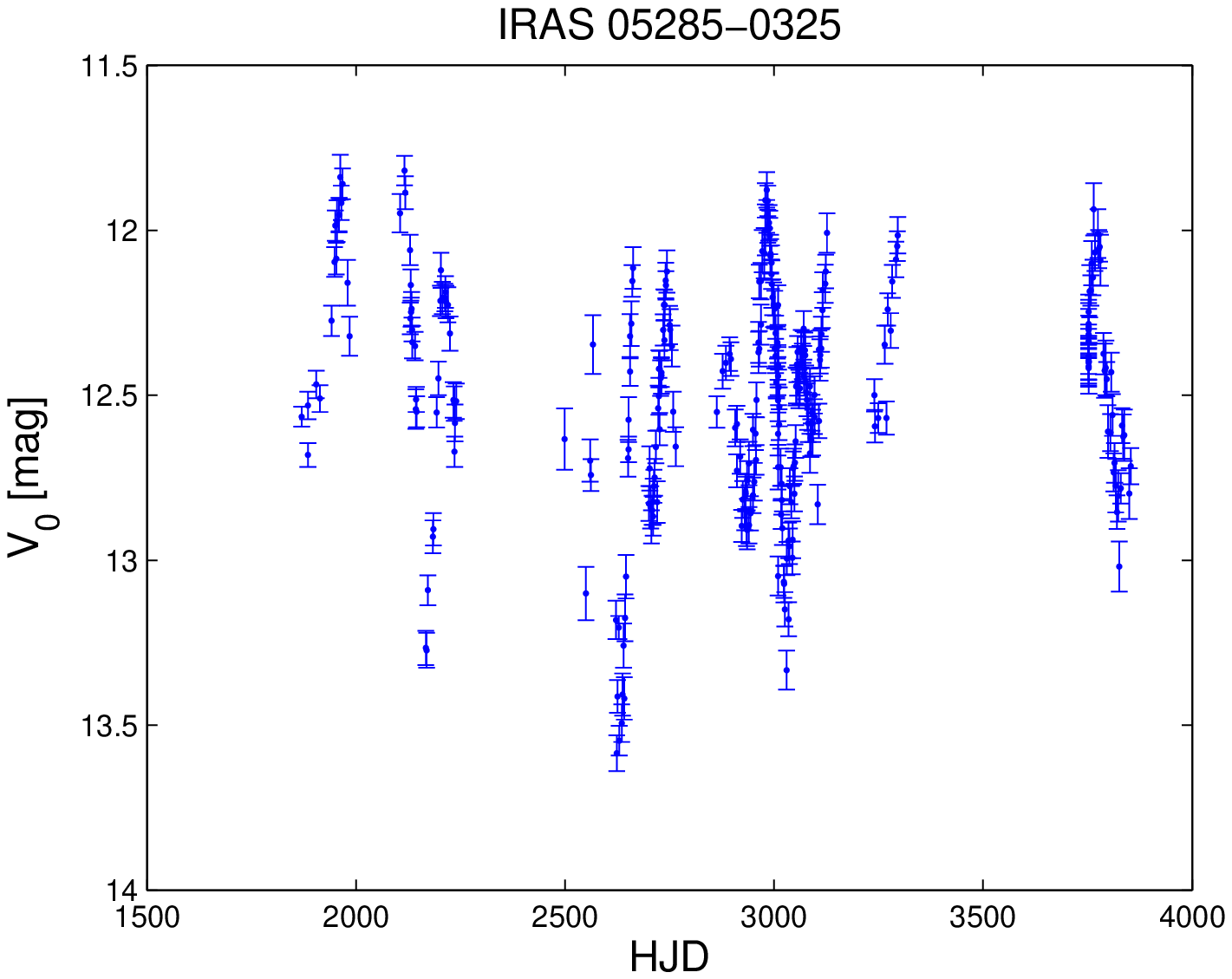}
\includegraphics[width=0.440\textwidth]{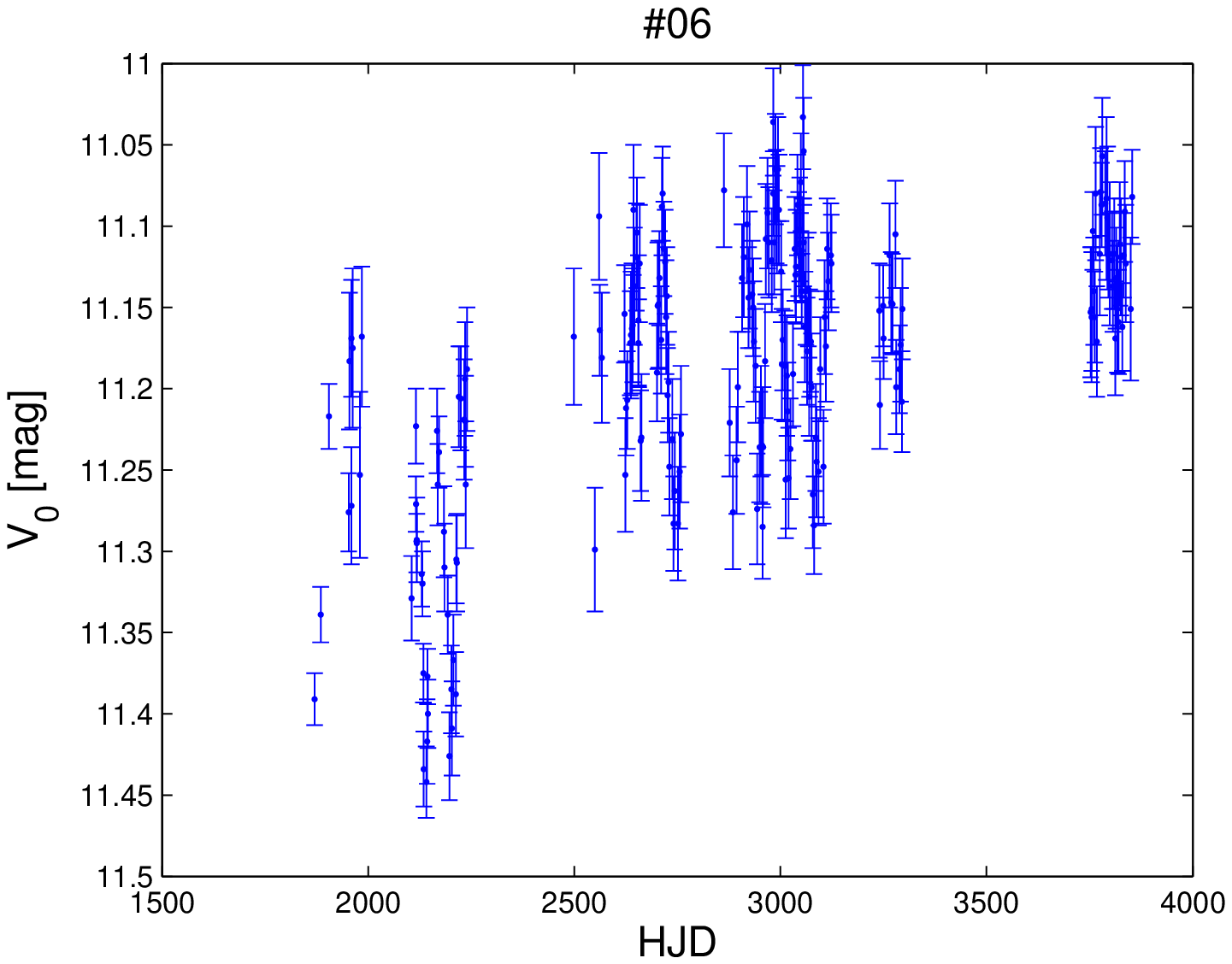}
\includegraphics[width=0.440\textwidth]{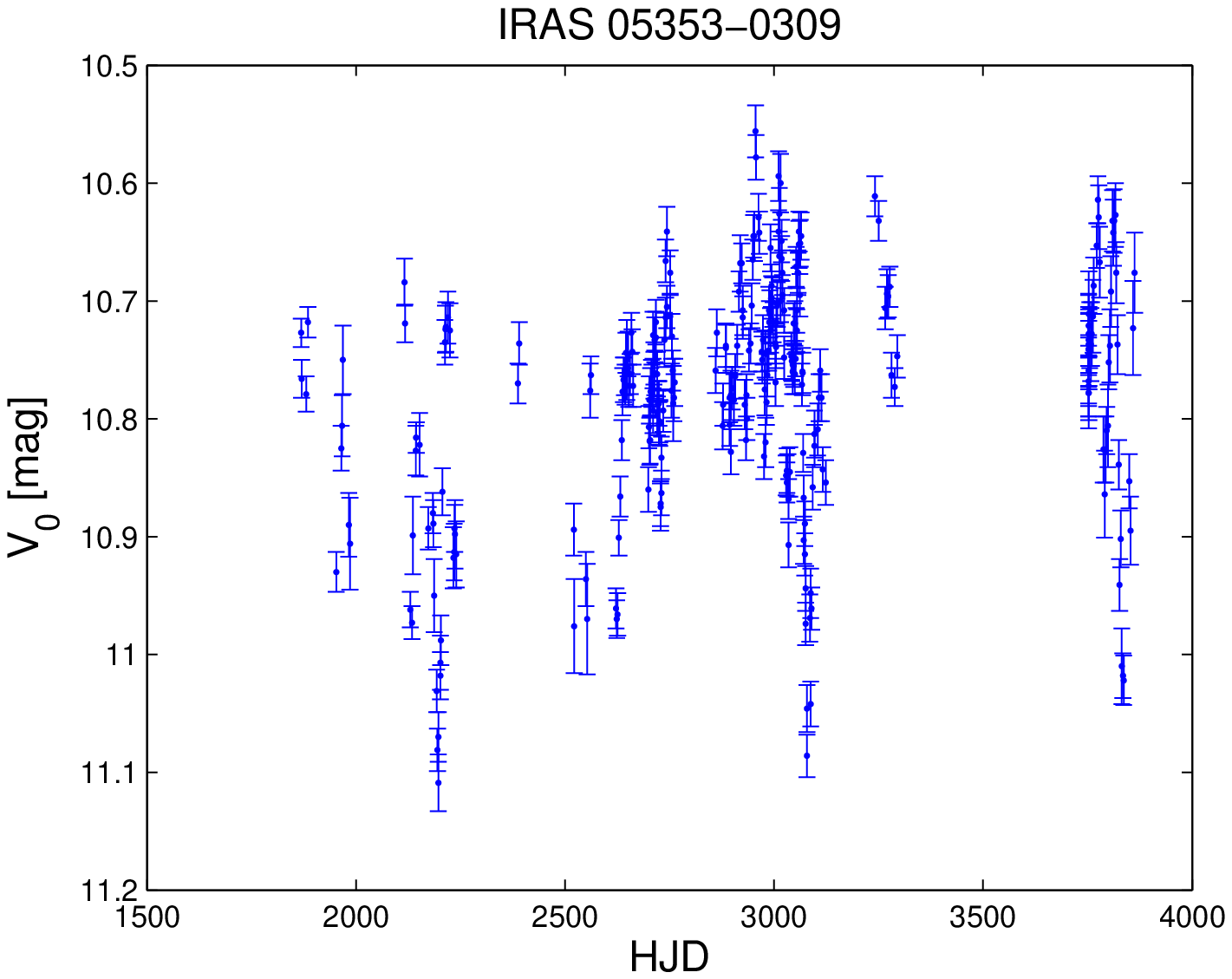}
\includegraphics[width=0.440\textwidth]{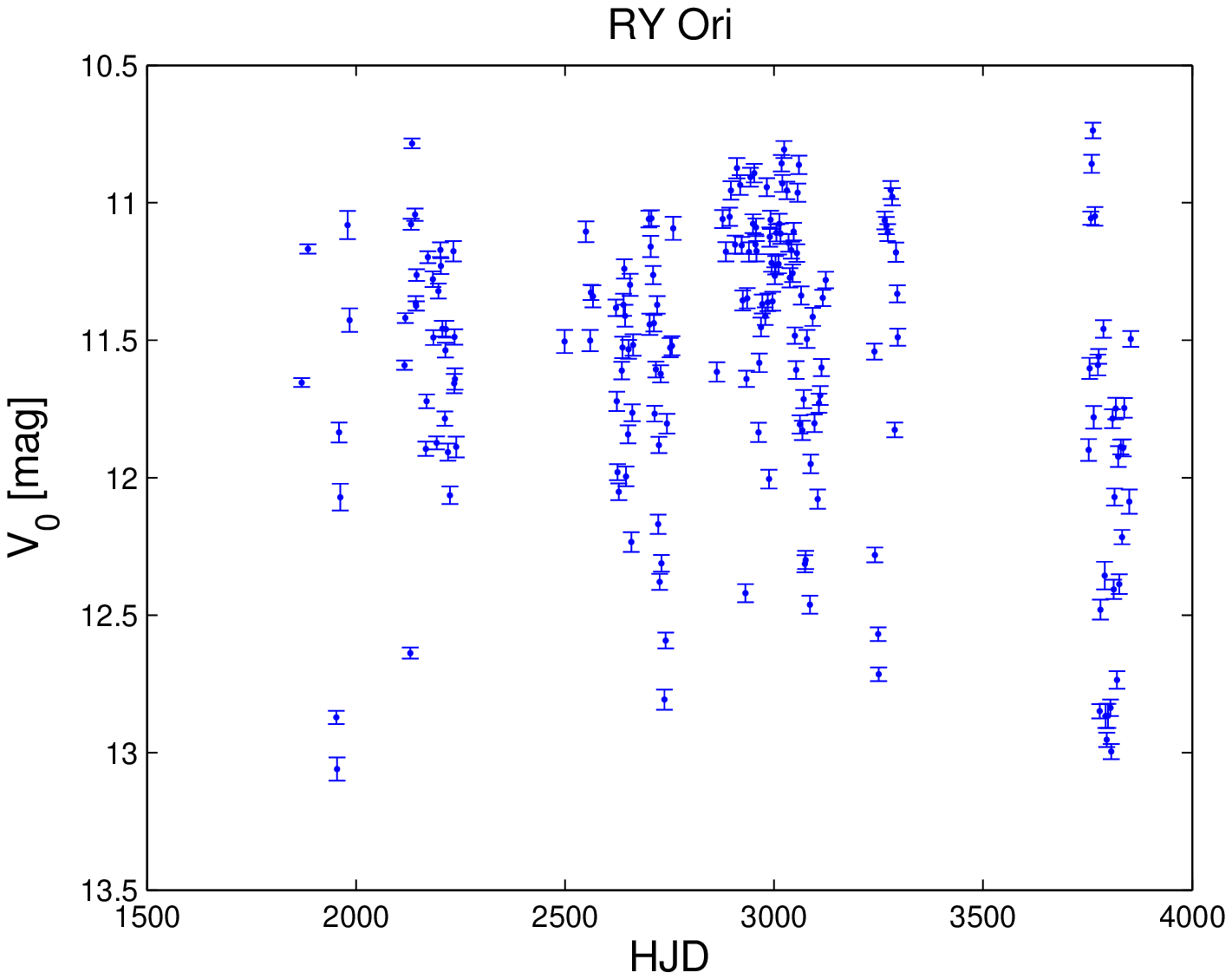}
\includegraphics[width=0.440\textwidth]{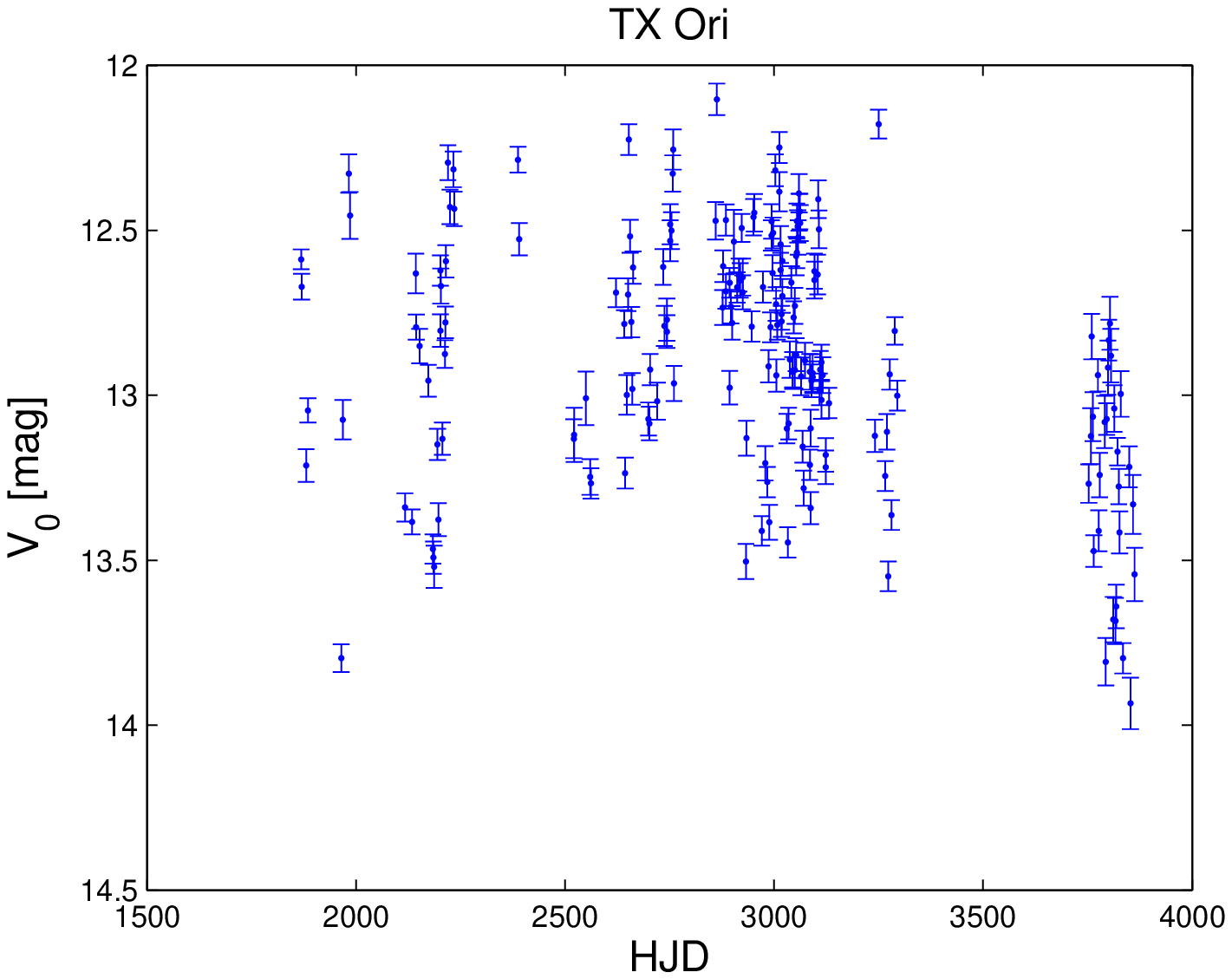}
\includegraphics[width=0.440\textwidth]{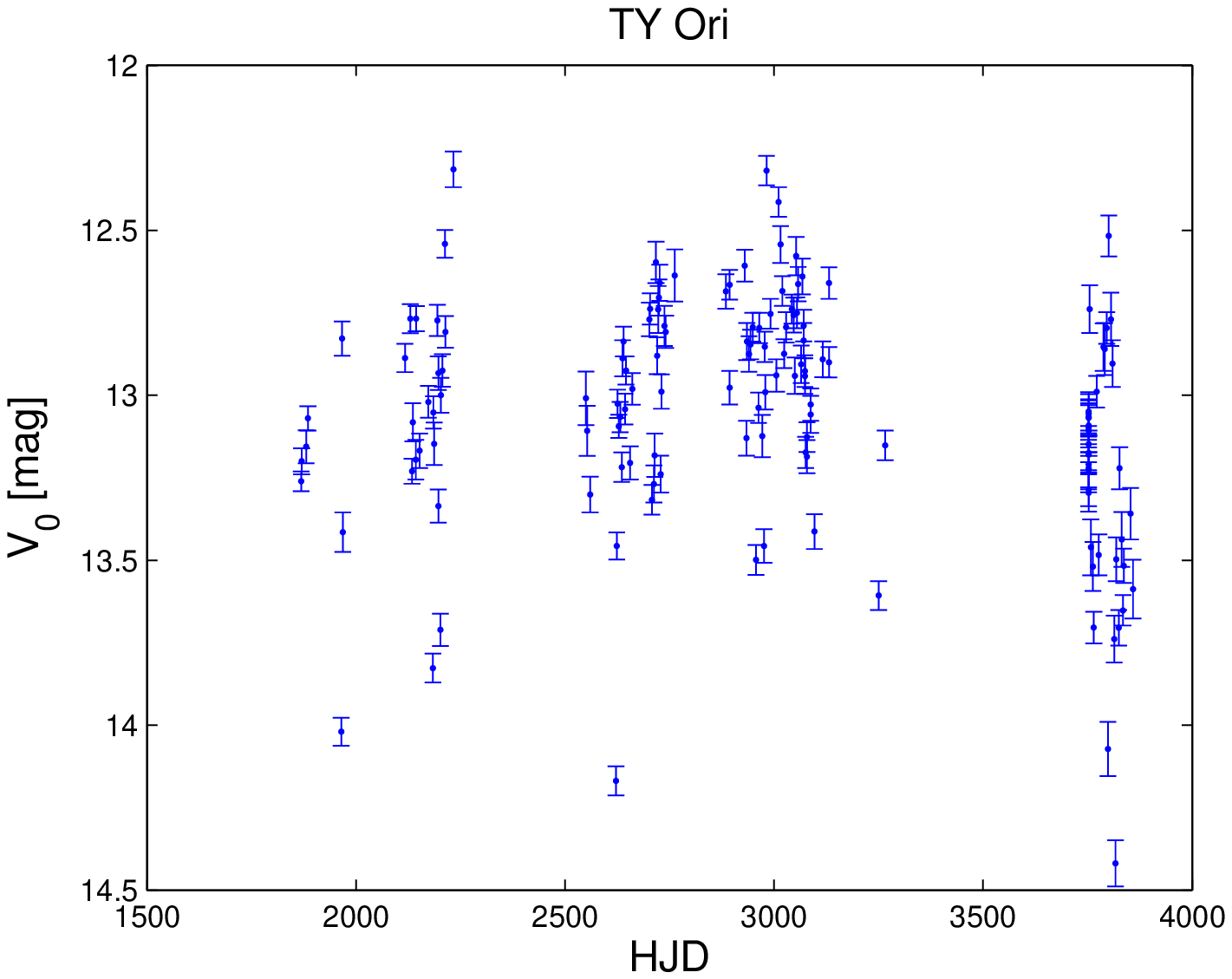}
\includegraphics[width=0.440\textwidth]{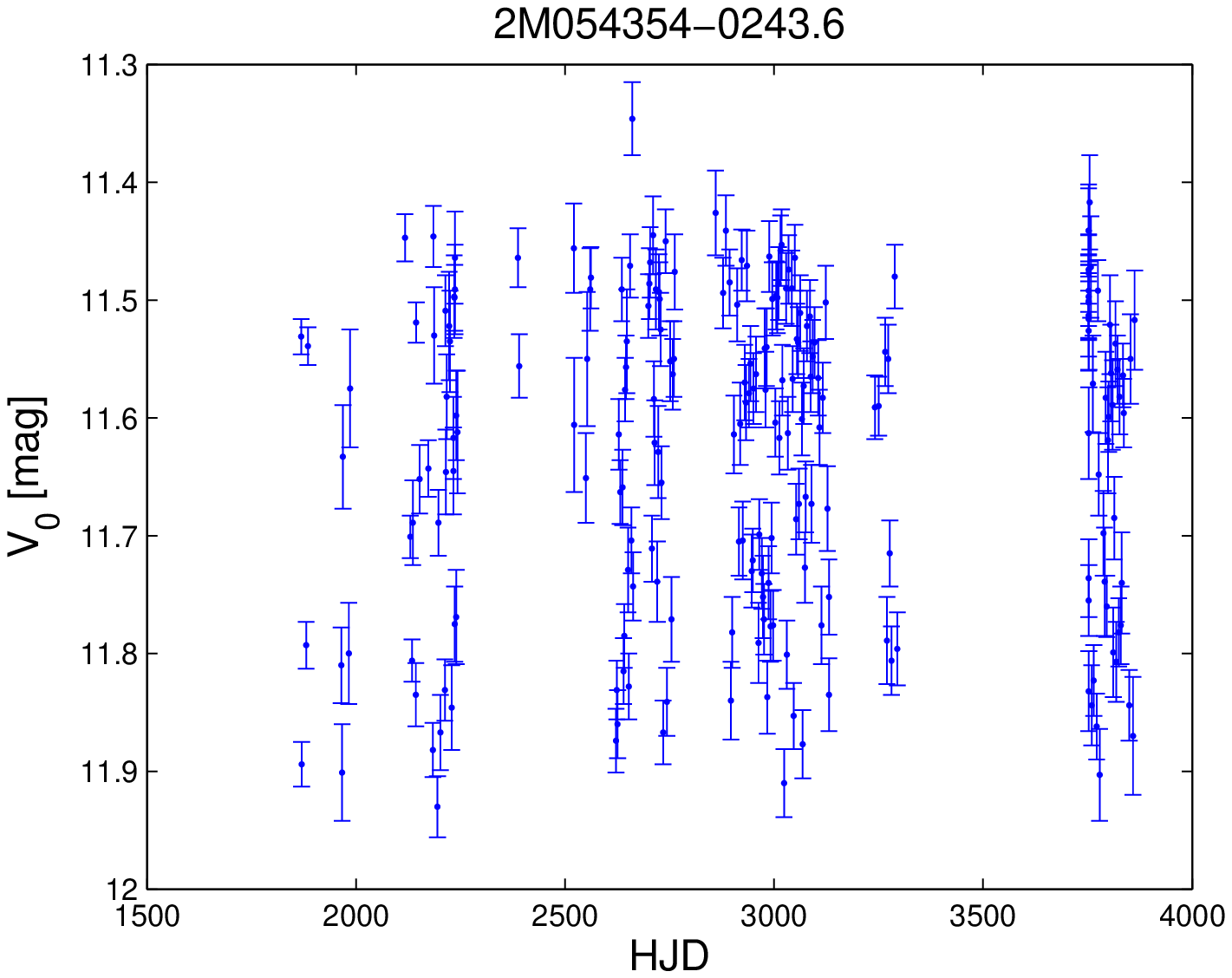}
\caption{ASAS ($V_0$) light curves of the identified variables No.~01 to~14 
(labeled).}
\label{lightsaber.01-14}
\end{figure*}
%

\begin{figure*}
\centering
\includegraphics[width=0.440\textwidth]{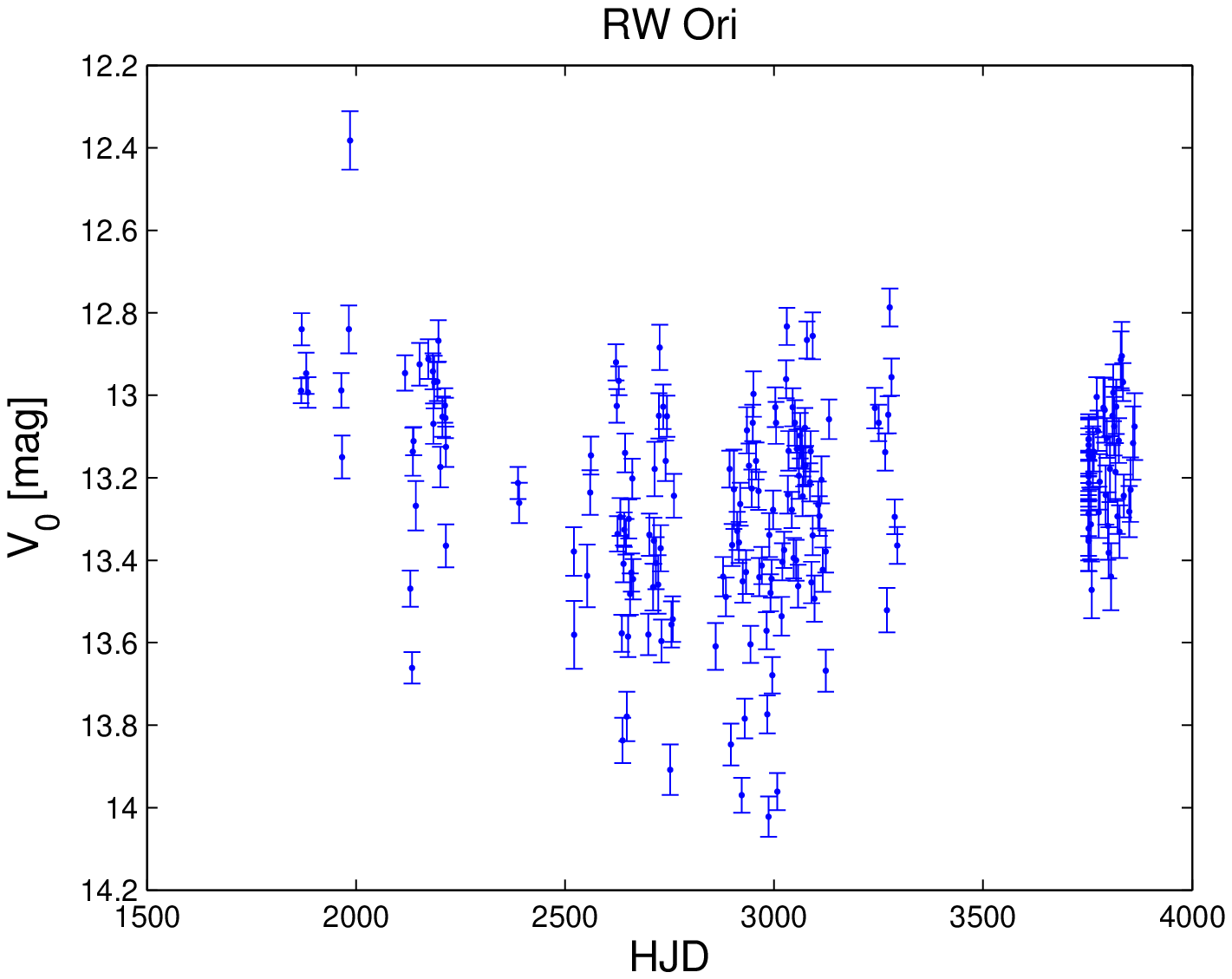}
\includegraphics[width=0.440\textwidth]{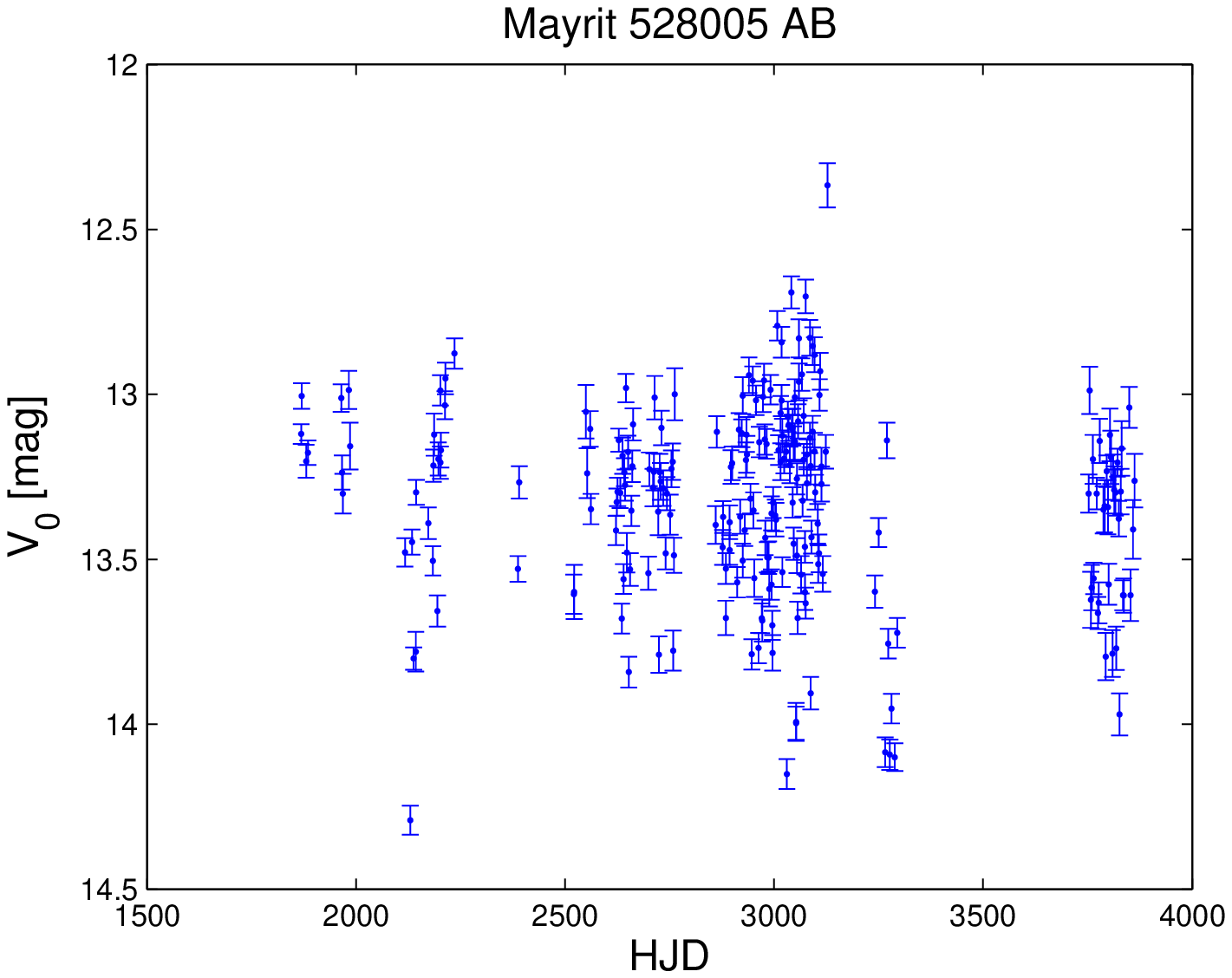}
\includegraphics[width=0.440\textwidth]{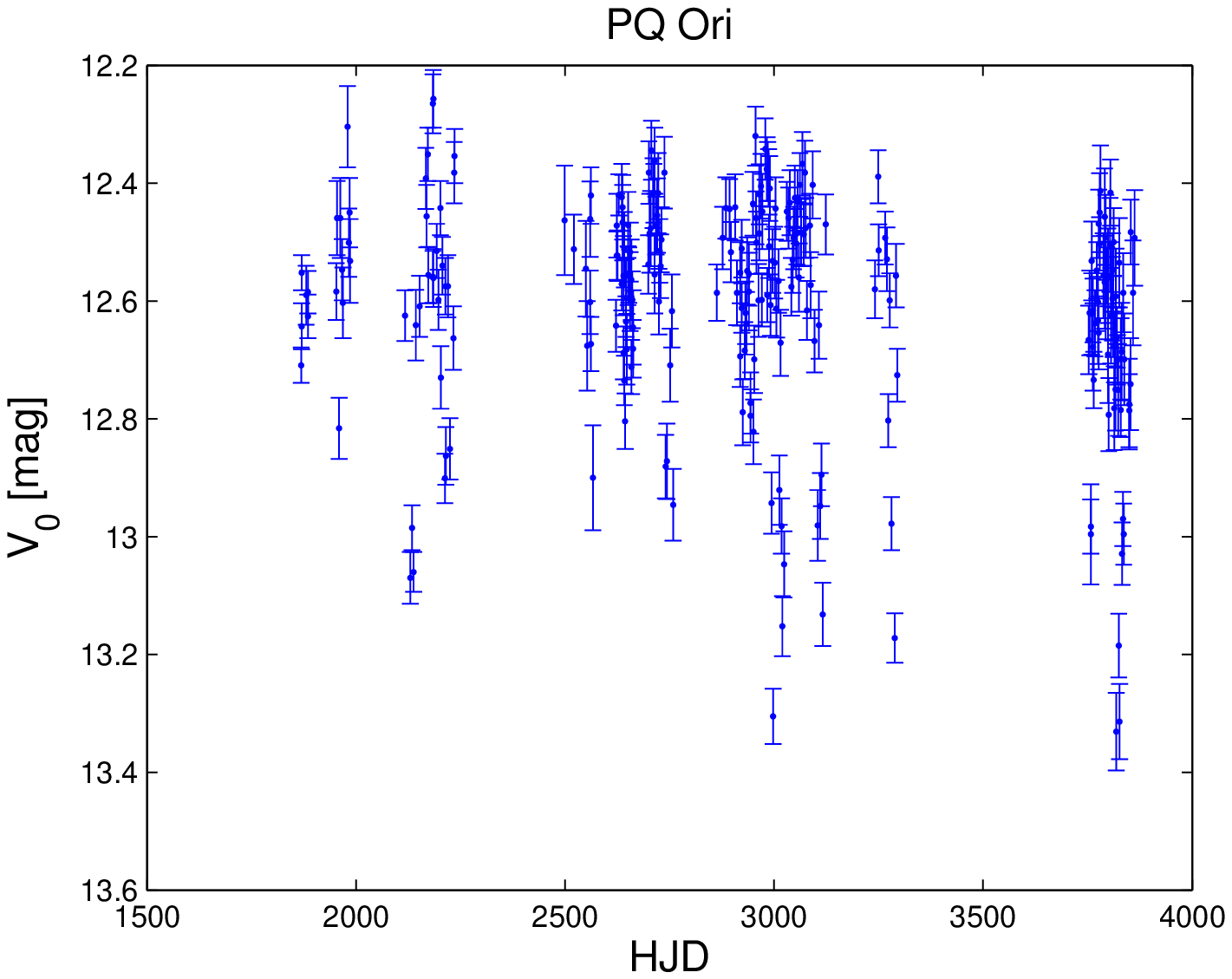}
\includegraphics[width=0.440\textwidth]{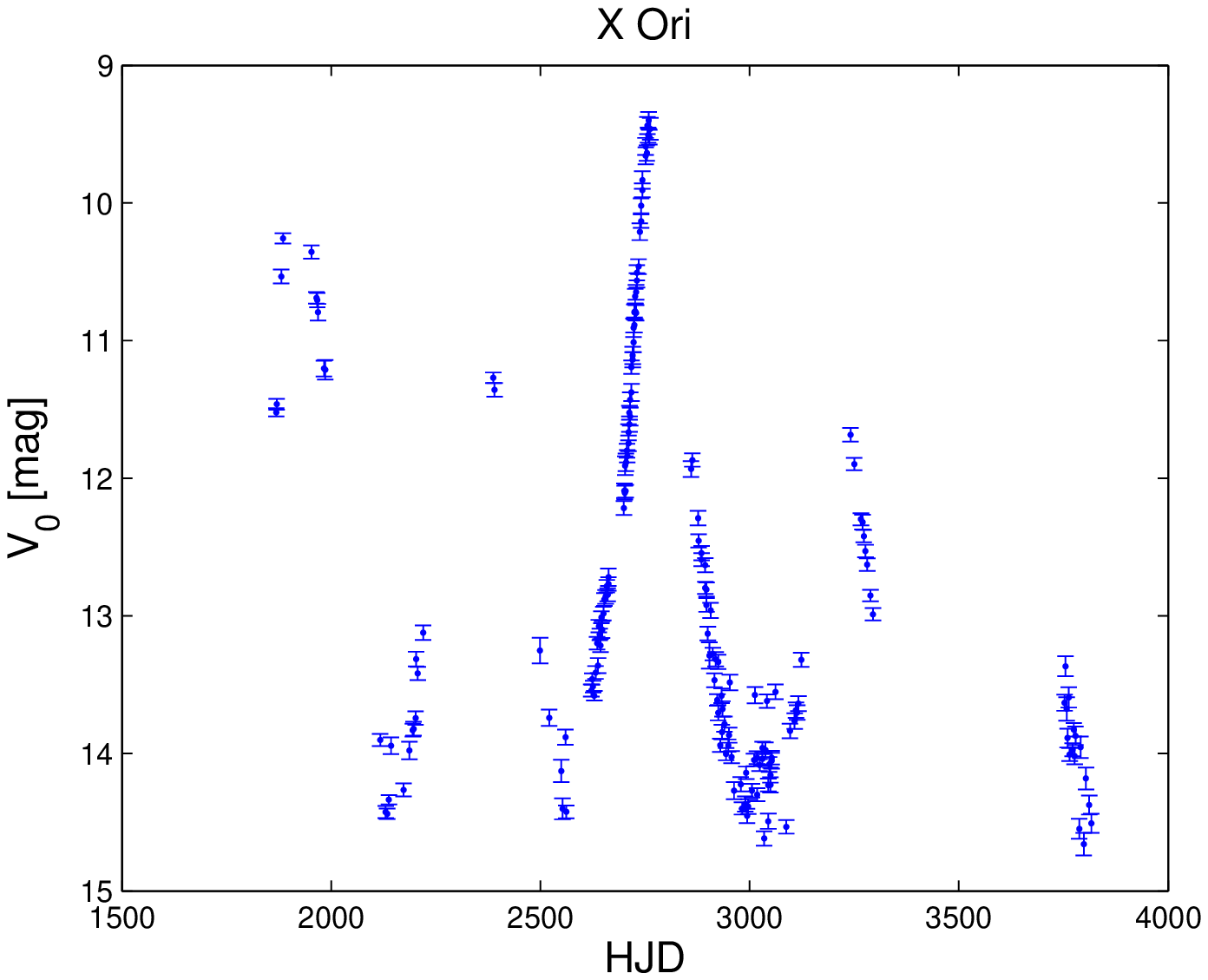}
\includegraphics[width=0.440\textwidth]{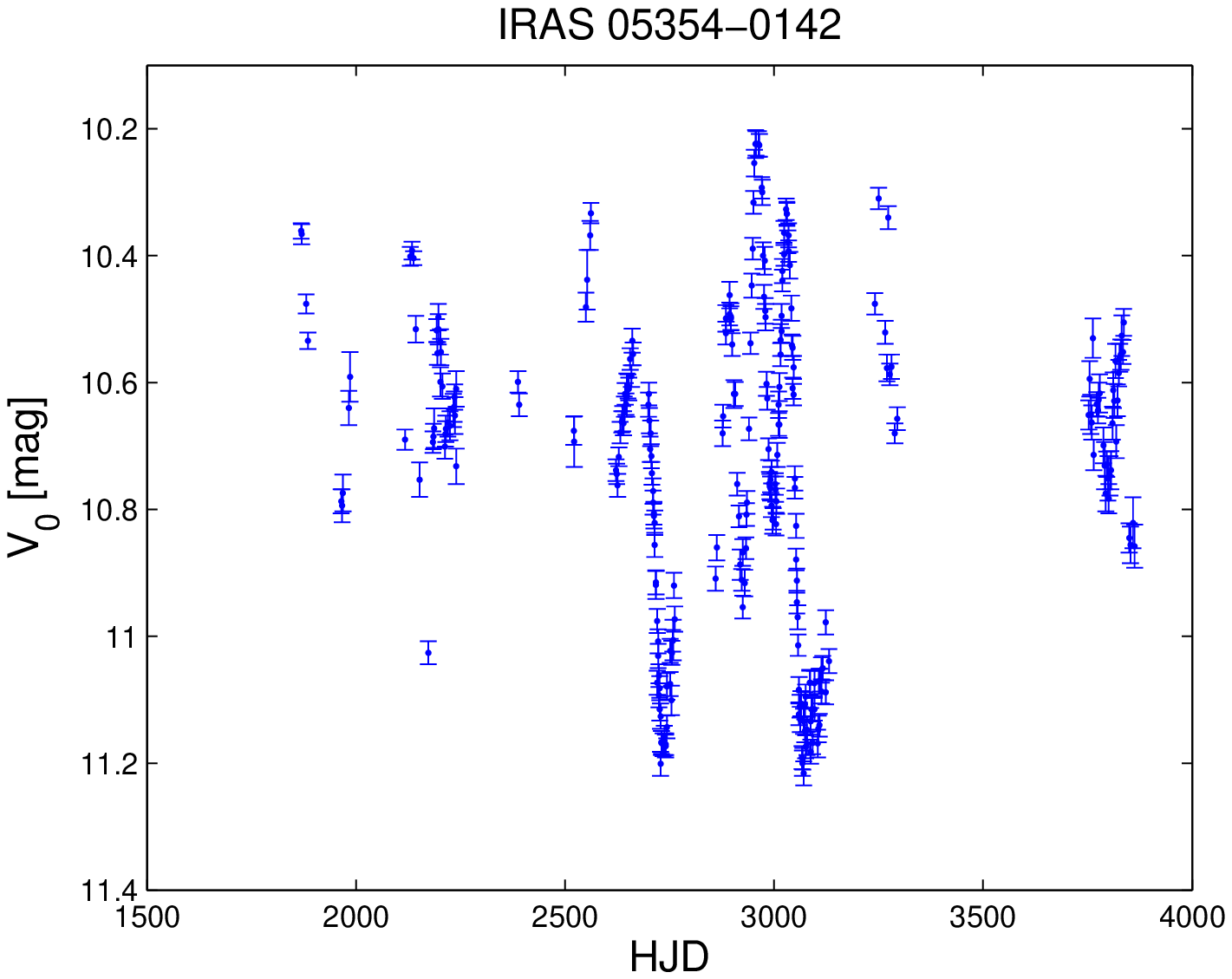}
\includegraphics[width=0.440\textwidth]{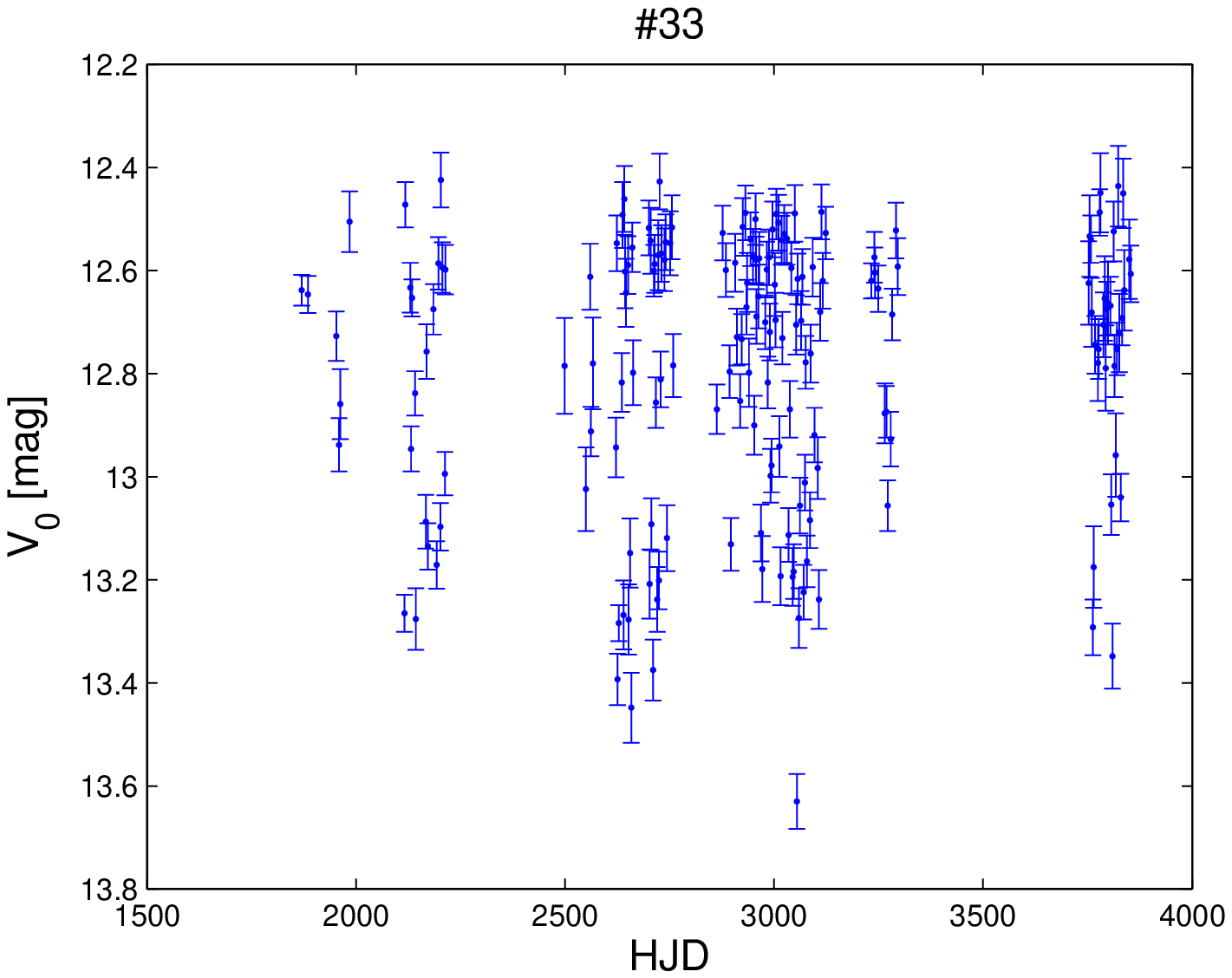}
\includegraphics[width=0.440\textwidth]{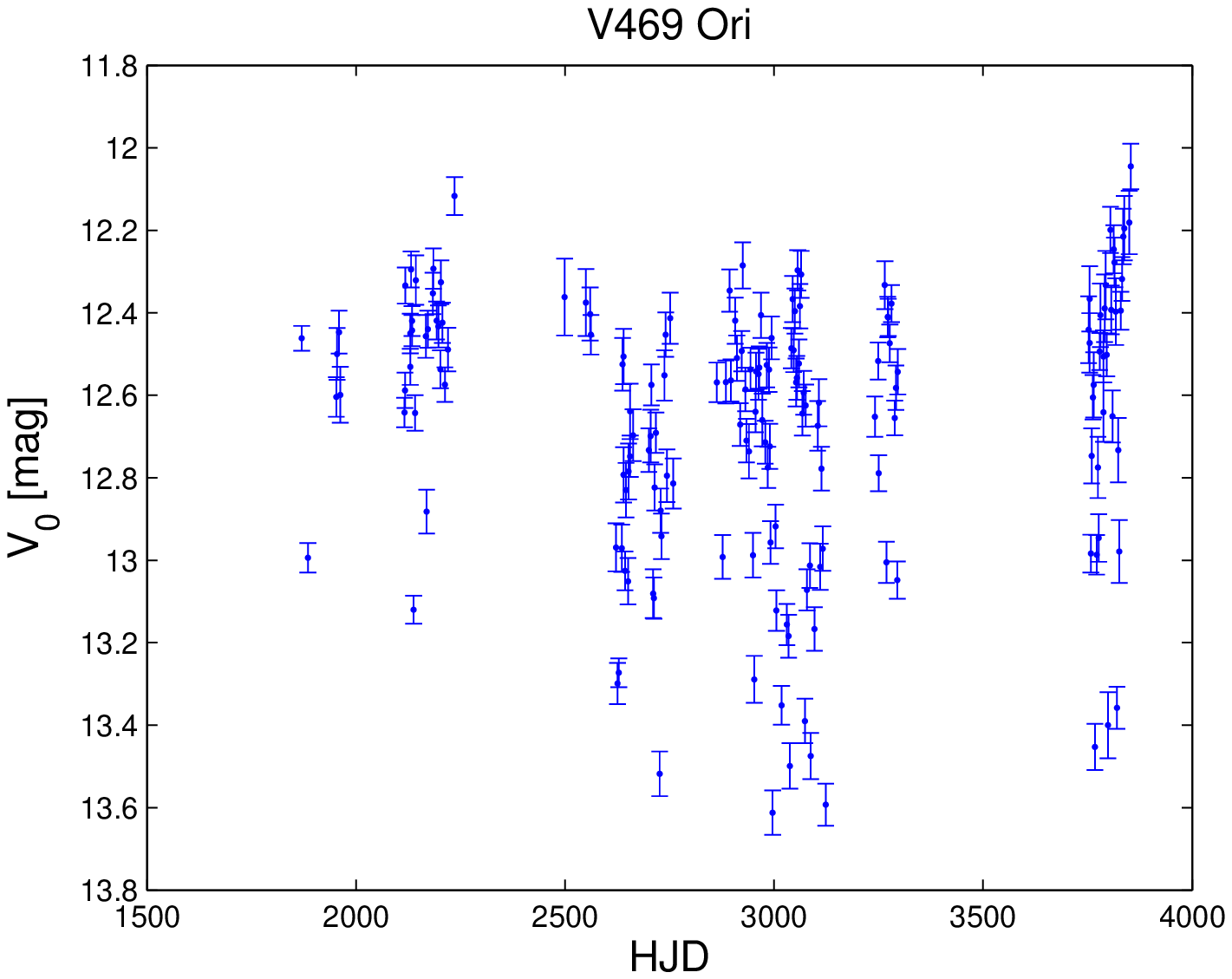}
\includegraphics[width=0.440\textwidth]{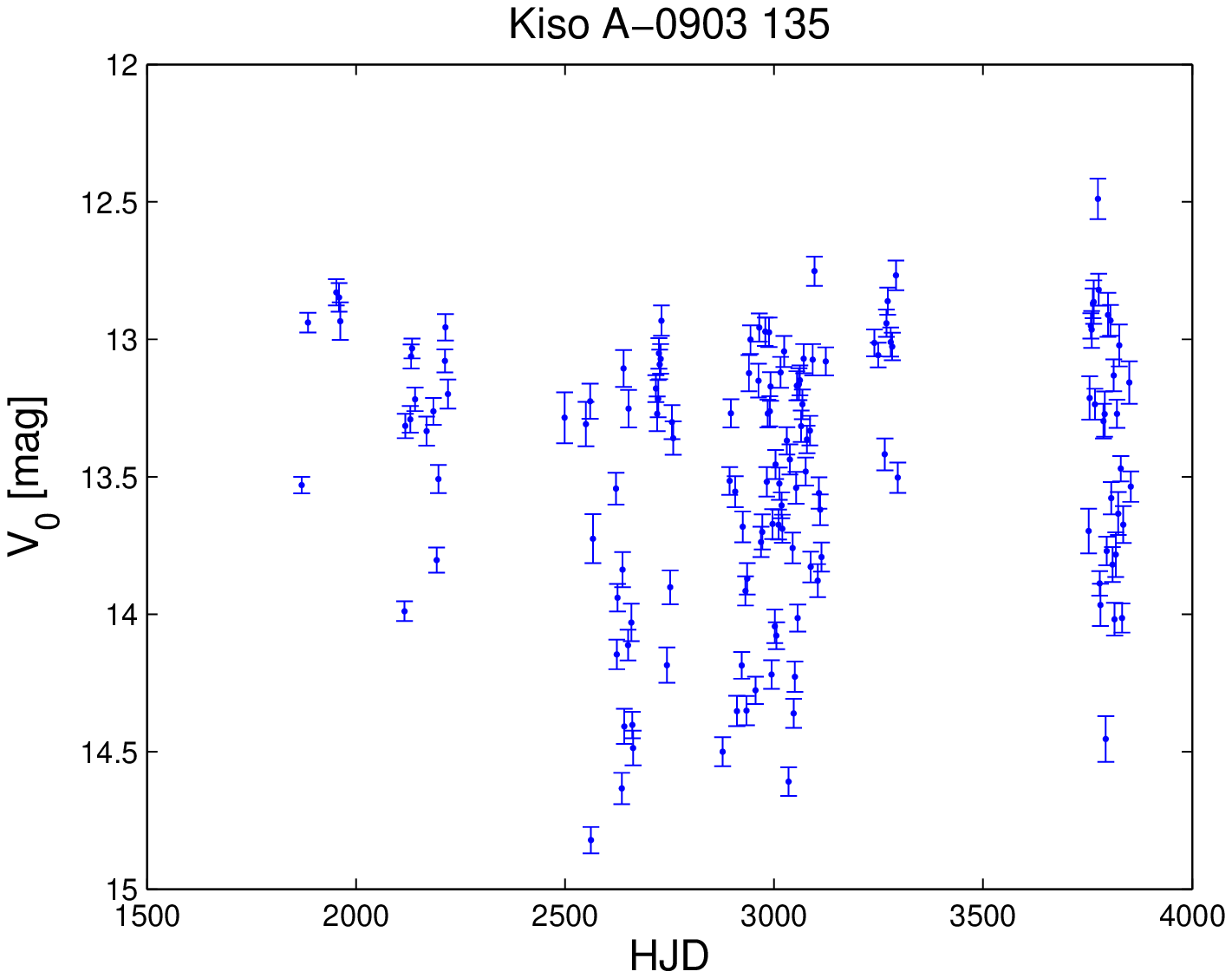}
\caption{Same as Fig.~\ref{lightsaber.01-14} but for identified variables No.~15 
to~41.}
\label{lightsaber.15-41}
\end{figure*}
%

\begin{figure*}
\centering
\includegraphics[width=0.440\textwidth]{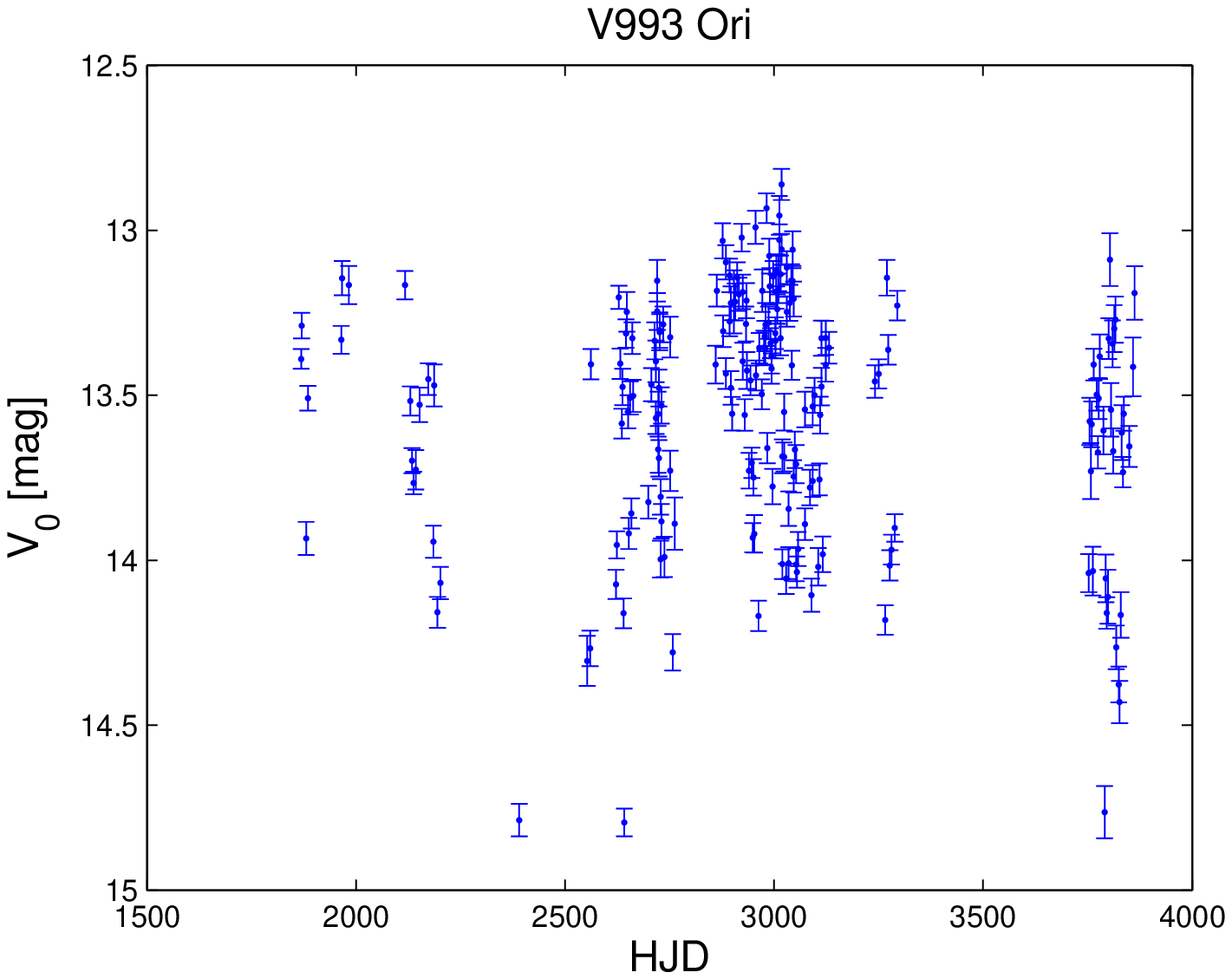}
\includegraphics[width=0.440\textwidth]{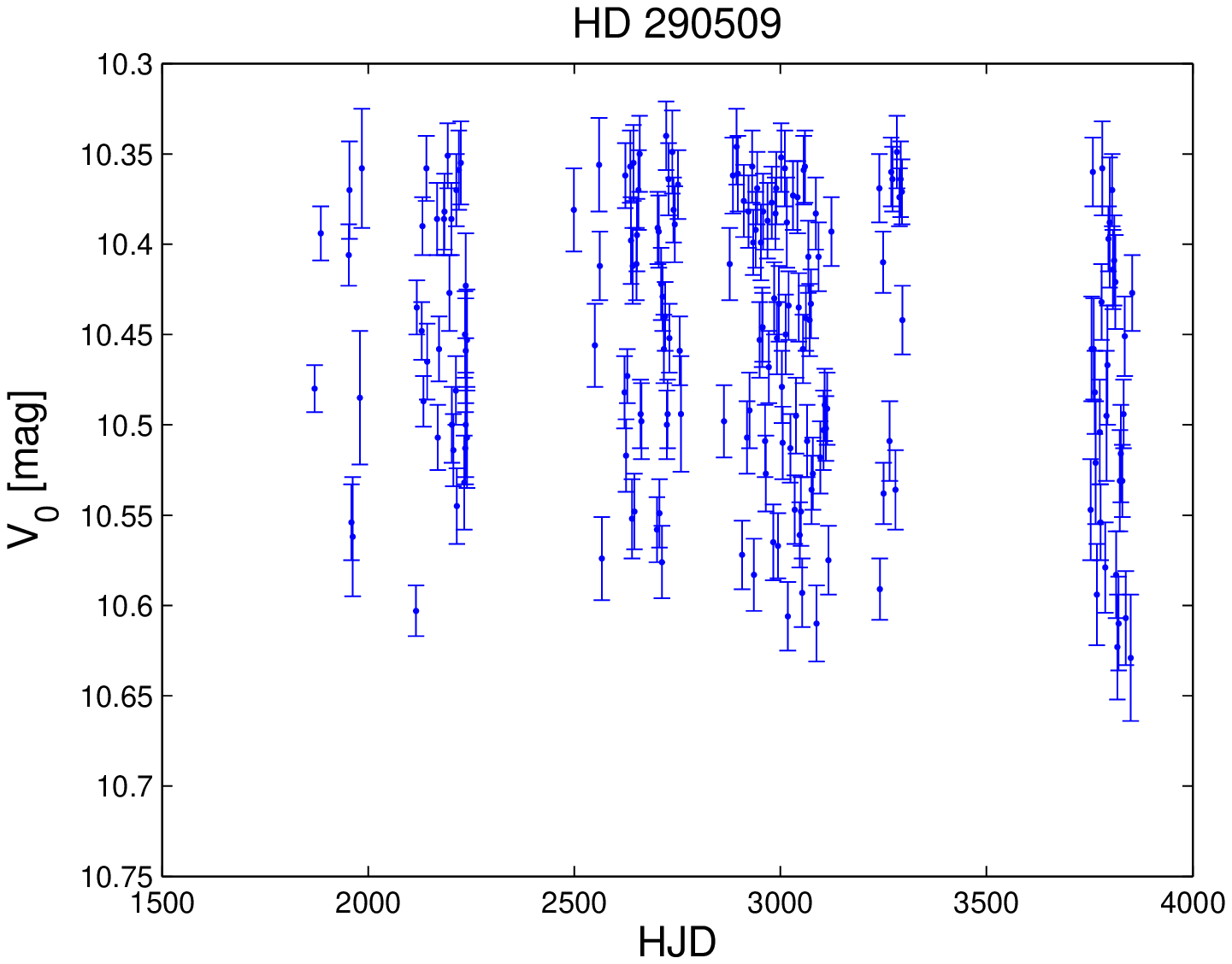}
\includegraphics[width=0.440\textwidth]{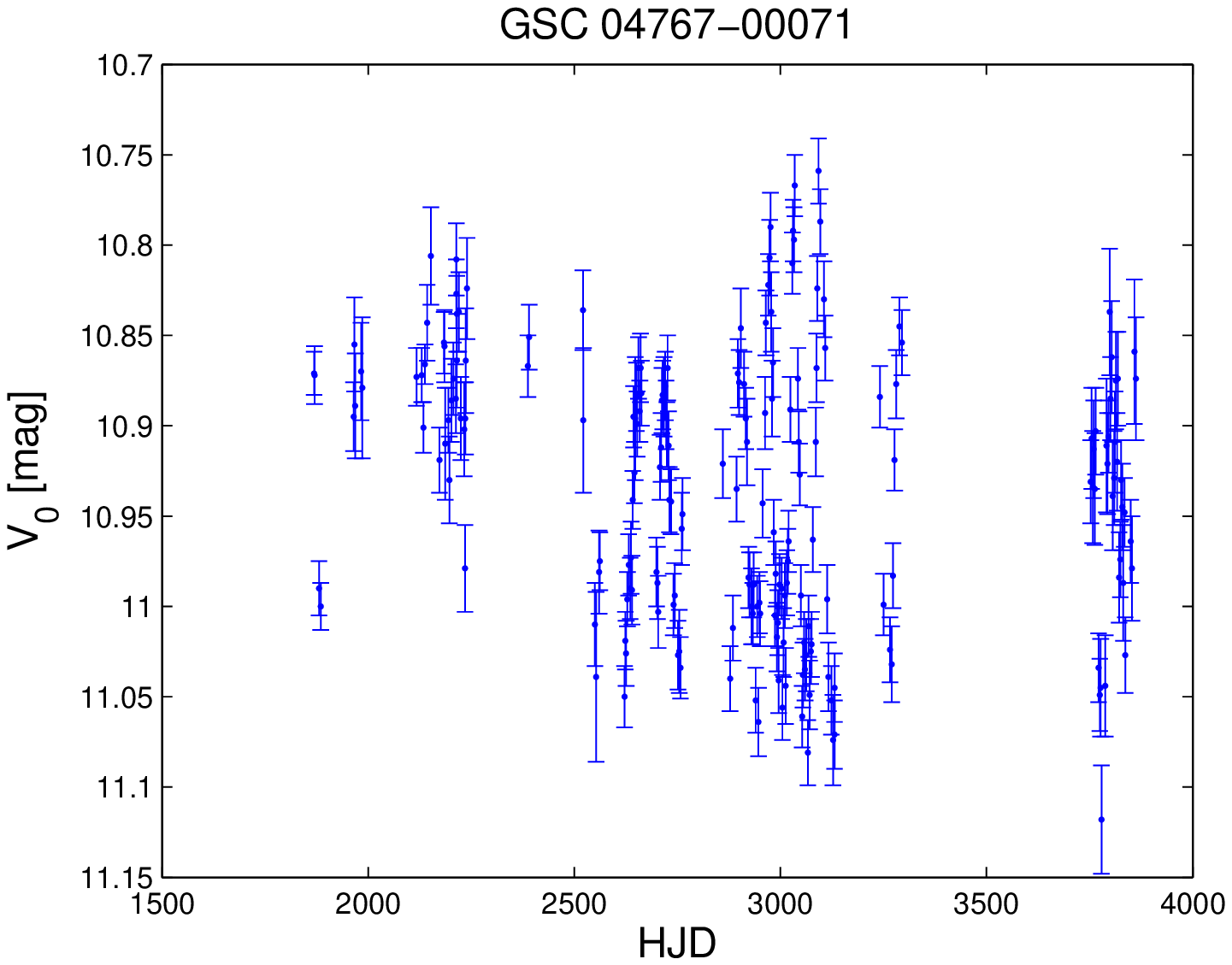}
\includegraphics[width=0.440\textwidth]{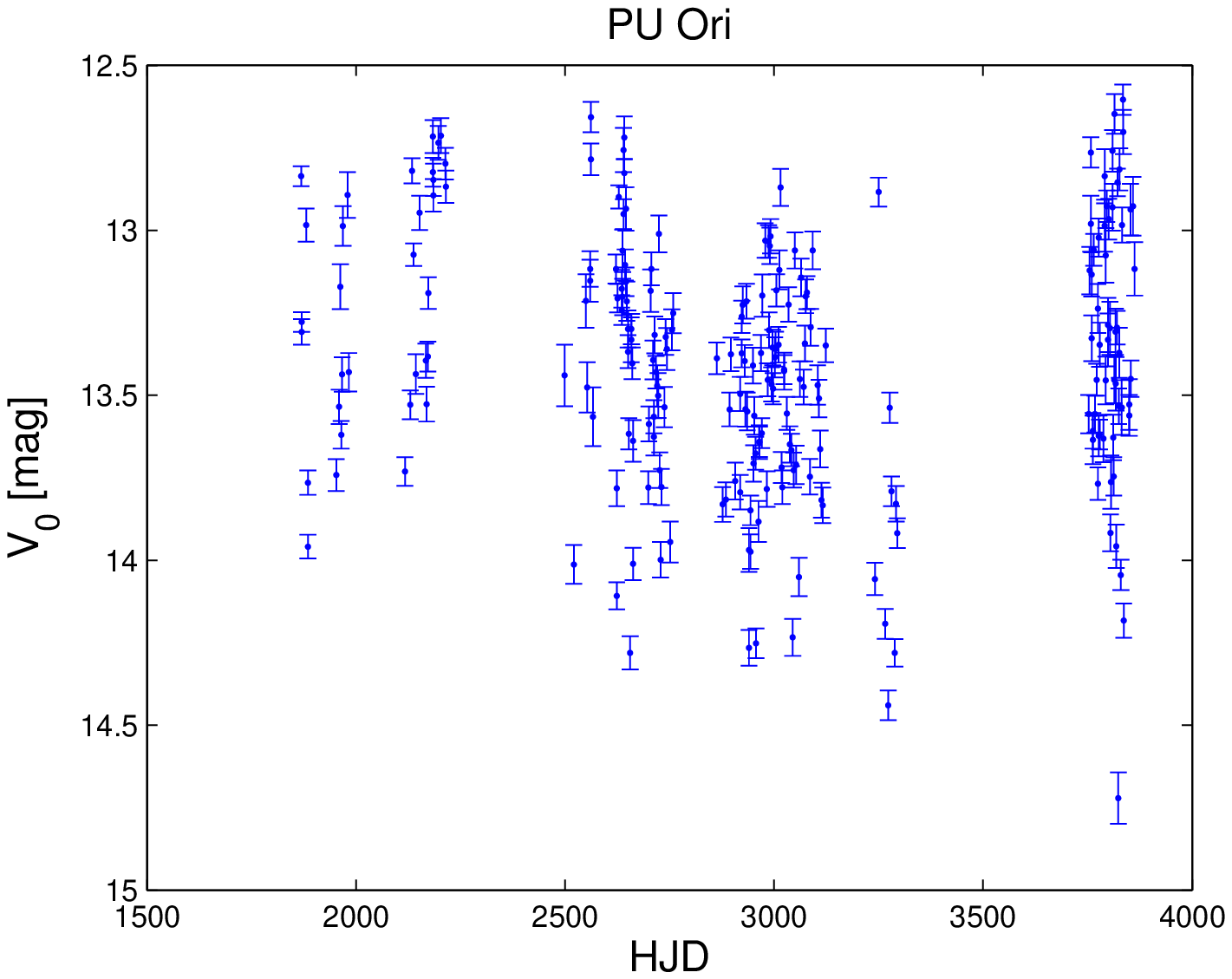}
\includegraphics[width=0.440\textwidth]{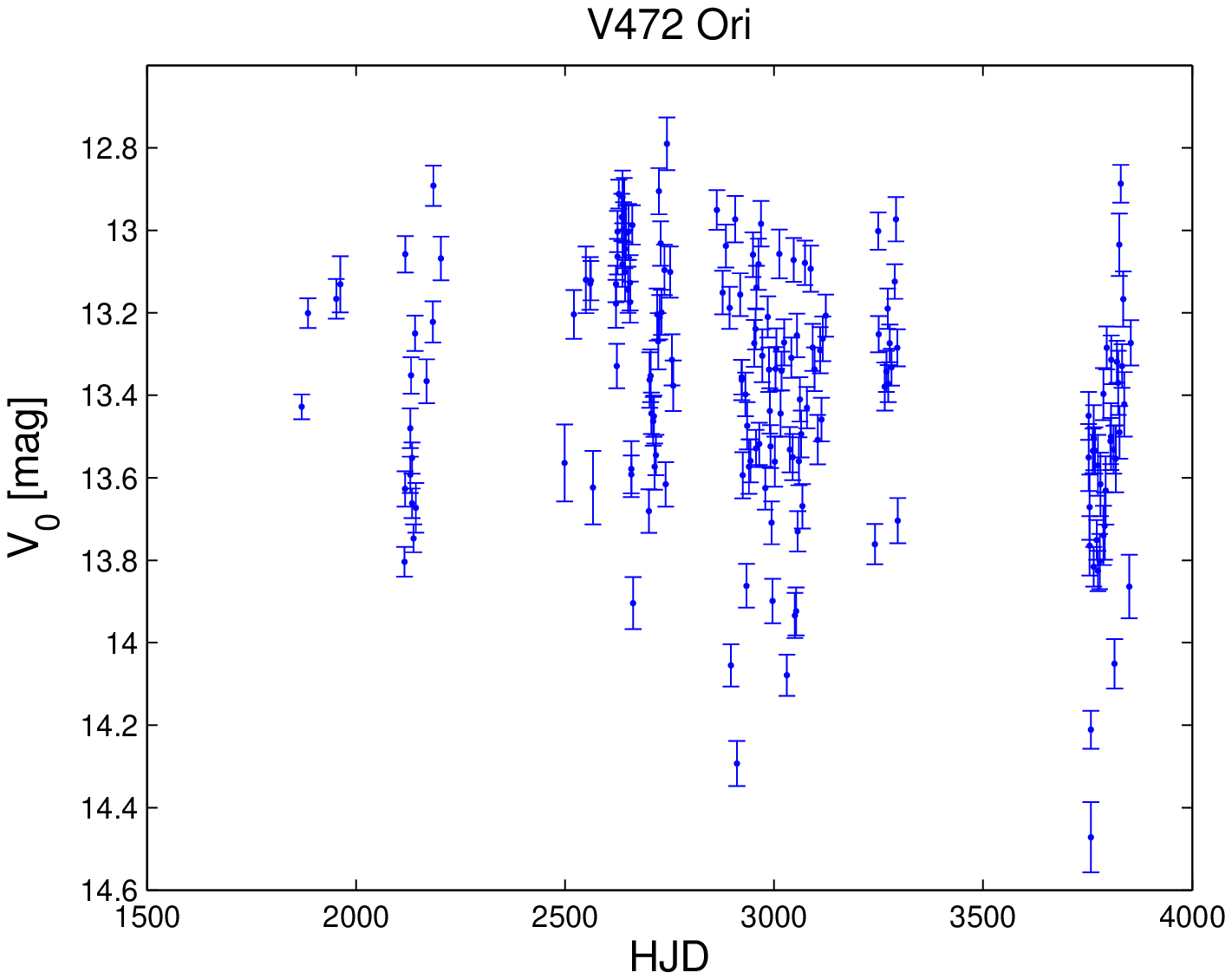}
\includegraphics[width=0.440\textwidth]{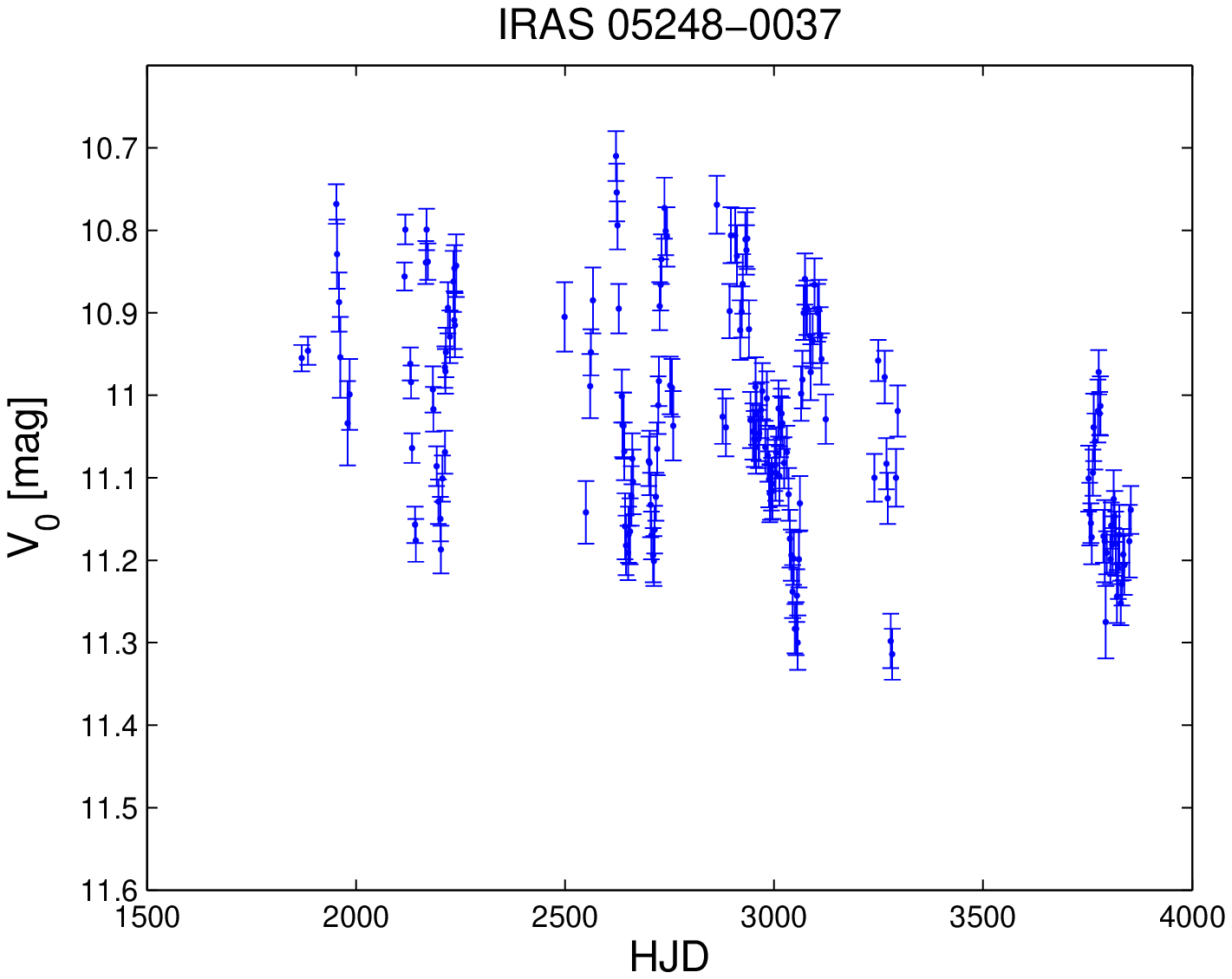}
\includegraphics[width=0.440\textwidth]{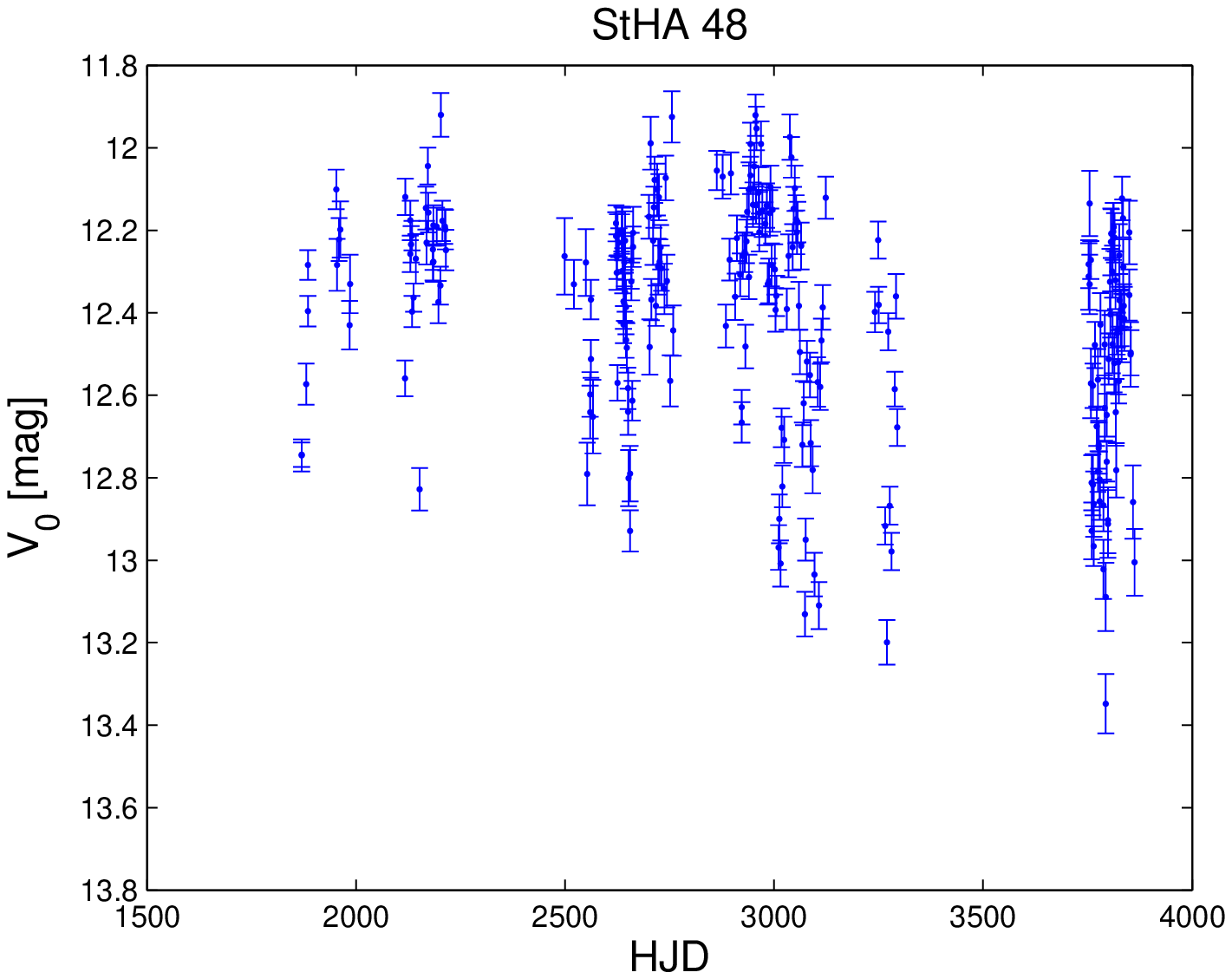}
\includegraphics[width=0.440\textwidth]{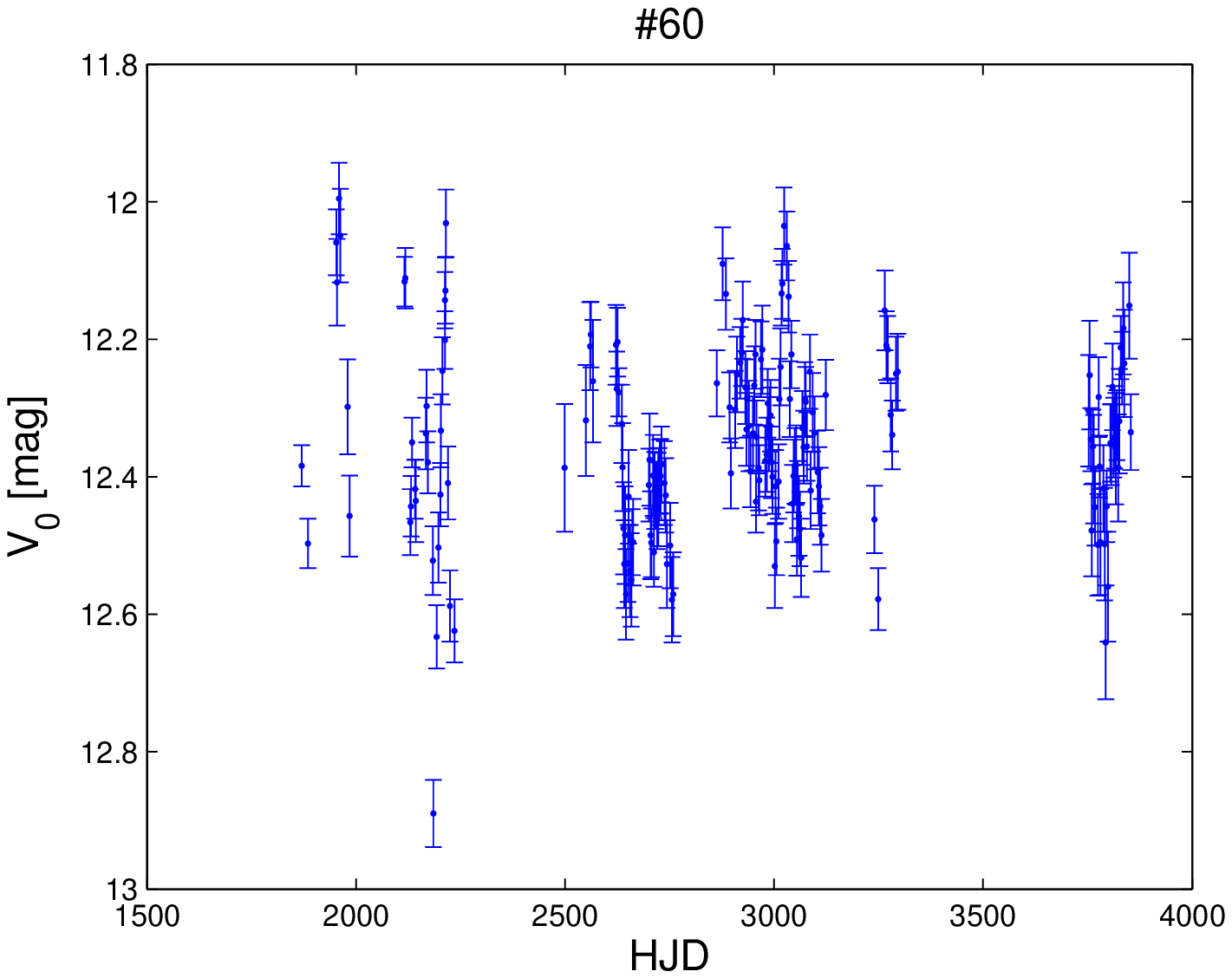}
\caption{Same as Fig.~\ref{lightsaber.01-14} but for identified variables No.~43 
to~60.}
\label{lightsaber.43-60}
\end{figure*}
%

\begin{figure*}
\centering
\includegraphics[width=0.440\textwidth]{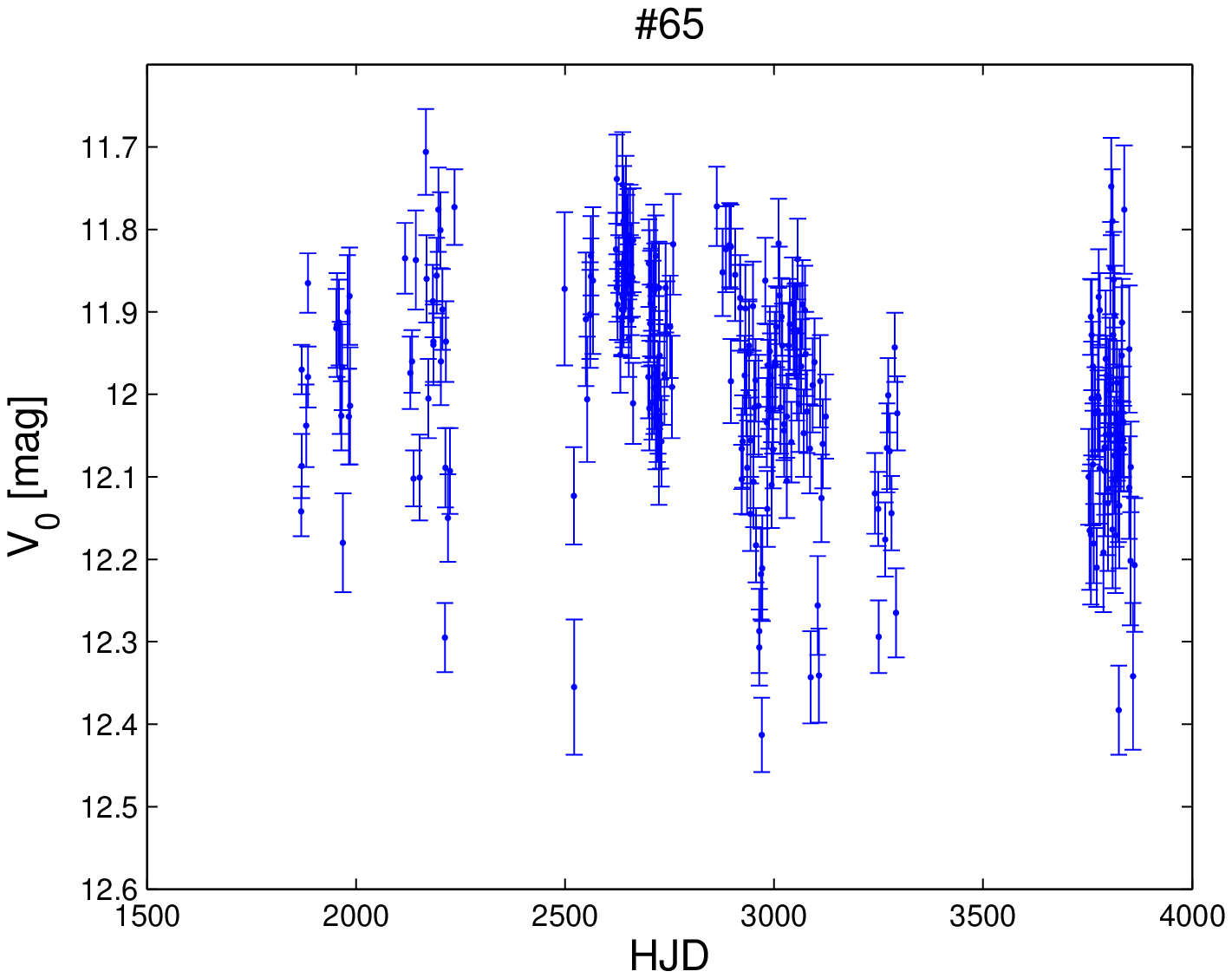}
\includegraphics[width=0.440\textwidth]{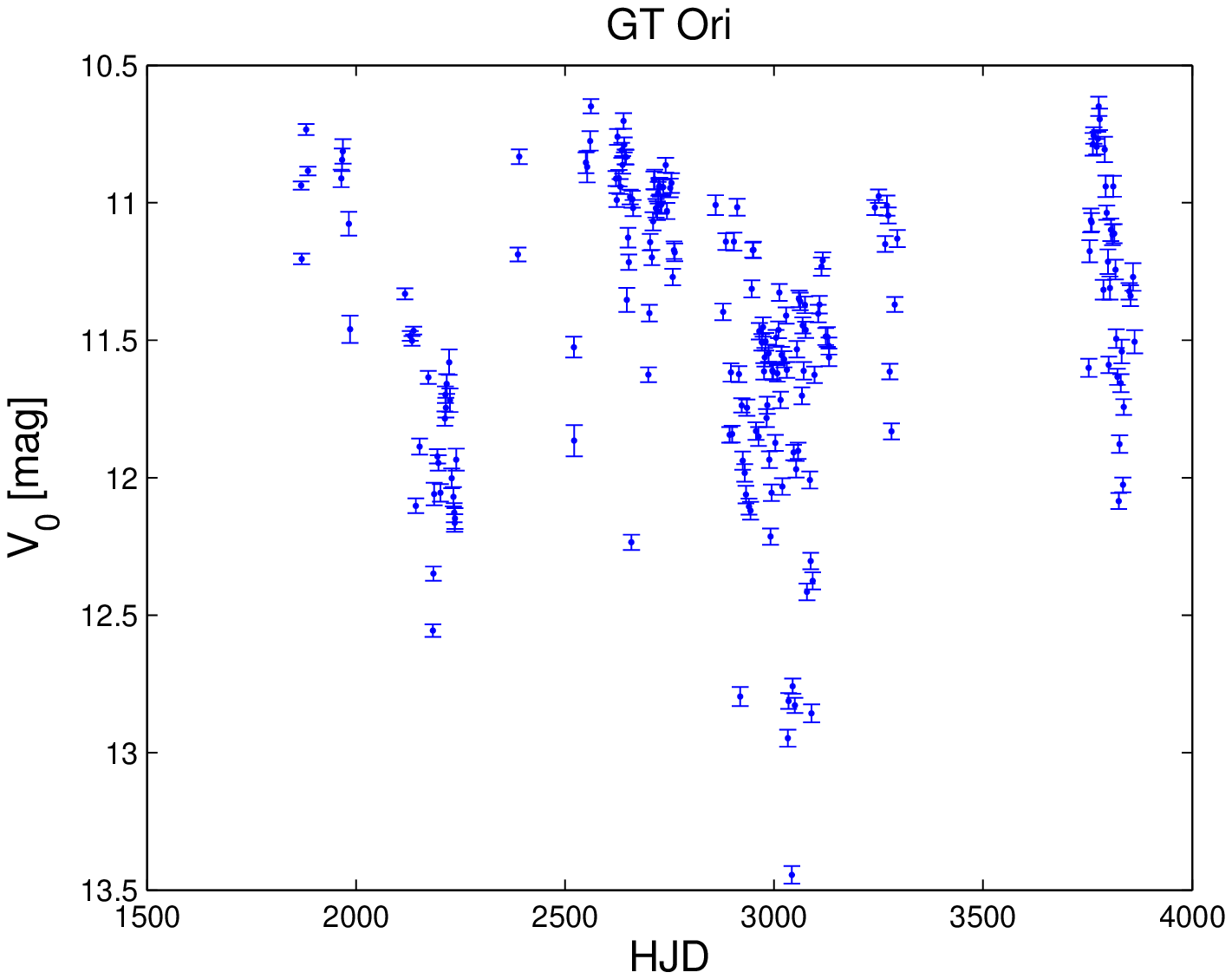}
\includegraphics[width=0.440\textwidth]{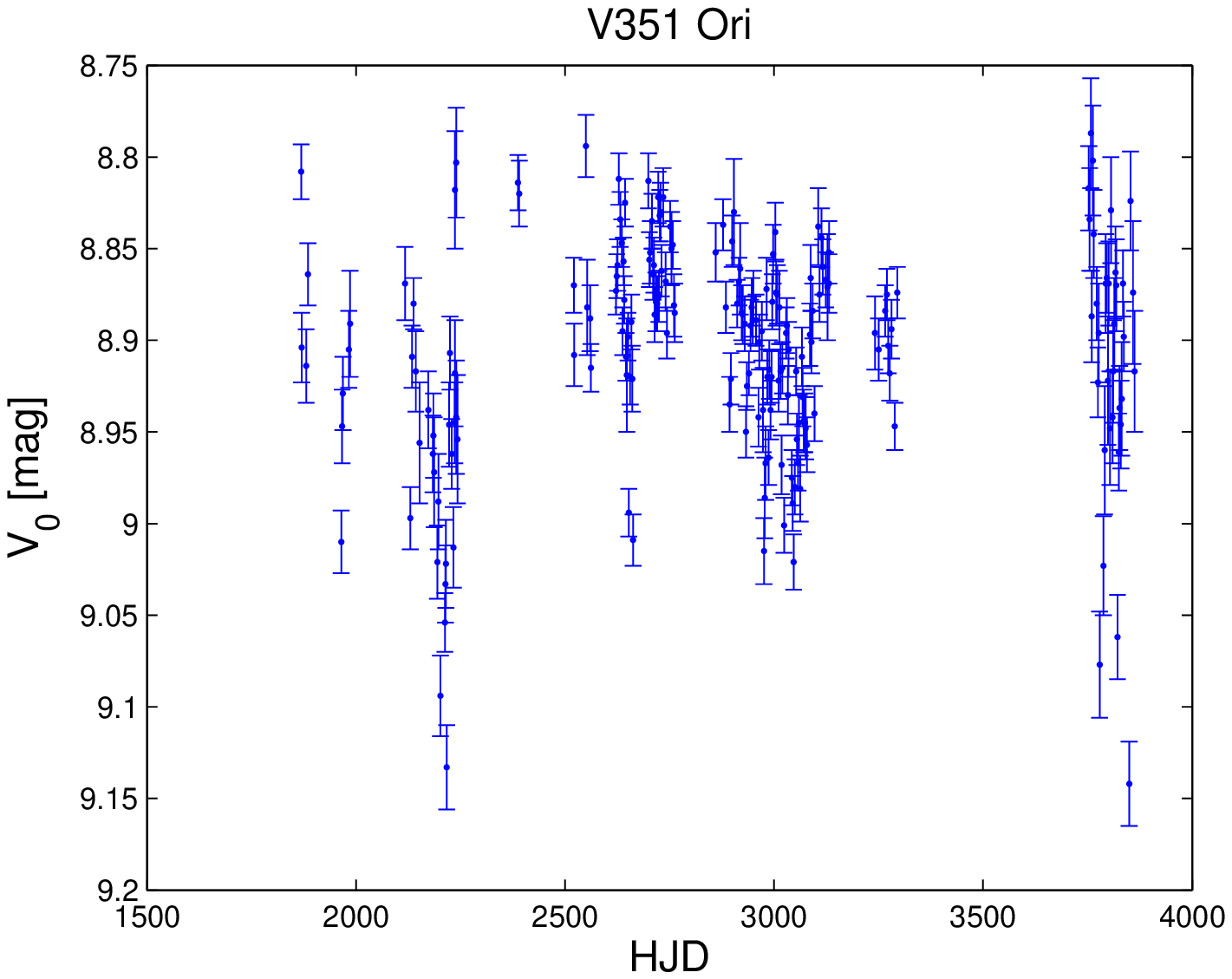}
\includegraphics[width=0.440\textwidth]{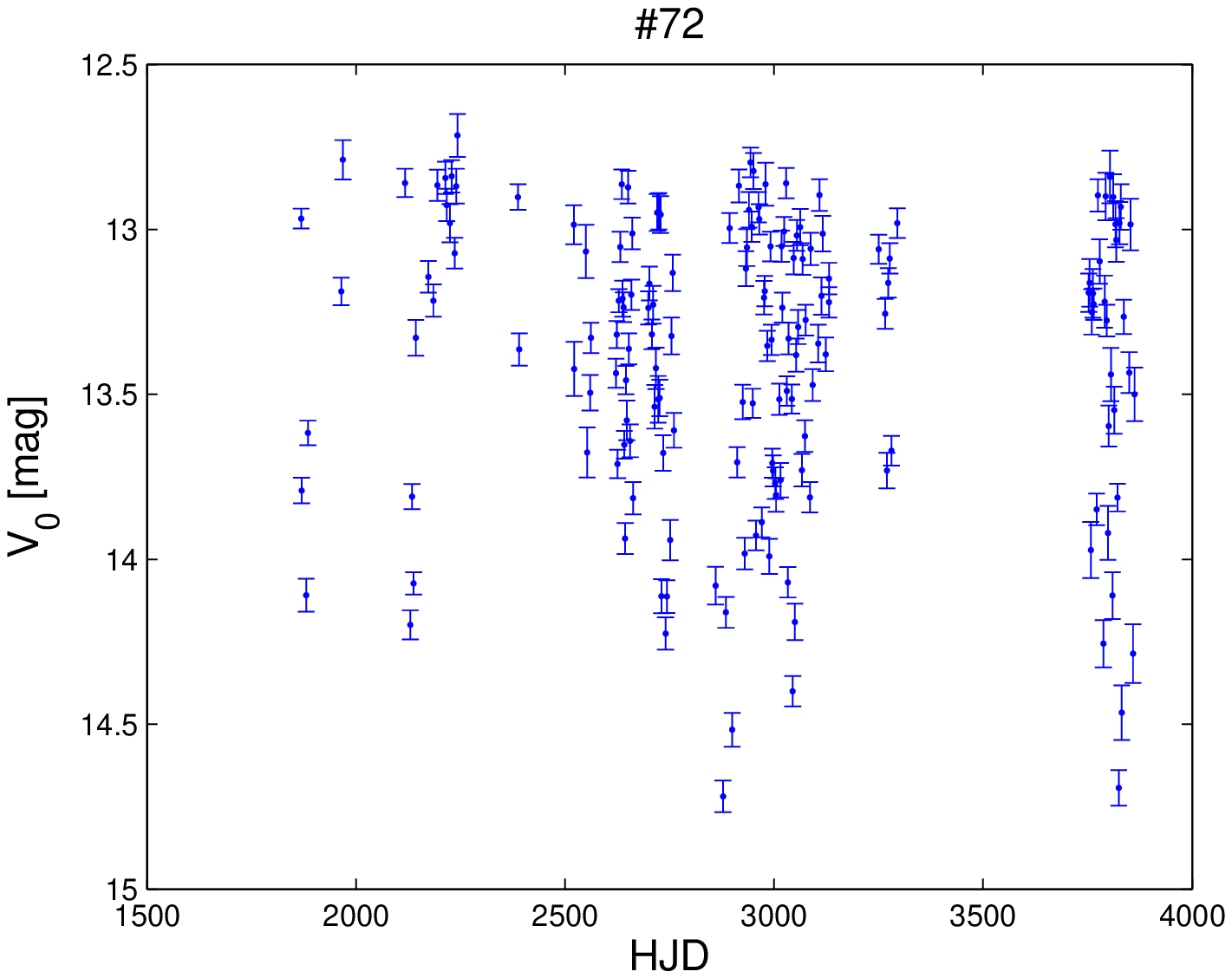}
\includegraphics[width=0.440\textwidth]{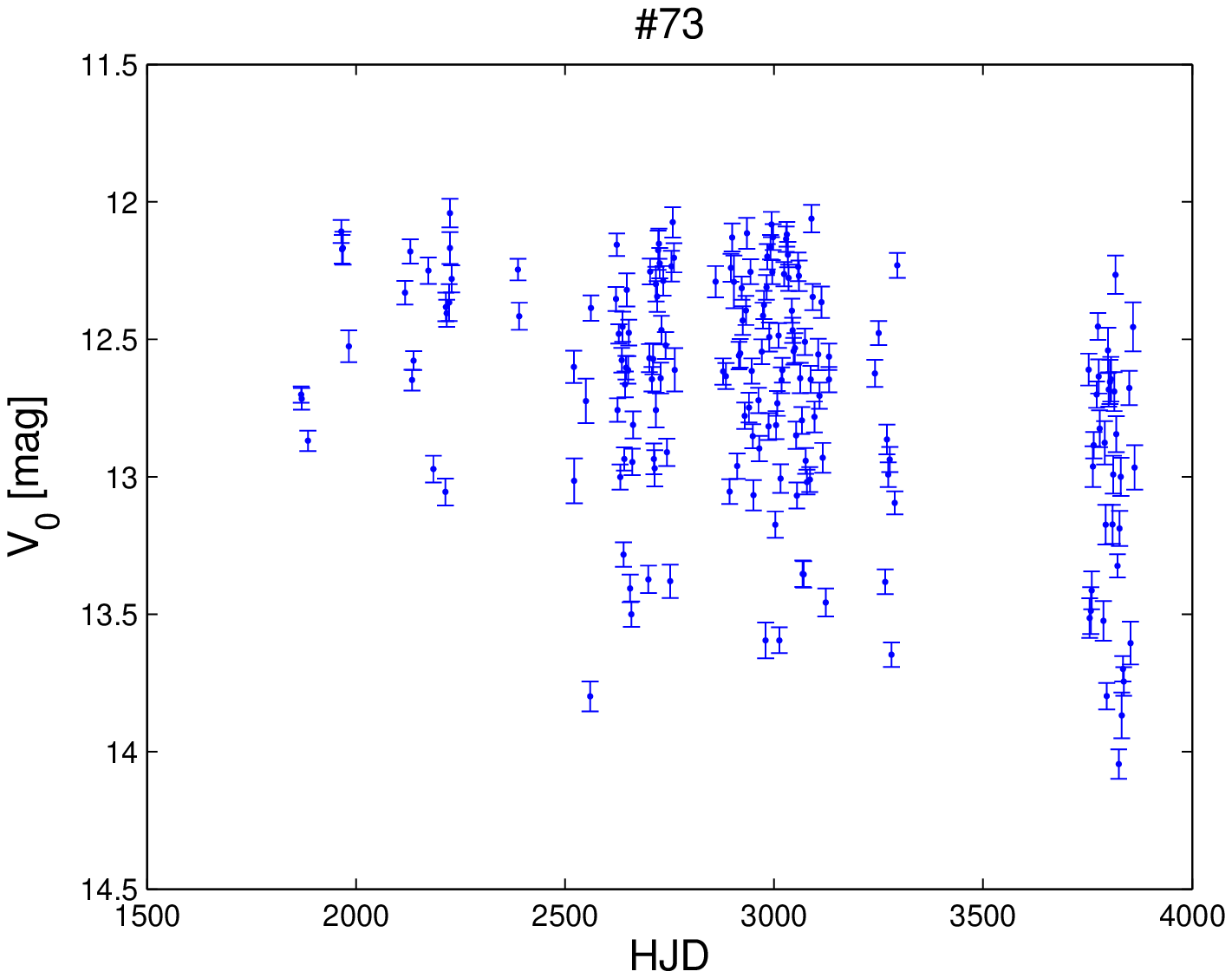}
\includegraphics[width=0.440\textwidth]{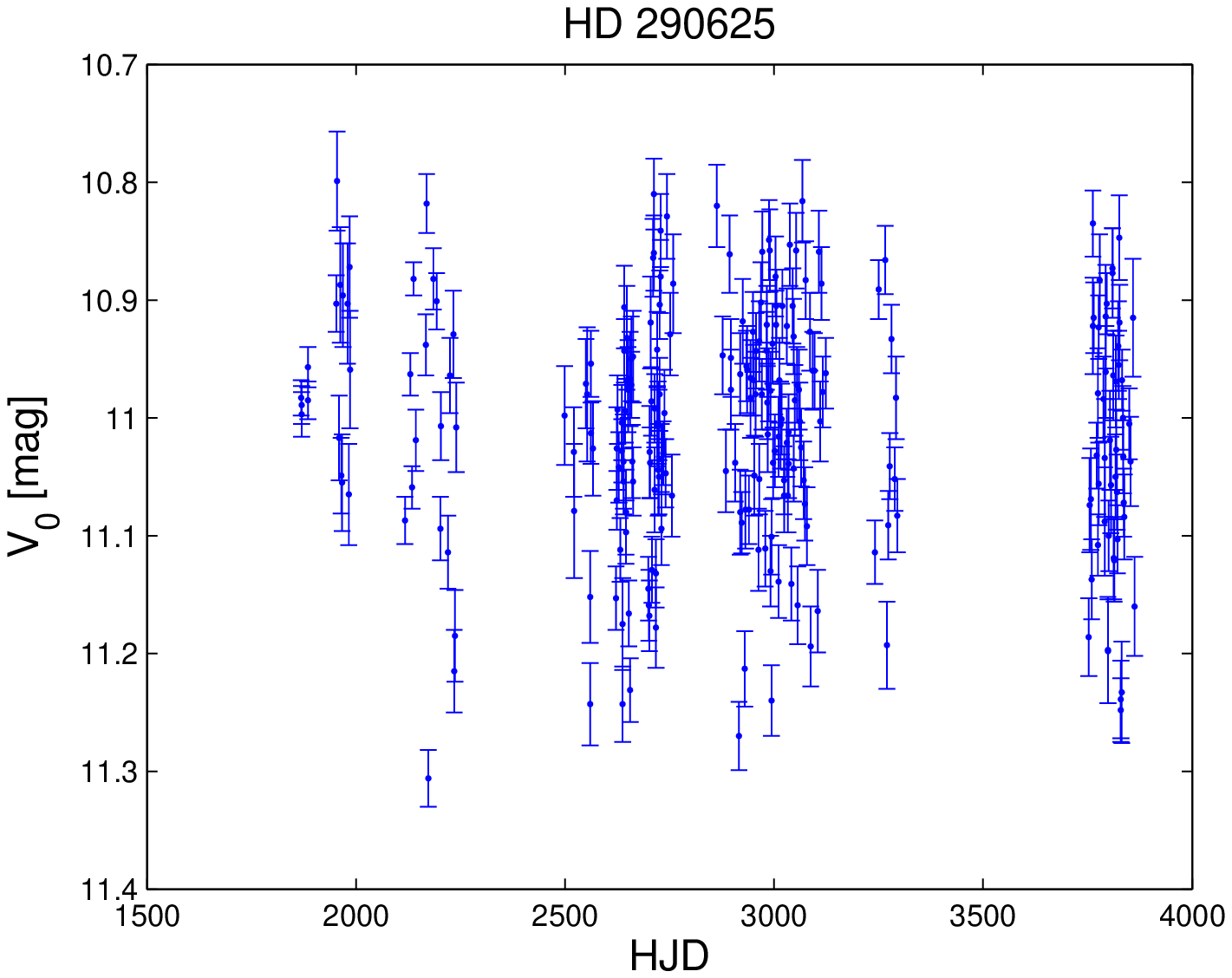}
\includegraphics[width=0.440\textwidth]{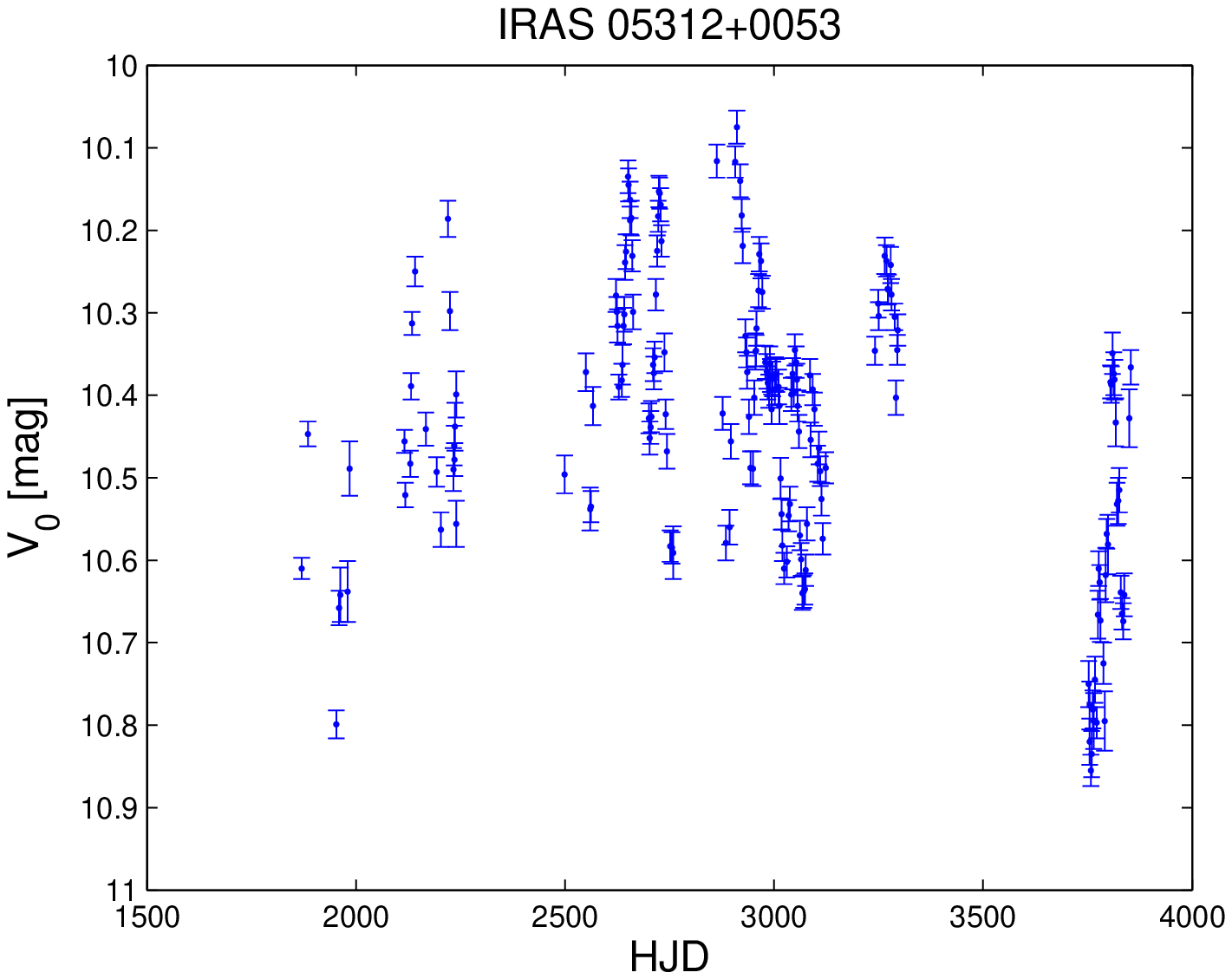}
\includegraphics[width=0.440\textwidth]{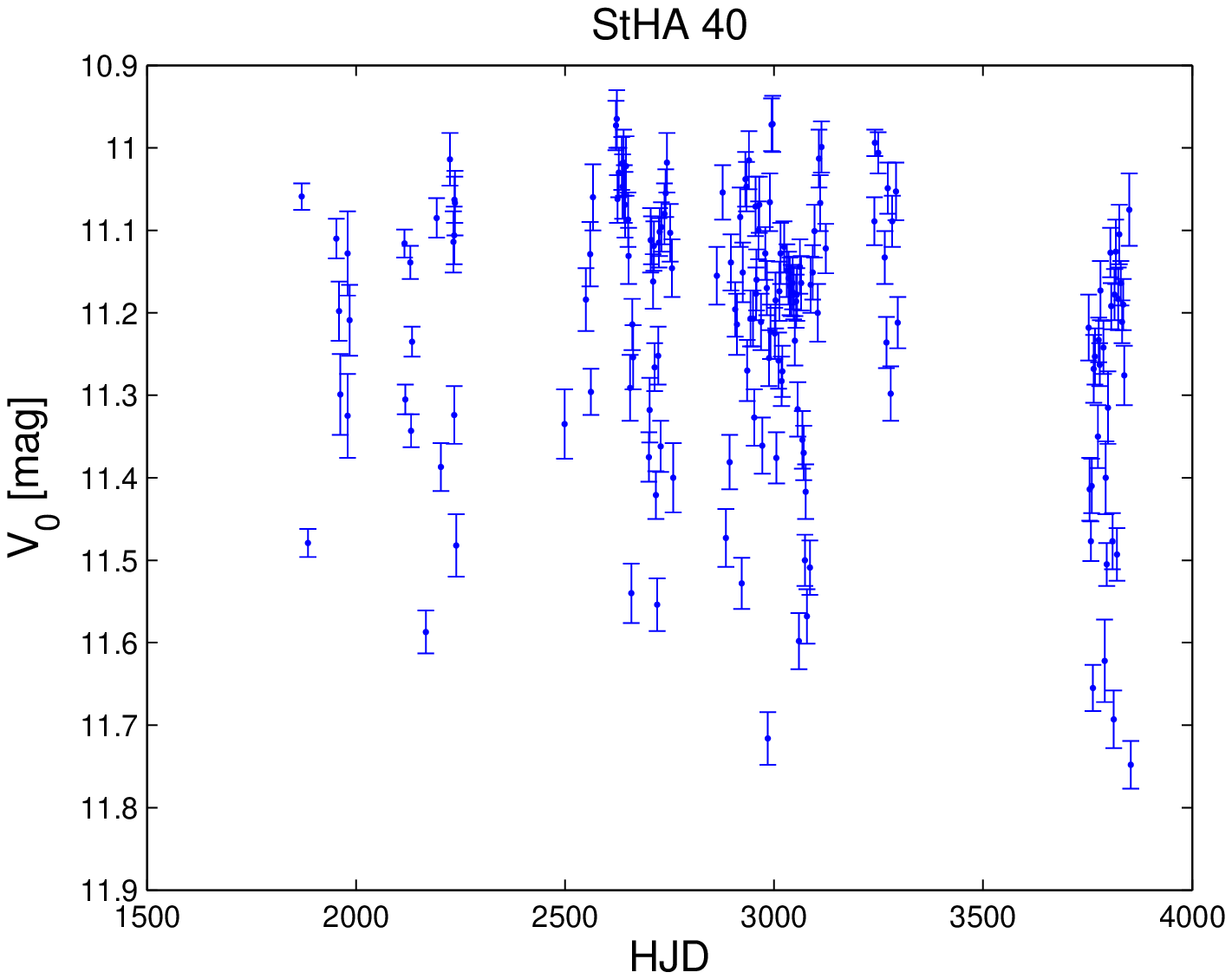}
\caption{Same as Fig.~\ref{lightsaber.01-14} but for identified variables No.~65 
to~78.}
\label{lightsaber.65-78}
\end{figure*}

\end{document}